\documentclass[12pt]{iopart}
\usepackage{iopams}  
\usepackage{hyperref}
\usepackage{doi}
\usepackage{mathrsfs}
\usepackage{enumerate} 

\usepackage{graphicx}
\usepackage{verbatim}
\usepackage{bm}
\usepackage{ntheorem}
\usepackage{float}
\usepackage{subcaption}
\usepackage{color}
\usepackage{url}
\usepackage{doi}
\usepackage{cite}

\newcommand{\n}{\bm{n}}

\newcommand{\ev}{\bm{e}}

\newcommand{\pot}{\Phi}
\newcommand{\oct}{\mathbf{A}}
\newcommand{\Oct}{\bm{\mathsf{A}}}
\newcommand{\Dev}{\bm{\mathsf{D}}}
\newcommand{\dev}{\bm{D}}
\newcommand{\Ddev}{\mathbf{D}}

\newcommand{\A}{\oct}
\newcommand{\sym}{\bm{S}}
\newcommand{\skw}{\bm{W}}
\newcommand{\p}{\bm{p}}
\newcommand{\dens}{\varrho}
\newcommand{\vt}{\vartheta}
\newcommand{\vv}{\bm{v}}
\newcommand{\x}{\bm{x}}
\newcommand{\y}{\bm{y}}
\newcommand{\z}{\bm{z}}
\newcommand{\nablas}{\nabla\!_\mathrm{s}}
\newcommand{\potor}{\Phi_\mathrm{o}}
\newcommand{\pott}{\Phi_\mathrm{t}}
\newcommand{\cyl}{\mathcal{C}}

\newcommand{\disk}{\mathscr{D}}
\newcommand{\axis}{\mathscr{A}}

\newcommand{\Centre}{\mathscr{C}}
\newcommand{\tetra}{\mathscr{T}}
\newcommand{\tspace}{\mathcal{T}}
\newcommand{\dd}{\mathrm{d}}
\newcommand{\ricci}{\epsilon}
\newcommand{\reals}{\mathbb{R}}
\newcommand{\sphered}{\mathbb{S}^1}
\newcommand{\sphere}{\mathbb{S}^2}
\newcommand{\spheren}{\mathbb{S}^{n-1}}
\newcommand{\framee}{(\ev_1,\ev_2,\ev_3)}
\newcommand{\trans}{^\mathsf{T}}

\newtheorem{proposition}{Proposition}
\theorembodyfont{\upshape}
\newtheorem{remark}{Remark}

\newlength{\irrl}
\newlength{\irrw}
\newcommand{\irr}[1]{
	\settowidth{\irrl}{\mbox{$\displaystyle #1$}}
	\setlength{\irrw}{0.12ex}
	\mbox{$\hspace{0.2em}
		\stackrel{
			\mbox{$\vphantom{\rule[-5\irrw]{\irrw}{6\irrw}}
				\rule[-4\irrw]{\irrw}{5\irrw}\hspace{-\irrw}
				\rule{\irrl}{\irrw}\hspace{-\irrw}
				\rule[-4\irrw]{\irrw}{5\irrw}$}}
		{\mbox{$\displaystyle #1$}}\hspace{0.2em}$}
}

\newcommand{\ave}[1]{\left\langle{#1}\right\rangle}

\newcommand{\avirt}[2]{\left\langle\irr{{#1}}\right\rangle_{#2}}
\newcommand{\avet}[2]{\left\langle{#1}\right\rangle_{#2}}
\newcommand{\aved}[1]{\avet{#1}{\dens}}

\newcommand{\avird}[1]{\avirt{#1}{\dens}}

\newcommand{\pn}[1]{\p^{\otimes{#1}}}
\newcommand{\xn}[1]{\x^{\otimes{#1}}}

\newcommand{\pk}{\p^{\otimes k}}

\newcommand{\areap}{\area{\p}}
\newcommand{\area}[1]{a({#1})}
\newcommand{\ptwo}{\p\otimes\p}
\newcommand{\pthree}{\p\otimes\p\otimes\p}
\newcommand{\vd}{\bm{d}}
\newcommand{\Q}{\mathbf{Q}}

\newcommand{\xc}{\widehat{\x}}
\newcommand{\xcc}{\widehat{x}}

\newcommand{\xthree}{\x\otimes\x\otimes\x}
\newcommand{\field}{\bm{u}}
\newcommand{\Space}{\mathsf{V}}
\newcommand{\complex}{\mathbb{C}}
\newcommand{\Field}{\mathbb{F}}
\newcommand{\spec}{\mathrm{sp}}
\newcommand{\sgn}{\mathrm{sgn}}
\newcommand{\av}{\bm{a}}
\newcommand{\vn}{\bm{n}}
\newcommand{\grad}{\nabla} 
\newcommand{\bsplay}{\mathbf{D}}
\newcommand{\bend}{\bm{b}}
\newcommand{\gradn}{\grad\n}
\newcommand{\vo}{\vn_{1}}
\newcommand{\curl}{\mathrm{curl}\,}
\newcommand{\tn}[1]{\mathbf{#1}}
\newcommand{\dv}{\mathrm{div}\,}
\newcommand{\dframe}{(\n_1,\n_2,\n)}

\def\({\left(}
\def\){\right)}
\def\[{\left[}
\def\]{\right]} 
\def\a{\alpha}
\def\b{\beta}
\def\ga{\gamma}
\def\de{\delta}
\def\la{\lambda}

\def\eps{\varepsilon}

\def\sse{\subseteq}

\newcommand{\xb}{\bm{x}}
\newcommand{\beq}{\begin{equation}}
\newcommand{\eeq}{\end{equation}}
\newcommand{\pa}{\partial}

\begin{document}

\title{A Review on Octupolar Tensors}

\author{Giuseppe Gaeta}
\address{Dipartimento di Matematica,
	Universit\`a degli Studi di Milano, via Saldini 50, I-20133 Milano
	(Italy)}
\ead{giuseppe.gaeta@unimi.it}
\author{Epifanio G. Virga}
\address{Dipartimento di Matematica, Universit\`a di Pavia,
	via Ferrata 5, I-27100 Pavia (Italy)}
\ead{eg.virga@unipv.it}
\vspace{10pt}

\begin{abstract}
In its most restrictive definition, an \emph{octupolar tensor} is a fully symmetric traceless third-rank tensor in three space dimensions. So great a body of works have been devoted to this specific class of tensors and their physical applications that a review would perhaps be welcome by a number of students. Here, we endeavour to place octupolar tensors into a broader perspective, considering non-vanishing traces and non-fully symmetric tensors as well. A number of general concepts are recalled and applied to either octupolar and higher-rank tensors. As a tool to navigate the diversity of scenarios we envision, we introduce the \emph{octupolar potential}, a scalar-valued function which can easily be given an instructive geometrical representation. Physical applications are plenty; those to liquid crystal science play a major role here, as they were the original motivation for our interest in the topic of this review.   
\end{abstract}

%
%

%
%
%

\section{Introduction}\label{sec:intro}
An \emph{octupolar tensor} $\oct$ usually designates a fully symmetric traceless tensor of rank $3$, possibly in three space dimensions. One may well wonder why such a specific topic should deserve an extended review. Granted that physical applications of such a class of tensors may indeed be many, the question would remain as to whether one should invest time reading such a review.

We offer (what we think are) two good reasons to continue reading. Both concern the perspective adopted here.

First, our perspective is broader than the title suggests. We  review properties of octupolar tensors as pertaining to general tensors of higher ranks and dimensions.
Second, our perspective is open to the many novel results that have been gathered in the last few decades, with an eye to their physical motivation.

Here is how our material is organized. Section~\ref{sec:prelim} contains all preliminary definitions and basic results that should make our presentation nearly self-contained, thus sparing the reader the hurdle of consulting respectable, but often opaque books on tensor algebra. The primary physical motivation behind our interest in the topic of this review rests with liquid crystal science and (especially) the new phases whose description calls for an octupolar tensor. This motivation is also recalled in \sref{sec:prelim}, but not divorced from those arising from other fields of physics.

In \sref{sec:geometric}, we present our geometric approach to octupolar tensors. It is based on the \emph{octupolar potential} $\pot$, a scalar-valued function on the unit sphere amenable to a geometric representation that we find instructive.

The characterization of a generic  octupolar tensor  $\oct$ afforded in \sref{sec:geometric} is backed by a different, fully algebraic approach presented in \sref{sec:algebraic}, where a polynomial of degree $6$ in a \emph{single} variable embodies all properties of $\oct$. Section~\ref{sec:algebraic} also contains new results; its development is  meticulous since a few, not totally irrelevant details were missed in the original literature. This section is finely articulated in minute computational items so as to ease the reader decide which details to skip and which to dwell in.

Section~\ref{sec:extensions} hosts our first extension: we consider the role of non-vanishing traces, mainly phrased in the language of the octupolar potential.

Section~\ref{sec:other_approaches} further widens our scope. We study third-rank non-symmetric tensors, trying to adapt to this general context the octupolar-potential formalism.

In \sref{sec:applications}, we briefly present a number of applications of the theory, ranging from gravitation to liquid crystals, as exemplary fields that could further benefit from the unified approach pursued here.

Finally, in \sref{sec:conclusions}, we outline issues that even a cursory glance at the different perspectives evoked in this review would suggest for future research.

\section{Preliminaries}\label{sec:prelim}
In this section we lay down the basis of our development. We start from a general decomposition of tensors of any rank and in any  dimension, with the aim of providing a solid mathematical justification for seeking special cases in our representations with a reduced number of parameters. Our primary interest lies in third-rank tensors in three dimensions. A noticeable subclass of these  are properly called the \emph{octupolar tensors}, but our terminology will be more flexible on this account.

\subsection{Invariant tensor decomposition}
The set  $\tspace(r,\Space)$  of tensors of rank $r$ in $n$-dimensional space $\Space$ over the field $\Field$ form a vector space of dimension $n^r$. If $G$ is a group of linear transformations in $\Space$, then $\tspace (r,\Space)$ provide a basis for a representation (in general, reducible) $T$ of $G$, $T \sse GL(n^r,\Field)$. By using, e.g., \emph{Young  diagrams} (which give rise to \emph{Young patterns}, or \emph{Young tableaux}), one can decompose such a representation of $GL(n,\Field)$ into irreducible ones. This decomposition is based on the decomposition of representation of the \emph{symmetry} group $S_n$ (the group of permutations of $n$ symbols); in turn, this decomposition can  also be performed with the technique of \emph{Yamanouchi symbols}. There is a one-to-one correspondence between Young diagrams  and Yamanouchi symbols; see, for example, \cite[p.\,221]{hamermesh:group} (general references on tensor algebra and irreducible representations are the classical books \cite{weyl:classical,naimark:theory,boerner:representations}).

A tensor $\Oct\in\tspace(r,\Space)$  transforms (under maps in the base space $\Space$) as the tensor  product  of $r$ vectors, $\x_1\otimes\x_2\otimes\cdots\x_r$. In studying the transformation properties of tensors in concrete terms under a given group action  $G$ in $\Space$, it is often convenient to consider the basis in $\tspace (r,\Space)$ built by taking the product of basis vectors in $\Space$,
\begin{equation}
	\label{eq:matrix_representation}
	\Oct=A_{i_1i_2\dots i_r}\ev_{i_1}\otimes\ev_{i_2}\otimes\cdots\otimes\ev_{i_r},
\end{equation}
where $(\ev_1,\ev_2,\dots,\ev_n)$ is a basis for $\Space$. In \eref{eq:matrix_representation}, and routinely below, we employ the convention of summing over repeated indices. Moreover, if the space $\Space$ is endowed with an inner product, the basis  $(\ev_1,\ev_2,\dots,\ev_n)$ can be taken to be orthonormal, in which case the corresponding scalar components $A_{i_1i_2\dots i_r}$ will also be referred to as Cartesian.

Covariance dictates that the matrix elements for the transformations of $\tspace (r,\Space)$ are homogeneous polynomials of degree $r$ in the matrix elements for the action of the group $G$ in $\Space$.

For second-rank tensors, any $\bm{L}\in\tspace(2,\Space)$ can be decomposed as $\bm{L}= \sym + \skw$, where $\sym\trans=\sym$, $\skw\trans= - \skw$, and a superscript $\trans$ denotes transposition.\footnote{Here $\sym$ stands for ``symmetric'' and $\skw$ for ``skew-symmetric'', synonymous with ``antisymmetric''.} A similar decomposition  exists for tensors $\Oct$ of arbitrary rank and can be  described with the aid of Young diagrams. In terms of the scalar components $A_{i_1i_2\dots i_r}$, with which we shall also identify $\Oct$, these are obtained by arranging $r$ boxes in all possible ways in a texture of rows and columns, with the constraint that each row should not be longer than the preceding one. The boxes represent tensor indices, and the corresponding tensor will be symmetric under permutations exchanging indices on different columns on the same row, and antisymmetric under permutations exchanging indices on different rows on the same column. The latter condition implies that there should not be more than $n$ rows, or the corresponding representation will be trivial (tensors fully antisymmetric in $r > n$ indices, having necessarily at least two equal indices, will automatically vanish). It should be noted that the representations corresponding to Young diagrams obtained from each other by an exchange of rows and columns are \emph{conjugated}; thus, in particular, the maximal number $n$ is such  for both rows and columns (see also Sect.\,7.4 of \cite{hamermesh:group}).

Thus, e.g., for $r=2$ we have 
\begin{equation}\label{eq:young_r=2}
\square \otimes \square  =  \left( \square \, \square \right)\  \oplus \ \left(\begin{array}{ccc} \square \\ \square \end{array}\right) ,
\end{equation}
while for $r=3$ we have 
\begin{equation}\label{eq:young_r=3}
\square \otimes \square \otimes \square = \( \square \, \square \, \square \)  \ \oplus \ \left(\begin{array}{cc}\square & \square \\ \square & \end{array}\right) \ \oplus \ \left(\begin{array}{c}\square \\ \square \\ \square \end{array}\right).
\end{equation}
Clearly, in the case $n=2$, the last diagram will correspond to null tensors.

Following \cite{schouten:ricci}, Weyl \cite{weyl:theory} initiated a fully general theory of decomposition of tensors into irreducible symmetry parts, having especially in mind its application to quantum mechanics. An early description of the role of both Young's diagrams and tableaux can be retraced in \cite{wade:tensor}; here we follow a more recent approach \cite{itin:decomposition}.

A Young tableau is obtained by filling the boxes of Young diagrams as in \eref{eq:young_r=2} or \eref{eq:young_r=3} with indices. Each diagram $\Lambda$ has a corresponding dimension, given by the following \emph{hook} formula:
\begin{equation}
	\label{eq:hook_formula}
	\dim\Lambda=\frac{r!}{\prod_{(\alpha,\beta)\in\Lambda}\mathrm{hook}(\alpha,\beta)}.
\end{equation}
Here $(\alpha,\beta)$ denotes the position of a cell in the diagram: $\alpha$ is the row index, while $\beta$ is the column index. For a cell $(\alpha,\beta)$ in the diagram $\Lambda$, the \emph{hook length} $\mathrm{hook}(\alpha,\beta)$ is the sum of the number of boxes that are in the same row on the right of the cell and the number of boxes in the same column below it plus $1$ (to account for the cell itself).

Denoting by $\Oct^{(p)}$ the tensorial component of $\Oct$ corresponding to the tableau generated by a diagram $\Lambda_p$, its dimension $\dim\Oct^{(p)}$, that is, the number of independent parameters needed to represent it, is given by
\begin{equation}
	\label{eq:hook_dimension}
	\dim\Oct^{(p)}=\prod_{(\alpha,\beta)\in\Lambda_p}\frac{n+\beta-\alpha}{\mathrm{hook}(\alpha,\beta)}.
\end{equation}

\subsubsection{Case of interest.}\label{sec:case_of_interest}
In the case where $r=3$, which will be  of special interest to us, letting $\Lambda_1$, $\Lambda_2$, and $\Lambda_3$ denote orderly the Young diagrams on the right-hand side of \eref{eq:young_r=3}, we easily see that
\begin{equation}
	\label{eq:Young_dimesnsions_r=3}
	\dim\Lambda_1=1,\quad \dim\Lambda_2=2,\quad \dim\Lambda_3=1,
\end{equation}
meaning  that $\Oct$ can be decomposed into three different types of tensors $\Oct^{(p)}$:
\begin{enumerate}[(i)]
	\item\label{case:i} A single $\Oct^{(1)}$, which is fully symmetric;
	\item\label{case:ii}  Two independent components of $\Oct^{(2)}$, $\Oct^{(2,1)}$ and $\Oct^{(2,2)}$, which are partly symmetric;
    \item\label{case:iii}  A single $\Oct^{(3)}$, which is fully antisymmetric.
\end{enumerate}
We shall denote by $\Oct$ a generic tensor of $\tspace(3,\Space)$ and by $A_{ijk}$ its scalar components in a basis $(\ev_1,\ev_2,\dots,\ev_n)$ of $\Space$. We shall reserve the symbol $\oct$ for the special, but important  case where $n=3$.

The three types of tensors outlined above correspond to coefficients $A_{ijk}$ having three types of symmetry under permutations. Specifically, case \eref{case:i} is characterized by 
\begin{equation}
	\label{eq:symmetry_case_i}
	A_{\pi (i,j,k)}^{(1)}=A_{ijk}^{(1)},
\end{equation}
for any permutation $\pi\in S_3$, and case \eref{case:iii} is characterized by
\begin{equation}
	\label{eq:symmetry_case_iii}
	A_{\pi (i,j,k)}^{(3)}=\sgn(\pi)A_{ijk}^{(3)},
\end{equation}
where $\sgn (\pi)$ is the \emph{sign} (or \emph{index}) of the permutation. Finally, tensors under case \eref{case:ii} are characterized by the following mixed symmetry relations (see also \cite{itin:decomposition})
\begin{equation}
	\label{eq:symmetry_case_ii}
	A_{ijk}^{(2,1)} = A_{jik}^{(2,1)}\quad\textrm{and}\quad A_{ijk}^{(2,2)}= A_{kji}^{(2,2)}.
\end{equation}

A general tensor $\Oct\in\tspace(3,\Space)$ can thus be written as the following sum of irreducible tensors with respect to $GL(n^3,\Field)$:
\begin{equation}
	\label{eq:A_as_sum}
	\Oct=\Oct^{(1)}+\Oct^{(2,1)}+\Oct^{(2,2)}+\Oct^{(3)}.
\end{equation}
The Cartesian components of these tensors can be expressed in terms of the components of $\Oct$ as follows,
\begin{eqnarray}
	A_{ijk}^{(1)}&=\frac16\left(A_{ijk}+A_{jki}+A_{kij}+A_{jik}+A_{kji}+A_{ikj}\right),\label{eq:components_A_1}\\
	A_{ijk}^{(2,1)}&=\frac13\left(A_{ijk}+A_{jik}-A_{kji}-A_{kij}\right),\label{eq:components_A_2_1}\\
	A_{ijk}^{(2,2)}&=\frac13\left(A_{ijk}-A_{jik}+A_{kji}-A_{jki}\right),\label{eq:components_A_2_2}\\
	A_{ijk}^{(3)}&=\frac16\left(A_{ijk}+A_{jki}+A_{kij}-A_{jik}-A_{kji}-A_{ikj}\right).\label{eq:components_A_3}
\end{eqnarray}
Moreover, it follows from \eref{eq:hook_dimension} that
\begin{eqnarray}
	\dim\Oct^{(1)}&=\frac16n(n+1)(n+2),\label{eq:dim_A_1}\\
	\dim\Oct^{(2,1)}&=\dim\Oct^{(2,2)}=\frac13n(n+1)(n-1),\label{eq:dim_A_2_2}\\
	\dim\Oct^{(3)}&=\frac16n(n-1)(n-2),\label{eq:dim_A_3}
\end{eqnarray}
which together with \eref{eq:A_as_sum} easily imply that 
\begin{equation}
	\label{eq:dim_A}
	\dim\Oct=n^3.
\end{equation}
While $\Oct^{(1)}$ and $\Oct^{(3)}$ are  (uniquely identified) irreducible components of $\Oct$, as pointed out in \cite{itin:decomposition}, the decomposition $\Oct^{(2)}=\Oct^{(2,1)}+\Oct^{(2,2)}$ is irreducible, but not unique. It is also worth noting that by \eref{eq:components_A_2_1} and \eref{eq:components_A_2_2} 
\begin{equation}
	\label{eq:components_A_2}
	A_{ijk}^{(2)}=\frac13\left(2A_{ijk}-A_{jki}-A_{kij}\right),
\end{equation}
which shows how both fully symmetric and fully antisymmetric parts of $\Oct^{(2)}$ (defined as in \eref{eq:components_A_1} and \eref{eq:components_A_3}, respectively) vanish, in agreement with \eref{eq:symmetry_case_i} and \eref{eq:symmetry_case_iii}.

With a tensor $\Oct$ we shall also associate the scalar field $\pot:\Space\to\Field$ defined as 
\begin{equation}\label{eq:proto_potential}
	\pot:=A_{ijk}x_ix_jx_k,
\end{equation}
where $x_i$ are the components of a vector $\x\in\Space$ in the basis $(\ev_1,\ev_2,\dots,\ev_n)$. 
$\pot$ will also be referred to as the \emph{potential} associated with $\Oct$. It is clear from the foregoing discussion that $\pot$ is only determined by the fully symmetric part $\Oct^{(1)}$ of $\Oct$,\footnote{A further characterization of $\pot$ for a fully symmetric tensor $\Oct$ will be given in \sref{sec:generalized_potential}.}
\begin{equation}
	\label{eq:proto_potential_symmetric}
		\pot=A_{ijk}^{(1)}x_ix_jx_k.
\end{equation}

We shall be especially interested in the case where $\Field=\reals$, $\Space$ is endowed with an inner product, and $r=n=3$; this is the case that identifies a general \emph{octupolar tensor} $\oct$.\footnote{The reason for this name will become clearer shortly below, see \sref{sec:motivation}.} Correspondingly, the potential in \eref{eq:proto_potential} will be called the \emph{octupolar potential}. In this special case, equations  \eref{eq:dim_A_1}, \eref{eq:dim_A_2_2}, and \eref{eq:dim_A_3} deliver
\begin{equation}
	\label{eq:dim_A_special}
	\dim\oct^{(1)}=10,\quad\dim\oct^{(2,1)}=\dim\oct^{(2,2)}=8,\quad\dim\oct^{(3)}=1
\end{equation}
and $\pot$ can be explicitly written as
	\begin{equation} \label{eq:A_components_full_symmetry}
		\eqalign{\pot &= A_{111} x_1^3+3(  A_{112}x_2+ 
		A_{113}x_3) x_1^2 \\
		&+3\left( A_{122} x_2^2+2
		 A_{123} x_2x_3+ 
		A_{133}x_3^2\right) x_1
		+A_{222}x_2^3 +3A_{223}x_2^2 x_3 \\
		&+ 3 A_{233}x_2x_3^2 +A_{333}x_3^3,}
	\end{equation}
which displays the  $10$ real parameters that represent $\oct^{(1)}$.

A recurrent case  is that of an octupolar tensor symmetric in all indices and with all vanishing partial traces. Strictly speaking, this is the case which the name \emph{octupolar tensor} should be reserved for, but here we shall adopt a more flexible terminology, occasionally denoting as \emph{genuine} the octupolar tensors in their strictest definition. Such  tensors feature $7$ independent; this is the simplest of all octupolar tensors with  a physical relevance and can be fully characterized by a variety of methods, elaborated upon in sections~\ref{sec:geometric} and \ref{sec:algebraic} below. The more general case of a fully symmetric tensor will be analyzed in \sref{sec:extensions}.

The octupolar potential $\pot$ is a homogeneous polynomial of degree $3$ over $\Space$; its values are thus completely determined by its restriction onto the \emph{unit} sphere $\sphere$, where $\pot$ can be properly defined. Occasionally, to reflect this restriction, we shall pass to
 spherical coordinates
\begin{equation}
	\label{eq:spherical_ccordinates}
	x_1 = r \cos \theta \cos \phi,\quad x_2 = r \sin \theta \cos \phi,\quad  x_3 = r \sin \phi, 
\end{equation}
where $r \in  (0,\infty)$, $\theta \in (0,2 \pi)$, $\phi \in [-\pi/2,\pi/2]$, or we shall  explicitly  represent one hemisphere of $\sphere$, writing, for example,
\begin{equation}
	\label{eq:hemisphere_representation}
	x_3 = \pm \sqrt{1 - x_1^2 - x_2^2}.
\end{equation}
This, however, is not the only decomposition of $\sphere$ in halves that shall be considered in the following.

\subsection{Orthogonal irreducible decomposition}
An alternative way to represent a tensor $\Oct\in\tspace(r,\Space)$ is by decomposing it in orthogonal irreducible tensors of rank $r$. It is known from the theory of group representation (see, for example, \cite{boerner:representations}) that $\Oct$ can be expanded as a direct sum of traceless symmetric tensors. Following \cite{zou:orthogonal}, we can formally write
\begin{equation}
	\label{eq:A_decomposition}
	\Oct=\Dev^{(r)}+J_1\Dev^{(r-1)}+J_2\Dev^{(r-2)}+\dots+J_{r-1}\Dev^{(1)}+J_r\Dev^{(0)},
\end{equation}
where $J_r\Dev^{(0)}$ collectively denotes the direct sum of $J_r$ scalars $\Dev^{(0)}$, $J_{r-1}\Dev^{(1)}$ the direct sum of $J_{r-1}$ vectors $\Dev^{(1)}$, and $J_{r-m}\Dev^{(m)}$ denotes the direct sum of $J_{r-m}$ traceless symmetric (deviatoric) tensors $\Dev^{(m)}$ of rank $m$.\footnote{\emph{Deviatoric} also means deprived of all traces, made traceless.} In particular,
\begin{equation}
	\label{eq:D_r_defonition}
	\Dev^{(r)}:=\irr{\Oct},
\end{equation}
where the brackets $\irr{\cdots}$ denote the irreducible, fully symmetric and completely traceless part of the tensor they are applied to, a notation that we borrow from \cite{hess:tensors}.\footnote{We also learn from \cite{hess:tensors} (see p.~34) that the symbol $\irr{\cdots}$, used to indicate the irreducible part of a tensor, was first introduced by Waldmann in \cite{waldmann:formale} for second-rank tensors and later extended to higher-rank tensors in \cite{hess:kinetic}.}

Ways for obtaining explicitly the decomposition in \eref{eq:A_decomposition} for tensors of any given finite rank can be found in \cite{spencer:generating,zheng:irreducible,zou:orthogonal}. A special case arises when $\Oct$ is fully symmetric, as \eref{eq:A_decomposition} simplifies into
\begin{eqnarray}
	\label{eq:A_decomposition_symmetric}
	\Oct=\irr{\Oct}+\Dev^{(r-2)}+\Dev^{(r-4)}+\dots,
\end{eqnarray}
meaning that there is a single deviatoric tensor for each rank $r-m$ that takes part in the decomposition of $\Oct$.
\begin{remark}
	\label{rmk:invariants}
	The decompositions in \eref{eq:A_decomposition} and \eref{eq:A_decomposition_symmetric} are especially instrumental to the search for the representation formulae of isotropic tensor functions (either scalar- or tensor-valued). A valuable review on this topic can be found in \cite{zheng:theory}. Although we are not especially concerned here with tensor invariants, it is worth heeding the result proved in \cite{smith:isotropic} to the effect that the isotropic integrity basis of a third-rank traceless symmetric tensor $\oct$ in three space dimensions consist of $4$ invariants. We shall see in \sref{sec:other_approaches} how this result can also be given a simple direct proof.
\end{remark}
\subsubsection{Case of interest.} In the case where $r=n=3$, which is where our main focus lies, decomposition \eref{eq:A_decomposition} has a classical explicit form \cite{jerphagnon:invariants,jerphagnon:description}, which we now reproduce for completeness, although we shall mostly be concerned in the rest of this review with genuine octupolar tensors in three dimensions, for which \eref{eq:A_decomposition} reduces to a single term.

Let $A_{ijk}$ denote the Cartesian components of a generic third-rank tensor $\oct$ in the orthonormal basis $\framee$. Following \cite{coope:irriducible}, we introduce a scalar $A$ defined as
\begin{equation}
	\label{eq:A_scalar}
	A:=\ricci_{ijk}A_{ijk},
\end{equation}
where $\ricci_{ijk}$ is Ricci's alternator. Similarly, we define three vectors $\vv^{(1)}$, $\vv^{(2)}$, and $\vv^{(3)}$, whose Cartesian components, denoted as $v_i^{(1)}$, $v_i^{(2)}$, and $v_i^{(3)}$, are given by
\begin{equation}
	\label{eq:v_s_components}
	v_i^{(1)}:=A_{ijj},\quad v_i^{(2)}:=A_{jij},\quad v_i^{(3)}:=A_{jji}.
\end{equation}
Two symmetric traceless (deviatoric) second-rank tensors $\dev^{(1)}$ and $\dev^{(2)}$ are also identified, whose Cartesian components are expressed as follows in terms of $A_{ijk}$,
\begin{eqnarray}
	D^{(1)}_{ij}&:=\frac12\left(\ricci_{iml} A_{mlj}+\ricci_{jml} A_{mli}\right)-\frac13A\delta_{ij},\label{eq:D_1_components}\\
	D^{(2)}_{ij}&:=\frac12\left(A_{iml}\ricci_{mlj}+A_{jml}\ricci_{mli}\right)-\frac13A\delta_{ij},\label{eq:D_2_components},
\end{eqnarray}
where $\delta_{ij}$ is Kronecker's symbol. The decomposition in \eref{eq:A_decomposition} can then be written as
\begin{equation}
	\label{eq:A_decomposition_special}
	\oct=\Ddev^{(3)}+\Ddev_1^{(2)}+\Ddev_2^{(2)}+\Ddev_1^{(1)}+\Ddev_2^{(1)}+\Ddev_3^{(1)}+\Ddev^{(0)},
\end{equation} 
where the Cartesian components of the third-rank tensors $\Ddev_i^{(j)}$ ar explicitly given by
\begin{eqnarray}
	D^{(0)}_{ijk}&=\frac16A\ricci_{ijk},\label{eq:D_0_components}\\
	D^{(1)}_{1,ijk}&=\frac{1}{10}\left(4v_i^{(1)}\delta_{jk}-\delta_{ik}v_j^{(1)}-\delta_{ij}v_k^{(1)}\right),\label{eq:D_1_1_components}\\
	D^{(1)}_{2,ijk}&=\frac{1}{10}\left(-v_i^{(2)}\delta_{jk}+4\delta_{ik}v_j^{(2)}-\delta_{ij}v_k^{(2)}\right),\label{eq:D_1_2_components}\\
	D^{(1)}_{3,ijk}&=\frac{1}{10}\left(-v_i^{(3)}\delta_{jk}-\delta_{ik}v_j^{(3)}+4\delta_{ij}v_k^{(3)}\right),\label{eq:D_1_3_components}\\
	D^{(2)}_{1,ijk}&=\frac13\left(2\ricci_{ijl}D^{(1)}_{lk}+D^{(1)}_{il}\ricci_{ljk}\right),\label{eq:D_2_1_components}\\
	D^{(2)}_{2,ijk}&=\frac13\left(2\ricci_{ijl}D^{(2)}_{lk}+D^{(2)}_{il}\ricci_{ljk}\right),\label{eq:D_2_2_components}\\
	D^{(3)}_{ijk}&=\irr{A}_{ijk}.\label{eq:D_3_components}
\end{eqnarray} 
\begin{remark}\label{rmk:direct_sum}
It is clear from this explicit representation of the third-rank tensors $\Ddev_i^{(j)}$ how they may fail to be symmetric, although they result from the direct sum of traceless symmetric tensors of lower rank.
\end{remark}	
\begin{remark}
	\label{rmk:D_3_components}
	By letting 
\begin{eqnarray}
	A_{(ijk)}:=\frac16\left(A_{ijk}+A_{jki}+A_{kij}+A_{kji}+A_{jik}+A_{ikj}\right),	\label{eq:A_symmetrized_components}\\
	V_i:=\frac13\left(v_i^{(1)}+v_i^{(2)}+v_i^{(3)}\right),\label{eq:V_components}
\end{eqnarray}
we can easily give $D^{(3)}_{ijk}$ in \eref{eq:D_3_components} the explicit form
\begin{equation}
	\label{eq:D_3_components_explicit}
	D^{(3)}_{ijk}=A_{(ijk)}-\frac15V_it_{ijk}\quad(\textrm{no sum on\ } i),
\end{equation}
where the symbol $t_{ijk}$ is defined as follows
\begin{equation}
	\label{eq:t_symbol_definition}
	t_{ijk}:=\cases{3&for $i=j=k$,\\1&when two indices are equal,\\0&otherwise.}
\end{equation}
\end{remark}
\begin{remark}
	\label{rmk:independent_parameters}
	It is easily seen that the representation of $\oct$ in \eref{eq:A_decomposition_special} depends on $27$ independent parameters, as it should: one is $A$, $9$ come from the components of the vectors $\vv$'s and $10$ from the components of the symmetric traceless second-rank tensors $\dev$'s; finally, only $7$ are hidden in $\Ddev^{(3)}$.
\end{remark}
\begin{remark}
	\label{rkm:independent_parameters_potential}
	Since $D^{(0)}_{ijk}$ is antisymmetric in the exchange of all indices,  $D^{(0)}_{ijk}x_ix_jx_k=0$ for all $\x\in\sphere$, and similarly vanish both $D^{(2)}_{1,ijk}x_ix_jx_k$ and $D^{(2)}_{2,ijk}x_ix_jx_k$. Moreover,
	\begin{equation}
		(D^{(1)}_{1,ijk}+D^{(1)}_{2,ijk}+D^{(1)}_{3,ijk})x_ix_jx_k=\frac35V_ix_i.
	\end{equation}
Thus, building the potential $\pot$ defined in \eref{eq:proto_potential} for the tensor $\oct$ expressed as in \ref{eq:A_decomposition_special} results in a function depending only on $10$ independent parameters out of the $27$ present in $\oct$, in accord with \eref{eq:proto_potential_symmetric}: here $7$ are needed for $\Ddev^{(3)}$ and $3$ for $\bm{V}$.
\end{remark}
\begin{remark}\label{rmk:A_partial_simmetries_1}
If $A_{ijk}$ enjoys the partial symmetry $A_{ijk}=A_{ikj}$, it follows from \eref{eq:A_scalar}, \eref{eq:v_s_components}, and \eref{eq:D_2_components} that $A=0$, $\vv^{(2)}=\vv^{(3)}$, and $\dev^{(2)}=\bm{0}$. The number of independent parameters in the decomposition \eref{eq:A_decomposition_special} then reduces to $18$: $6$ are the components of the $\vv$'s, $5$ the components of $\dev^{(1)}$, and $7$ those of $\Ddev^{(3)}$. We can also write explicitly the Cartesian components of $\oct$ as follows:
\begin{equation}
	\label{eq:A_partial_symmetries_1}
	\eqalign{A_{ijk}&=\irr{A}_{ijk}+\frac13\left(2\ricci_{ijl}D_{lk}+D_{il}\ricci_{ljk}\right)\\
	&+\frac{1}{10}\left(4u_i\delta_{jk}-u_j\delta_{ik}-u_k\delta_{ij}\right)
	+\frac{1}{10}\left(-2v_i\delta_{jk}+3v_j\delta_{ik}+3v_k\delta_{ij}\right),}
\end{equation}
where $u_i=A_{ikk}$ and $v_i=A_{jij}=A_{jji}$ are the components of $\vv^{(1)}$ and $\vv^{(2)}=\vv^{(3)}$, respectively, and
\begin{equation}
	\label{eq:D_components}
	D_{ij}=\frac12\left(\ricci_{iml}A_{mlj}+\ricci_{jml}A_{mli}\right)
\end{equation}
are the components of $\dev^{(1)}$. Similar expressions for $A_{ijk}$ in this case can also be found in \cite{zou:orthogonal,zou:maxwell}.
\end{remark}
\begin{remark}
	\label{rmk:A_partial_symmetries_2}
In the fully symmetric case, where $A_{ijk}=A_{ikj}=A_{jik}$, $A=0$, all vectors $\vv^{(i)}$ are one and the same $\vv$, and both $\dev^{(1)}$ and $\dev^{(2)}$ vanish, so that \eref{eq:A_decomposition_special} reduces to
\begin{equation}
	\label{eq:A_partial_symmetries_2}
	\oct=\Ddev^{(3)}+\Ddev^{(1)}.
	\end{equation}
Here $\Ddev^{(1)}=\Ddev^{(1)}_1+\Ddev^{(1)}_2+\Ddev^{(1)}_3$ and, in explicit components,
\begin{equation}
	\label{eq:A_partial_symmetries_2_components}
	A_{ijk}=\irr{A}_{ijk}+\frac15\left(v_i\delta_{jk}+v_j\delta_{ik}+v_k\delta_{ij}\right),
\end{equation}
where $v_i=A_{ijj}$ are the components of $\vv$, in agreement with the general  \emph{detracer} operator introduced in \cite{applequist:fundamental,applequist:traceless}.
\end{remark}

\subsection{Generalized eigenvectors and eigenvalues}\label{sec:generalized}
For tensors of rank higher than $2$, the very notion of eigenvectors and eigenvalues is \emph{not} universally accepted and, what is worse for our purposes, for these tensors no analogue is known of the Spectral Theorem, which characterizes symmetric second-rank tensors in terms of their eigenvectors and eigenvalues. Different notions of generalized eigenvectors and eigenvalues have been proposed for not necessarily symmetric  tensors of rank $r>2$ in a general $n$-dimensional space $\Space$. The one we adopt below has been put forward and studied in \cite{qi:eigenvalues,qi:eigenvalues_2007,ni:degree}; it has also been enriched by a theorem \cite{cartwright:number} that estimates the cardinality of eigenvalues.\footnote{An approach alternative to the one followed here has been pursued in \cite{zheng:eigenvalue}.}

For definiteness, here we shall take $\Space$ endowed with an inner product (denoted by the symbol $\cdot$) over the field $\Field=\complex$. 
Letting $\xn{r}$ be the member of $\tspace(r,\Space)$ defined by the multiple tensor product
 \begin{equation}\label{eq:x_r_definition}
 	\xn{r}:=\underbrace{\x\otimes\cdots\otimes\x}_{r\ \mathrm{times}}
 \end{equation}
and following \cite{qi:eigenvalues,qi:eigenvalues_2007}, for a tensor $\Oct\in\tspace(r,\Space)$, we define $\Oct\x^{r-1}:=\Oct\cdot\xn{(r-1)}$, which is the vector in $\Space$ with Cartesian components
\begin{equation}\label{eq:Ax_m-1_definition}
	\left(\Oct\x^{r-1}\right)_i:=A_{ii_2\dots i_r}x_{i_2}\dots x_{i_r},
\end{equation}
where $A_{i_1i_2\dots i_r}$ are the Cartesian components of $\Oct$ relative to a prescribed, orthonormal  basis $(\ev_1,\ev_2,\dots,\ev_n)$ of $\Space$, as in \eref{eq:matrix_representation}.
The solutions $\x\in\Space$ and $\lambda\in\complex$ of the non-linear problem
\begin{equation}\label{eq:generalized_eigenvalue_problem}
	\Oct\x^{r-1}=\lambda\x
\end{equation}
such that
\begin{equation}\label{eq:unit_sphere_constraint}
	\x\cdot\x=1
\end{equation}
are a (generalized) \emph{eigenvector} $\x$ of $\Oct$ and the associated (generalized) eigenvalue $\lambda$.\footnote[1]{In the traditional case, where $r=2$ and equation \eref{eq:generalized_eigenvalue_problem} becomes linear, the normalization condition \eref{eq:unit_sphere_constraint} is virtually superfluous. It is far from being so in the present non-linear context.}

$\Oct$ is said to be \emph{real} if all its Cartesian components are real.
A solution $(\lambda,\xc)$ of \eref{eq:generalized_eigenvalue_problem} and \eref{eq:unit_sphere_constraint} is also collectively called a (generalized) \emph{eigenpair} of $\Oct$.\footnote{Often the (generalized) eigenvectors  defined above are also said to be \emph{normalized}, as they are required to satisfy the constraint \eref{eq:unit_sphere_constraint}. We do not consider here non-normalized eigenvectors, as others do, and so we need not that appellation. Similarly, whenever no ambiguity can arise, we also omit the adjective ``generalized'' in referring to the solutions of \eref{eq:generalized_eigenvalue_problem} and \eref{eq:unit_sphere_constraint} and we simply call them the eigenvectors and eigenvalues of $\Oct$.}

A number of facts have been established about the eigenvectors of a generic tensor $\Oct$. Below, we recall from \cite{cartwright:number} those which are more relevant to our pursuit.
\begin{enumerate}[(1)]
	\item It should be noted that eigenpairs $(\lambda,\xc)$ come in equivalence classes. Letting $\x'=t\xc$ and $\lambda'=t^{r-2}\lambda$ with $t^2=1$, it is readily seen that  $(\lambda',\x')$ is an eigenpair whenever $(\lambda,\xc)$ is so. We shall consider both $(\lambda,\xc)$ and $(\lambda',\x')$ as members of one and the same equivalence class.
	\item The \emph{spectrum} $\spec(\Oct)$ of all eigenvalues of $\Oct$ is either finite or it consists of all complex numbers in the complement of a finite set. If $\spec(\Oct)$ is finite and $r>2$, then the number $d$ of equivalence classes of eigenvalues in $\spec(\Oct)$ (counted with their multiplicity) is given by
	\begin{equation}\label{eq:d_formula}
		d=\frac{(r-1)^n-1}{r-2}.
	\end{equation}
	\item If $\Oct$ is \emph{real} and either $r$ or $n$ is odd, then $\A$ has at least one real eigenpair.
	\item Every fully \emph{symmetric} tensor $\Oct$ (as under case \eref{case:i} above) has \emph{at most} $d$ distinct (equivalence classes of) eigenvalues. Moreover, this bound is indeed attained for \emph{generic} fully symmetric tensors $\Oct$.\footnote{Here as in \cite{cartwright:number}, \emph{generic} is meant in the sense of algebraic geometry, that is, ``there exists a polynomial in the components of $\Oct$ such that the asserted conclusion holds for all tensors $\Oct$ at which that polynomial does not vanish.'' This is a way of making precise expressions such as ``in most cases'' or ``nearly always.'' }
\end{enumerate}

\subsubsection{Generalized potential.}\label{sec:generalized_potential}
A potential $\pot$ that generalizes \eref{eq:proto_potential_symmetric} can be defined for a fully symmetric tensor $\Oct\in\tspace(r,\Space)$ as
\begin{equation}
	\label{eq:generalized_potential}
	\pot(\x):=\Oct\cdot\xn{r}=A_{i_1i_2\dots i_r}x_{i_1}x_{i_2}\dots x_{i_r},
\end{equation}
which is a (complex) homogeneous polynomial of degree $r$. Differentiating $\pot$ with respect to $\x$, we easily see from \eref{eq:Ax_m-1_definition} that
\begin{equation}
	\label{eq:nabla_pot}
	\nabla\pot(\x)=r\Oct\x^{r-1}.
\end{equation}
If $\xc$ is a generalized eigenvector of $\Oct$ with eigenvalue $\lambda$, then it follows from \eref{eq:nabla_pot} that 
\begin{eqnarray}
	\label{eq:nabla_pot_crit}
	\nabla\pot(\xc)=r\lambda\xc,
\end{eqnarray}
which, by \eref{eq:unit_sphere_constraint} and Euler's theorem on homogeneous functions, implies that
\begin{equation}
	\label{pot_crit_value}
	\pot(\xc)=\lambda.
\end{equation}
Thus, the eigenvalues of $\Oct$ are the values taken by the potential $\pot$ on the corresponding eigenvectors in the unit sphere $\spheren$ of $\Space$. 
Conversely, the critical points of $\pot$ in $\spheren$ satisfy the parallelism condition
\begin{equation}
	\nabla\pot\parallel\x,
\end{equation} 
which by \eref{eq:nabla_pot} is equivalent to  \eref{eq:generalized_eigenvalue_problem}. Thus, all the eigenvectors of $\Oct$ are characterized as critical points of $\pot$ in $\spheren$ and the corresponding eigenvalues are given by the values attained there by $\pot$.

\begin{remark}\label{rem:potential}
For $\Field=\reals$, if $\Oct$ is real and symmetric, then $\pot$ is a real-valued polynomial. Its critical values and critical points in $\spheren$ are all the generalized real eigenpairs of $\Oct$, whose number can be far less than $d$ given in \eref{eq:d_formula}.
\end{remark}
\begin{remark}
Although a potential $\pot$ can also be associated with a partly symmetric tensor $\Oct$ as in \eref{eq:generalized_potential}, its critical points in $\spheren$ can no longer be interpreted as generalized eigenvectors of $\Oct$ according to definition \eref{eq:generalized_eigenvalue_problem}, but just as those of the fully symmetric part $\Oct^{(1)}$ of $\Oct$ defined by extending  \eref{eq:components_A_1}. In the rest of this review, we shall lay special emphasis on fully symmetric tensors, so that their eigenvectors can  be identified with the critical points of $\pot$ (and their generalized eigenvalues with the corresponding critical values).
\end{remark}	

\subsubsection{Case of interest.}We shall tackle in detail the case where $r=3$ and $n=3$, so that by \eref{eq:d_formula} $d=7$. If a tensor $\A\in\tspace(3,\Space)$ with $\textrm{dim}\,\Space=3$ is both real and symmetric, we are assured that it possesses at most $7$ distinct (equivalence classes of) complex eigenvalues, of which at least $1$ is real. The analysis performed in sections~\ref{sec:geometric} and \ref{sec:algebraic} will actually reveal more than  the general facts recalled above would lead us to expect. For example, we shall see that the distinct real eigenvalues of $\A$ are never less than $5$, but they can be less than $7$ in a generic fashion.

In the following section, we  pause briefly to illustrate the physical meanings that a general octupolar tensor $\oct$ can have, both in the symmetric and non-symmetric cases.

\subsection{Physical motivation}\label{sec:motivation}
\emph{Octupolar order} in soft matter physics is not just an exotic mathematical curiosity.
 Our main physical motivation for this review lies in the theory of liquid crystals, especially in connection with  the recently discovered \emph{polar} nematic phases \cite{lavrentovich:ferroelectric,madhusudana:simple,mandle:new,sebastian:ferroelectric}. 
 This is why we start from liquid crystals to illustrate the physical background of the mathematical theory. 

\subsubsection{Generalized nematic phases.} Liquid crystals  provide a noticeable case of soft ordered materials for which a \emph{quadrupolar order} tensor may not suffice to capture the complexity of the condensed phases they can exhibit.

\begin{figure}[h]
	\centering
	\includegraphics[width=.4\linewidth]{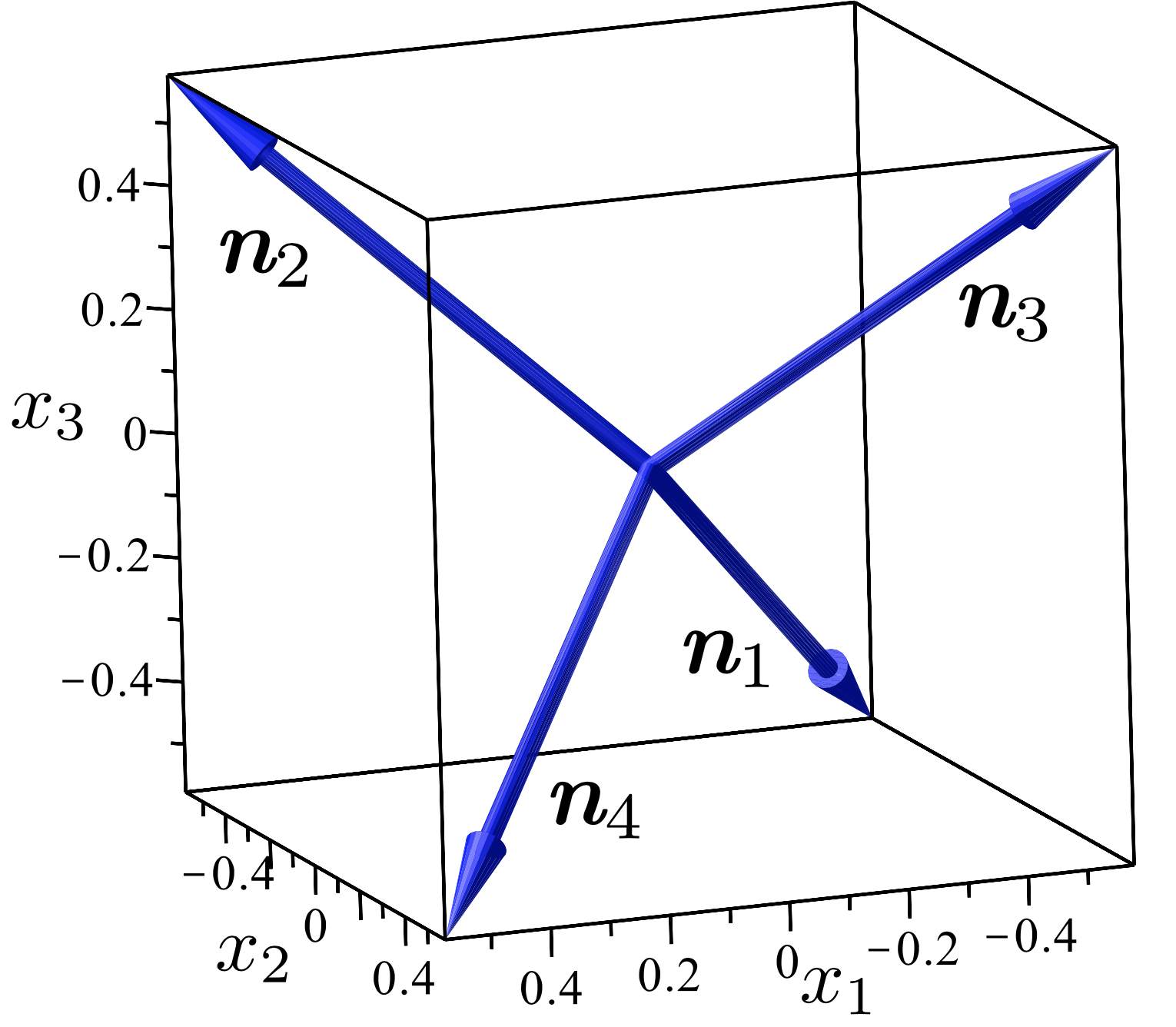}
	\caption{The tetrahedral unit vectors $\n_\alpha$ defined in \eref{eq:tetrahedral_vectors} and featuring in \eref{eq:A_proto_definition}.}
	\label{fig:tetrahedron}
\end{figure}

After some earlier theoretical attempts to describe \emph{tet\-ra\-he\-dratic} nematic phases \cite{fel:tetrahedral,fel:symmetry}, it was established \cite{radzihovsky:fluctuation,lubensky:theory,brand:flow} that the phases observed experimentally in liquid crystals composed of bent-core molecules \cite{niori:distinct,link:spontaneous} could be described by means of an additional fully symmetric, completely traceless, third-rank order tensor $\oct$. 
\footnote{An excluded-volume theory to this effect is presented in \cite{bisi:excluded}. For the role played by octupolar tensors in representing steric interactions and ``shape polarity'', the reader could also consult the works \cite{piastra:octupolar,piastra:onsagerian,piastra:explicit}.}
Intuition was rooted in representing $\A$ as the following ensemble average,
\begin{equation}\label{eq:A_proto_definition}
	\A=\ave{\sum_{\alpha=1}^4\n_\alpha\otimes\n_\alpha\otimes\n_\alpha},
\end{equation}
where the \emph{tetrahedral vectors} $\n_\alpha$ are the unit vectors directed from the centre of a (microscopic) tetrahedron to its vertices, as shown in \fref{fig:tetrahedron} \cite{pleiner:low,pleiner:tetrahedral},
\begin{equation}
	\label{eq:tetrahedral_vectors}
		\cases{
	\n_1=-\frac{1}{\sqrt{3}}(\ev_1+\ev_2+\ev_3),\quad\n_2=\frac{1}{\sqrt{3}}(\ev_1-\ev_2+\ev_3),\\
	\n_3=\frac{1}{\sqrt{3}}(-\ev_1+\ev_2+\ev_3),\quad\n_4=\frac{1}{\sqrt{3}}(\ev_1+\ev_2-\ev_3),
}
\end{equation}
where $(\e_1,\e_2,\e_3)$ is a Cartesian frame.
\begin{remark}
Since $\sum_{\alpha=1}^4\n_\alpha=\bm{0}$, it is easy to see that $\oct$ in \eref{eq:A_proto_definition} is a symmetric traceless octupolar tensor.
\end{remark}
\begin{remark}\label{rmk:A_longa}
Alternatively, in a series of papers \cite{longa:chiral,longa:generalized,trojanowski:tetrahedratic,longa:ambidextrous,trojanowski:modulated} on generalized nematic phases (both achiral and chiral) the octupolar order tensor $\oct$ was defined as
\begin{equation}
	\label{eq:A_longa}
	\oct=\frac{1}{\sqrt{6}}\ave{\sum_{\pi\in S_3}\ev_{\pi(1)}\otimes\ev_{\pi(2)}\otimes\ev_{\pi(3)}},
\end{equation}		
where the sum is extended to all permutations in $S_3$. It is a simple exercise to show that, despite appearances, the tensors in \eref{eq:A_longa} and \eref{eq:A_proto_definition} are proportional to one another.
\end{remark}
This would suggest that $\A$ should partly preserve the parent tetrahedral symmetry and be somehow associated with \emph{four} directions in space. Such a supposition would also be supported by the analysis in \cite{virga:octupolar_2D}, which showed that in two space dimensions $\A$ is indeed geometrically fully described by an equilateral triangle. We shall show in the following sections how this expectation is indeed illusory.

An octupolar tensor arises as an \emph{order tensor} in the description of the orientational distribution of a microscopic polar axis $\p$. This is especially relevant to the study of generalized liquid crystals, including polar nematic phases.

A probability density $\dens$ over the unit sphere $\sphere$ can be represented by Buckingham's formula \cite{buckingham:angular} as
\begin{equation}\label{eq:buckingham_formula}
	\dens(\p)=\frac{1}{4\pi}\left(1+\sum_{k=1}^\infty\frac{(2k+1)!!}{k!}\avird{\pn{k}}\cdot\pn{k} \right),
\end{equation}
where, much in the spirit of \cite{zannoni:distribution}, $\avird{\pn{k}}$ is the \emph{multipole average} corresponding to the multiple tensor product $\pk$ (see \cite{turzi:cartesian}). A combinatoric proof of \eref{eq:buckingham_formula} can be found in \cite{gaeta:octupolar}. Collectively, the multipole averages are \emph{order tensors} of increasing rank that decompose $\dens$. In \eref{eq:buckingham_formula},  $\otimes$ denotes (as above) tensor product, and $\aved{\cdots}$ is the \emph{ensemble} average associated with $\dens$,
\begin{equation}\label{eq:ensemble_average_definition}
	\aved{\cdots}:=\frac{1}{4\pi}\int_{\sphere}(\cdots)\dens(\p)\dd\areap.
\end{equation}
Especially, the first three multipole averages play a role in resolving the characteristic features of $\dens$: they are the \emph{dipolar}, \emph{quadrupolar}, and \emph{octupolar} order tensors defined by
	\begin{eqnarray}
		\vd:=&\aved{\p},\quad \Q:=\avird{\ptwo},\label{eq:dipole_quadrupole}\\
		\A:=&\avird{\pthree},\label{eq:octupole}
	\end{eqnarray}
respectively.\footnote{Other computational definitions of scalar order parameters for both tetrahedral and cubatic symmetries can also be found in \cite{romano:computer:06,romano:computer}.}

Here, we shall focus on the octupolar order tensor $\A$. In accordance with \eref{eq:matrix_representation}, in a Cartesian frame $\framee$, the tensor $\A$ is represented as
\begin{equation}\label{eq:A_tensor_representation}
	\A=A_{ijk}\,\ev_i\otimes\ev_j\otimes\ev_k,
\end{equation}
where by \eref{eq:octupole} the coefficients $A_{ijk}$ fall under case \eref{case:i} above and obey the following properties, see \eref{eq:symmetry_case_i}:
\begin{equation}
		A_{ijk}=A_{jik}=A_{ikj},\ \forall\ i,j,k,\qquad
		A_{iik}=A_{iki}=A_{kii}=0,\ \forall\ k.
\end{equation}
As already remarked, combined together, these properties reduce to $7$ the number of independent parameters needed to represent in a generic frame all possible octupolar order tensors $\A$. For definiteness, we shall adopt the following definitions:
	\begin{eqnarray}
		\cases{
		\alpha_0:=A_{123},\\
		\alpha_1:=A_{111},\quad\alpha_2:=A_{222},\quad\alpha_3:=A_{333}, \\
		\beta_1:=A_{122},\quad\beta_2:=A_{233},\quad\beta_3:=A_{311},}
	\label{eq:parametrization_alpha_0_beta}
	\end{eqnarray}
	so that
	\begin{equation}\label{eq:parametrization_alpha_beta}
		A_{133}=-(\alpha_1+\beta_1),\quad A_{211}=-(\alpha_2+\beta_2),\quad A_{322}=-(\alpha_3+\beta_3).
	\end{equation}

Given the number of scalar coefficients needed to represent $\A$ in a generic Cartesian frame, one may think to absorb three by selecting a convenient \emph{orienting} frame and let the remaining four describe scalar order parameters with a direct
physical meaning, in complete analogy with what is customary for the second-rank, symmetric and traceless quadrupolar order tensor $\Q$, which is described by five scalar coefficients in a generic frame and characterized by only two scalar order parameters. For $\Q$, the reduction of the scalar coefficients to the essential scalar order parameters is performed by representing $\Q$ in its eigenframe, where only two eigenvalues suffice to characterize it. 

Now the definition of generalized eigenvectors and eigenvalues for $\oct$ recalled in \sref{sec:generalized} above comes in handy. Here, we shall take the equivalent route of representing $\oct$ through the critical points of the \emph{octupolar potential} $\pot$, the scalar-valued function defined  on the unit sphere $\sphere$ as in \eref{eq:generalized_potential}; in this setting,  $\pot$ is nothing but  the octupolar component of Buckingham's formula \eref{eq:buckingham_formula}. Thus, in particular, maxima and minima of $\pot$, with their relative values, would designate the directions in space along which a microscopic polar axis $\p$ is more and less likely to be retraced, respectively, according with the octupolar component of $\dens$. We shall see in \sref{sec:oriented_potential} how to employ the properties of the octupolar potential to reduce the number of independent parameters that represent $\oct$ in the orienting frame.
\begin{remark}\label{rmk:octupolar_vs_quadrupolar}
Such a reduction is meaningful as long as the octupolar component of the probability density $\dens$ can be isolated from the quadrupolar component, so as to be treated independently. Allegedly, this is seldom the case for ordinary liquid crystals, where the quadrupolar component is expected to be dominant. If that is the case, the natural frame for $\oct$ would be the eigenframe of $\Q$, which need not coincide with the orienting frame. In our applications of $\oct$ to liquid crystal science (which are not the only ones considered here), we shall consistently presume that quadrupolar and octupolar effects are separable.		
\end{remark}

The physical motivation illustrated here will primarily guide our intuition below, to the point that we shall often picture the maxima of the octupolar potential as designating an ordered condensed \emph{phase} on its own. Other interpretations are also possible, which do not require $\oct$ to be fully symmetric and traceless, and so cannot uniquely rely on the octupolar potential $\pot$ as defined in \eref{eq:proto_potential}. They are briefly recalled for completeness in the following.

\subsubsection{Non-linear optics.}The optical properties of crystals are described by the constitutive laws linking electromagnetic fields and induced polarizations. In the linear theory, for example, the induced polarization $\bm{P}$ is related to the electric field $\bm{E}$ through the formula
\begin{equation}
	\label{eq:polarization}
	\bm{P}(\omega)=\bm{\chi}^{(1)}\bm{E}(\omega),
\end{equation}
where $\omega$ is the oscillation frequency of the fields and  the linear susceptibility $\bm{\chi}^{(1)}$ is in general represented by a symmetric second-rank tensor.

The lower-order optical non-linearity, such as frequency mixing, arises when the polarization $\bm{P}(\omega_3)$ at frequency $\omega_3=\omega_1+\omega_2$ is related to the electric fields $\bm{E}(\omega_1)$ and $\bm{E}(\omega_2)$ oscillating at frequencies $\omega_1$ and $\omega_2$ through the following quadratic law (see, for example, \cite{jerphagnon:invariants} and Sect.\,1.5 of \cite{boyd:nonlinear}),
\begin{equation}
	\label{eq:frequency_mixing}
	\bm{P}(\omega_1+\omega_2)=\oct(\omega_1,\omega_2)[\bm{E}(\omega_1)\otimes\bm{E}(\omega_2)],
\end{equation} 
where the generic third-rank tensor $\oct$ represents a non-linear susceptibility. In Cartesian components, \eref{eq:frequency_mixing} reads as 
\begin{equation}
	\label{eq:frequency_mixing_components}
	P_i(\omega_1+\omega_2)=A_{ijk}(\omega_1,\omega_2)E_j(\omega_1)E_k(\omega_2).
\end{equation}
In general, for $\omega_1\neq\omega_2$, $A_{ijk}$ need not enjoy any symmetry, as $E_j(\omega_1)$ may differ from $E_k(\omega_2)$. However, for $\omega_1=\omega_2$, which is the case of \emph{second harmonic} generation, we may take $A_{ijk}=A_{ikj}$  in \eref{eq:frequency_mixing_components} with no loss of generality, and $18$ parameters suffice to represent $\oct$.

Moreover, often non-linear optical interactions involve waves with frequency much smaller than the lowest resonance frequency of the material. If this is the case, the non-linear susceptibility $\oct$ is virtually independent of frequency and we can permute all indices in $A_{ijk}$ leaving the response of the material unaltered. This is often called the Kleinman symmetry condition for the tensor $\oct$ \cite{kleinman:nonlinear}. When it applies, $\oct$ is represented by $10$ independent parameters.

\subsubsection{Linear piezoelectricity.}\label{sec:piezoelectricity} In a crystal, polarization can also arise in response to stresses; this is called the \emph{piezoelectric effect} and was discovered by the Curie brothers \cite{curie:development_CR,curie:development}. In a linear constitutive theory, the induced polarization $\bm{P}$ is related to the Cauchy stress tensor $\bm{T}$ by
\begin{equation}
	\label{eq:piezoelectricity}
	\bm{P}=\oct[\bm{T}],
\end{equation} 
where $\oct$ is now the piezoelectric tensor. The component form of \eref{eq:piezoelectricity} is
\begin{equation}
	\label{eq:piezoelectricity_components}
	P_i=A_{ijk}T_{jk}.
\end{equation}
In classical elasticity, $T_{ij}=T_{ji}$, and so $\oct$ enjoys the symmetry 
\begin{equation}
	\label{eq:A_piezoelectric_symmetry}
	A_{ijk}=A_{ikj}
\end{equation}
and is represented by $18$ independent parameters.

The invariant decomposition of the piezoelectric tensor can help to classify piezoelectric crystals; its algebraic properties  have recently received a renewed interest  (see, for example, \cite[Chapt.\,7]{qi:tensor} and \cite{lazard:irreducible,geymonat:classes,itin:decomposition}). The decomposition of $\oct$ as in \eref{eq:A_as_sum} is affected by the extra symmetry requirement \eref{eq:A_piezoelectric_symmetry}. Clearly, $\oct^{(3)}$ vanishes, but neither $\oct^{(2,1)}$ nor $\oct^{(2,2)}$ does. These two latter do not enjoy the symmetry \eref{eq:A_piezoelectric_symmetry}, whereas $\oct^{(2)}=\oct^{(2,1)}+\oct^{(2,2)}$ does. Moreover, as shown in \cite{itin:decomposition},
\begin{equation}
	\label{eq:piezoelectric_decomposition}
	\oct=\oct^{(1)}+\oct^{(2)}
\end{equation} 
is the unique irreducible invariant decomposition of the piezoelectric tensor.
\subsubsection{Couple-stresses.}Cauchy's stress tensor $\bm{T}$ is symmetric to guarantee the balance of moments, but it has long been known that non-symmetric stress tensors may occur in mechanics \cite[Sect.\,98]{truesdell:non-linear}. The symmetry of Cauchy's stress tensor actually amounts to the assumption that all torques come from moments of forces.

The presence of internal contact couples was already hypothesized in the early theory of the Cosserat brothers \cite{cosserat:theorie,cosserat:theorie_livre}, although in the special context of rods and shells. Toupin \cite{toupin:elastic} put forward a non-linear theory of elastic materials with couple-stresses, which was soon found to be equivalent to Grioli's \cite{grioli:elasticita}. In Toupin's theory, the contact couple $\bm{c}$ is represented by the second-rank skew-symmetric tensor $\bm{C}$ that has $\bm{c}$ as its axial vector. The couple stress is then the third-rank tensor $\oct$ that delivers $\bm{C}$ when applied to the outer unit normal $\bm{\nu}$ designating the orientation of the contact surface,
\begin{equation}
	\label{eq:couple-stress}
	\bm{C}=\oct[\bm{\nu}].
\end{equation}
In components, \eref{eq:couple-stress} reads as
\begin{equation}
	\label{eq:couple-stress_components}
	C_{ij}=A_{ijk}\nu_k
\end{equation}
and, since $C_{ij}=-C_{ji}$, 
\begin{equation}
	\label{eq:couple-stress_symmetry}
	A_{ijk}=-A_{jik},
\end{equation}
which shows that there are only $9$ independent components of $\oct$.

The reader is referred to  \cite{ericksen:anisotropic,ericksen:conservation,ericksen:transversely,ericksen:theory} for the connection between Toupin's theory and the early mechanical theories of Ericksen for liquid crystals.

In a way similar to that enacted in \sref{sec:piezoelectricity}, also the symmetry property \eref{eq:couple-stress_symmetry} affects the representation of a couple-stress tensor $\oct$ (see  \cite{toupin:elastic} and \cite{liu:isotropic}, the latter also referring to $\oct$ as the \emph{Hall} tensor for the role an octupolar tensor with the symmetry \eref{eq:couple-stress_symmetry} plays in describing the Hall effect in crystals). Clearly, in this case $\oct^{(1)}=\bm{0}$, while $\oct^{(2)}$ enjoys the symmetry \eref{eq:couple-stress_symmetry}. As shown in \cite{itin:decomposition},
\begin{equation}
	\label{eq:couple-stress_decomposition}
	\oct=\oct^{(2)}+\oct^{(3)}
\end{equation}
is a unique irreducible invariant decomposition of $\oct$.

\subsection{Octupolar potential}\label{sec:octupolar_potential}
For the octupolar order tensor $\A$ in \eref{eq:A_tensor_representation},
the octupolar potential $\pot$ is given by  \eref{eq:proto_potential}, which we reproduce here for the reader's ease, 
\begin{equation}\label{eq:octupolar_potential_definition}
	\pot(\x):=\A\cdot\xthree=A_{ijk}x_ix_jx_k.
\end{equation}
Given the symmetries enjoyed by $\oct$, the octupolar potential $\pot$ identifies it uniquely.
The critical points $\xc$ of $\pot$  constrained to $\sphere$ have Cartesian components $(\xcc_1,\xcc_2,\xcc_3)$ that solve the equations
\begin{equation}\label{eq:A_equilibrium_equations}
	A_{ijk}x_jx_k=\lambda x_i,\quad i=1,2,3,
\end{equation}
where $\lambda$ is a Lagrange multiplier associated with the constraint
\begin{equation}\label{eq:unit_sphere_constraint_text}
	x_ix_i=1.
\end{equation}
Comparing \eref{eq:A_equilibrium_equations} and \eref{eq:generalized_eigenvalue_problem}, we readily realize that $(\lambda,\xc)$ is a real eigenpair of $\A$. Moreover, it follows from \eref{eq:A_equilibrium_equations} and \eref{eq:unit_sphere_constraint_text} that
\begin{equation}\label{eq:F_equals_lambda}
	\pot(\xcc_1,\xcc_2,\xcc_3)=\lambda,
\end{equation}
which is a specialization of \eref{pot_crit_value}.  Since each real eigenpair $(\lambda,\xc)$ is accompanied by its opposite $(-\lambda,-\xc)$, we see that maxima and minima of $\pot$ are conjugated by a parity transformation.

As $\A$ is real and symmetric, we know from the general results recalled in \sref{sec:generalized} that, \emph{modulo} the parity conjugation, there are generically $7$ distinct eigenvalues of $\A$ one of which at least is real. However, we have no clue as to whether all other eigenvalues are real or not. We are exclusively interested in the real eigenvalues of $\A$, as, by \eref{eq:F_equals_lambda}, they are extrema attained by $\pot$ and so they possibly bear a statistical interpretation whenever $\A$ can be regarded as the collective representation of the third moments of a probability density distribution over $\sphere$.

Since $\pot$ is a polynomial, real-valued mapping on $\sphere$, its critical points are \emph{singularities} for the \emph{index} field $\field_\pot$ defined on $\sphere$ by
\begin{equation}\label{eq:index_field}
	\field_\pot:=\frac{\nablas \pot}{|\nablas \pot|},
\end{equation}
where $\nablas$ denotes the surface gradient on $\sphere$. Each \emph{isolated} singularity of $\field_\pot$ can be assigned an \emph{index}, which is a signed integer $\iota$ \cite[section~VIII.10]{stoker:differential}. Assuming that $\field_\pot$ possesses a finite number $N$ of isolated singularities, by a theorem of Poincar\'e and Hopf \cite[pp.\,239-247]{stoker:differential}, the sum of all their indices must equal the Euler characteristic of the sphere, that is,
\begin{equation}\label{eq:total_index}
	\sum_{i=1}^N\iota_i=2.
\end{equation}
Now, both maxima and minima of $\pot$ are critical points with index $\iota=+1$, whereas its non-degenerate saddle points are critical points with $\iota=-1$.\footnote{A non-degenerate critical point of $\pot$ is one for which the product  of the \emph{tangential} (ordinary) eigenvalues of the Hessian of $\pot$ does \emph{not} vanish.} Thus, were the eigenvalues of a generic, symmetric traceless tensor $\A$ all real (so that according to \eref{eq:d_formula} they occur in $7$ distinct pairs), letting $M$ be the number of eigenvalues corresponding to the maxima of $\pot$ (which equal in number the minima of $\pot$) and $S$ the number of  eigenvalues of $\A$ corresponding to saddle points of $\pot$ (which equal in number the saddles with negative eigenvalues), if the critical points of $\pot$ have all either index $\iota=+1$ or $\iota=-1$, we easily obtain from \eref{eq:total_index} that
\begin{equation}\label{eq:maxima_saddle_numbers}
	M-S=1\quad\mathrm{and}\quad S+M=7,
\end{equation}
whence it follows that $M=4$ and $S=3$.\footnote[1]{Under precisely these assumptions, equation \eref{eq:maxima_saddle_numbers} had already been established by Maxwell~\cite{maxwell:hills} in 1870, elaborating on earlier qualitative considerations of Cayley~\cite{cayley:countour}.}

We shall see below that the complete picture is indeed far more complicated that this, for two reasons: first, not all eigenvalues of $\A$ are real; second, not all critical points of $\pot$ have index $\iota=\pm1$.

\subsubsection{Oriented potential.}\label{sec:oriented_potential}
Making use of \eref{eq:parametrization_alpha_0_beta} and \eref{eq:parametrization_alpha_beta} in \eref{eq:octupolar_potential_definition}, the octupolar potential can be written in the following explicit form,
\begin{equation}
	\label{eq:octupolar_potential}
\eqalign{\pot(x_1,x_2,x_3)&= 6  \alpha_0 x_1 x_2 x_3  +   \alpha_1 x_1 
	\left(x_1^2-3 x_3^2\right)  \\ &+ \alpha_2  x_2  \left(x_2^2-3 x_1^2
	\right)  +\alpha_3x_3   \left(x_3^2-3 x_2^2 \right) \\
	&+  3  \left[\beta_1  x_1  \left(x_2^2-x_3^2\right)  + 
	\beta_2  x_2 
	\left(x_3^2-x_1^2\right)  +  \beta_3  x_3  \left(x_1^2 - x_2^2
	\right) \right],}
\end{equation}
which is described by $7$ scalar parameters. To reduce these, we choose a special \emph{orienting} Cartesian frame $\framee$. The octupolar potential $\pot$ cannot be constant on $\sphere$, lest it trivially vanishes. It will then have at least a local maximum (accompanied by its antipodal minimum). For now, we choose $\ev_3$ such that $\pot$ attains a critical point at the North pole $(0,0,1)$ of $\sphere$, which requires, see section~5.2 of \cite{gaeta:symmetries},
\begin{equation}
	\label{eq:parameter_reduction_1}
	\alpha_1=-\beta_1,\qquad\beta_2=0.
\end{equation}
Later, we shall require $\pot$ to attain a local maximum at the North pole of $\sphere$, which will result in an inequality to be obeyed by a conveniently chosen parameter, see \eref{eq:rho_inequality}.
We can still choose the orientation of the pair $(\ev_1,\ev_2)$. Since $\pot$ is odd on $\sphere$, and so is also on the unit disk $\sphered$ on $\sphere$ orthogonal to $\ev_3$, there must be a point on $\sphered$ where $\pot$ vanishes. We further orient $\pot$ by requiring that $\pot(1,0,0)=0$, which implies that
\begin{eqnarray}
	\label{eq:parameter_reduction_2}
	\beta_1=0.
\end{eqnarray}
Finally, the potential can be scaled with no prejudice to its critical points. By requiring that $\pot(0,0,1)=1$, we obtain that
\begin{equation}
	\label{eq:parameter_reduction_3}
	\alpha_3=1.
\end{equation}

Combining equations \eref{eq:parameter_reduction_1}-\eref{eq:parameter_reduction_3}, we then define
the \emph{oriented} octupolar potential as
\begin{equation}\label{eq:oriented_potential}
\fl\quad	\potor(x_1,x_2,x_3):=6\alpha_0x_1x_2x_3 +\alpha_2(x_2^2-3x_1^2)x_2
+(x_3^2-3x_2^2)x_3
	+3\beta_3(x_1^2-x_2^2)x_3,
\end{equation}
which features only $3$ scalar parameters.

In sections~\ref{sec:geometric} and \ref{sec:algebraic}, two alternative, concurring methods will be presented that afford a complete characterization of the critical points of $\potor$.

\section{Geometric Approach}\label{sec:geometric}
 The oriented potential in \eref{eq:oriented_potential} enjoys a number of symmetries, which are better explained and represented if the parameter space is described by three new variables $(\rho,\chi,K)$ related to $(\alpha_0,\alpha_2,\beta_3)$ through the equations
\begin{equation}
	\label{eq:new_parameters}
	\alpha_0=\frac{1}{2}\rho\cos\chi,\quad\alpha_2=K,\quad\beta_3=\frac12(\rho\sin\chi-1),
\end{equation}
where
\begin{equation}
	\label{eq:new_parameter_bounds}
	\rho\geqq0,\quad\chi\in[-\pi,\pi],\quad K\in\reals.
\end{equation}
$\potor$ in \eref{eq:oriented_potential} is accordingly represented as
\begin{equation}
	\label{eq:oriented_potential_new_parameters}
	\eqalign{\potor&=3\rho\cos\chi x_1x_2x_3+K(x_2^2-3x_1^2)x_2\\
	&+(x_3^2-3x_2^2)x_3+\frac32(\rho\sin\chi-1)(x_1^2-x_2^2)x_3.}
\end{equation}
In the new parameters, the North pole of $\sphere$ is guaranteed to be a maximum for $\potor$ if
\begin{equation}
	\label{eq:rho_inequality}
	 0\leqq\rho\leqq2,
\end{equation}
see \cite{gaeta:symmetries}. This shows that the parameter space can be effectively reduced to a cylinder $\cyl$ with axis along $K$. 

The choice of freezing a maximum of $\potor$ along the $x_3$-axis preempts the action of rotations other than those preserving that axis as possible symmetries of the octupolar potential. However, a number of discrete symmetries survive; they are illustrated in detail in section~5.5 of \cite{gaeta:symmetries}, where it is shown in particular that changing $\chi$ into $\chi+2\pi/3$ simply induces a rotation by $2\pi/3$ about the $x_3$-axis in the graph of $\potor$ over $\sphere$, a symmetry that establishes a rotation covariance between parameter and physical spaces. Combining all discrete symmetries, one finally learns that the study of $\potor$ can be confined to a half-cylinder with $K\geqq0$ and any sector  delimited by the inequalities $\chi_0\leqq\chi\leqq\chi_0+\pi/3$, with any $\chi_0\in[-\pi,\pi]$. Extending the parameter space outside one such sector would add noting to the  octupolar potential landscape: the graph of $\potor$ over $\sphere$ would only be affected by rotations about the $x_3$-axis and mirror symmetries across planes through that axis, which leave all critical points unchanged \cite{gaeta:symmetries}.

For definiteness, we choose $\chi_0=-\pi/2$ and proceed to identify the mirror symmetry that involves both physical and parameter spaces. Subjecting $\potor$ in \eref{eq:oriented_potential_new_parameters} to the change of variables
\begin{equation}
	\label{eq:change_variables}
	x_1=\cos\vt x_1'-\sin\vt x_2'\quad x_2=-\sin\vt x_1'-\cos\vt x_2',\quad x_3=x_3',
\end{equation}
which represents a mirror reflection with fixed point $x_1=x_1'$, $x_2=x_2'$ along the plane 
\begin{equation}
	\label{eq:mirror_plane}
	x_2=-x_1\tan\frac{\vt}{2},
\end{equation}
one easily sees that $\potor$ remains  formally unchanged in the variables $(x_1',x_2',x_3')$, thus making \eref{eq:change_variables} a mirror symmetry for $\potor$, if $\vt=\pi/3$ and $\chi$ is changed into $\chi'=-\chi-\pi/3$, which has a fixed point for $\chi=-\pi/6$. Thus, by \eref{eq:mirror_plane}, a reflection of the sector $-\pi/2\leqq\chi\leqq-\pi/6$ across the plane $\chi=-\pi/6$ in  parameter space $(\rho,\chi,K)$ induces a reflection of $\potor$ across the plane through the $x_3$-axis that makes the angle $-\pi/6$ with the $x_1$-axis in the physical space $(x_1,x_2,x_3)$. 

Thus, we shall hereafter confine attention to the sector of $\cyl$ that is represented  in cylindrical coordinates $(\rho,\chi,K)$ as
\begin{equation}
	\label{eq:cylindrical_sector}
	0\leqq\rho\leqq2,\quad -\frac{\pi}{2}\leqq\chi\leqq-\frac{\pi}{6},\quad K\geqq0.
\end{equation}

Special subsets in $\cyl$ (and their intersection with the relevant sector in \eref{eq:cylindrical_sector}) make $\potor$ enjoy special symmetries in physical space. The corresponding symmetry groups (in the Schoenflies notation) are summarised in \tref{tab:symmetries}; the special subsets are the centre $\Centre$ ($\rho=K=0$), the disk $\disk$ ($K=0$), the axis $\axis$ ($\rho=0$), and the tetrahedral pair $\tetra\in\axis$ ($\rho=0$, $K\pm1/\sqrt{2}$). 
\begin{table}
	\caption{\label{tab:symmetries}Symmetry groups for the oriented octupolar potential (in the Schoenflies notation) corresponding to special sets in the reduced parameter space.} 
	
	\begin{indented}
		\lineup
		\item[]\begin{tabular}{@{}*{3}{c}}
			\br                              
			Group&Parameters&Subset\cr 
			\br
			$D_{\infty h}$&$\rho=K=0$ &centre $\Centre$\cr
			\mr
			$D_{2h}$&$K=0$&disk $\disk$\cr 
			\mr
			$D_{3h}$&$\rho=0$&axis $\axis$\cr 
			\mr
			$T_d$&$\rho=0$, $K=\pm1/\sqrt{2}$&$\tetra\in\axis$\cr 
			\br
		\end{tabular}
	\end{indented}
\end{table}

To illustrate these special cases, we shall draw the polar plot of $\potor$ and its contour plot in the plane $(x_1,x_3)$. The former is the surface in space spanned by the tip of the vector $\potor\ev_r$, where $\ev_r$ is the radial unit vector,
\begin{equation}
	\label{eq:radial_unit_vector}
	\ev_r:=\frac{1}{r}(x_1\ev_1+x_2\ev_2+x_3\ev_3),\quad r:=\sqrt{x_1^2+x_2^2+x_3^2}.
\end{equation}
Since $\potor$ is odd under reversal of the coordinates $(x_1,x_2,x_3)$, antipodal points on $\sphere$ are mapped into the same point on the polar plot of $\potor$, so that minima of $\potor$ are invaginated under its maxima, and the latter are the only ones to be shown by the polar plot of $\potor$. To resolve this ambiguity, we shall often supplement the polar plot of $\potor$ with the contour plot of the function in $(x_1,x_3)$ obtained by setting $x_2=\sqrt{1-x_1^2-x_3^2}$ in \eref{eq:oriented_potential_new_parameters}. This gives a view of the octupolar potential on a hemisphere based on a great circle passing through both North and South poles and culminating at the point $(0,1,0)$. If polar plots give a quite vivid representation of the maxima (and minima) of $\potor$, the contour plots in the $(x_1,x_3)$ plane give a side view of half its critical points.

Before showing the illustrations for the symmetric cases in \tref{tab:symmetries}, we must warn the reader that whereas maxima, minima, and genuine saddles (either degenerate or not, but with index $\iota\neq0$) are easily discerned from a contour plot, degenerate saddles with index $\iota=0$ may easily go unnoticed.

We illustrate in the following subsections the special symmetries listed in \tref{tab:symmetries}.

\subsection{$D_{\infty h}$}\label{sec:D_infinity_h}
\begin{figure}[h]
	\centering
	\begin{subfigure}[b]{0.5\linewidth}
		\centering
		\includegraphics[width=\linewidth]{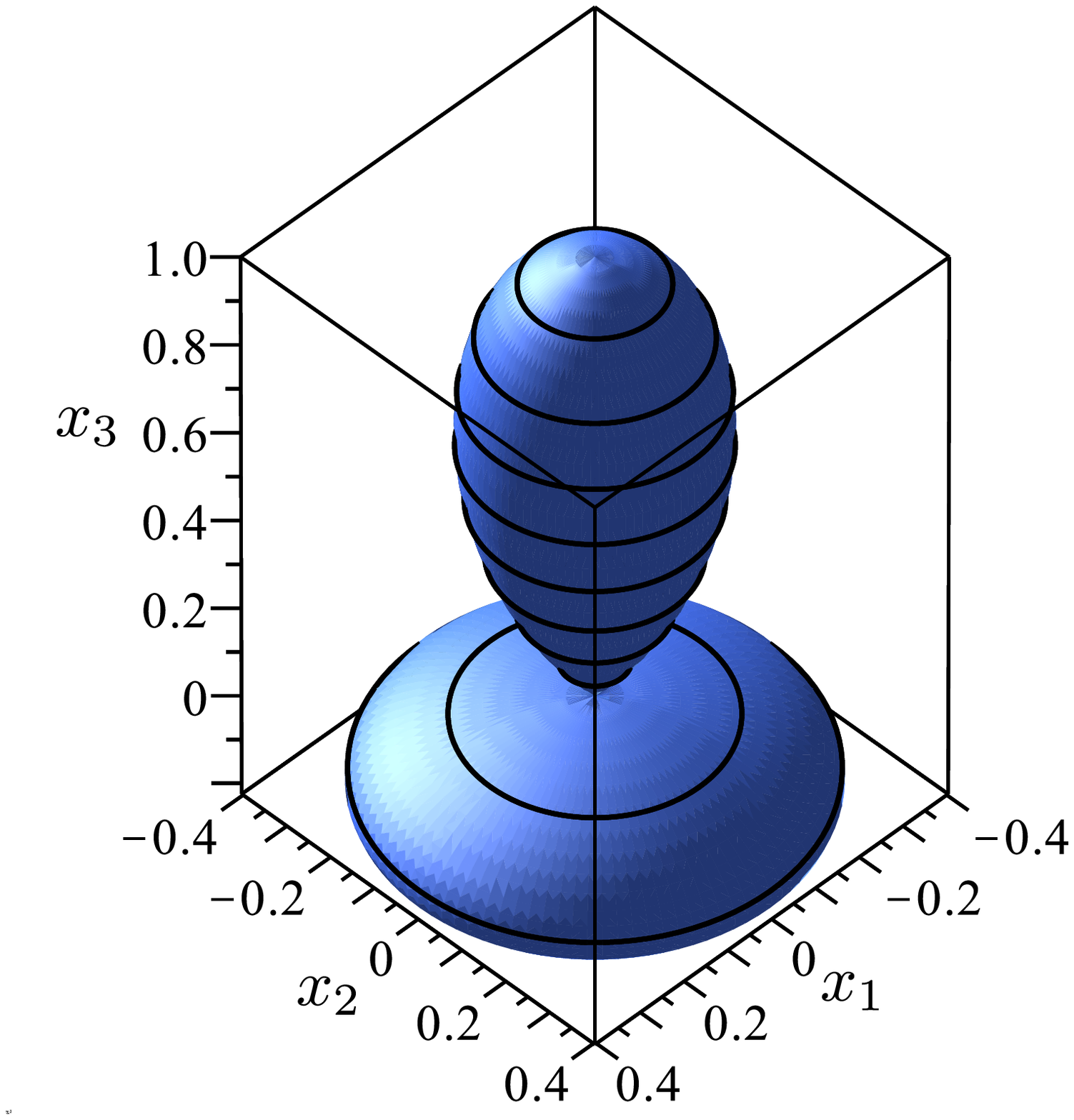}
		\caption{Polar plot.}
		\label{fig:centre_polar_plot}
	\end{subfigure}
	\quad
	\begin{subfigure}[b]{0.4\linewidth}
		\centering
		\includegraphics[width=\linewidth]{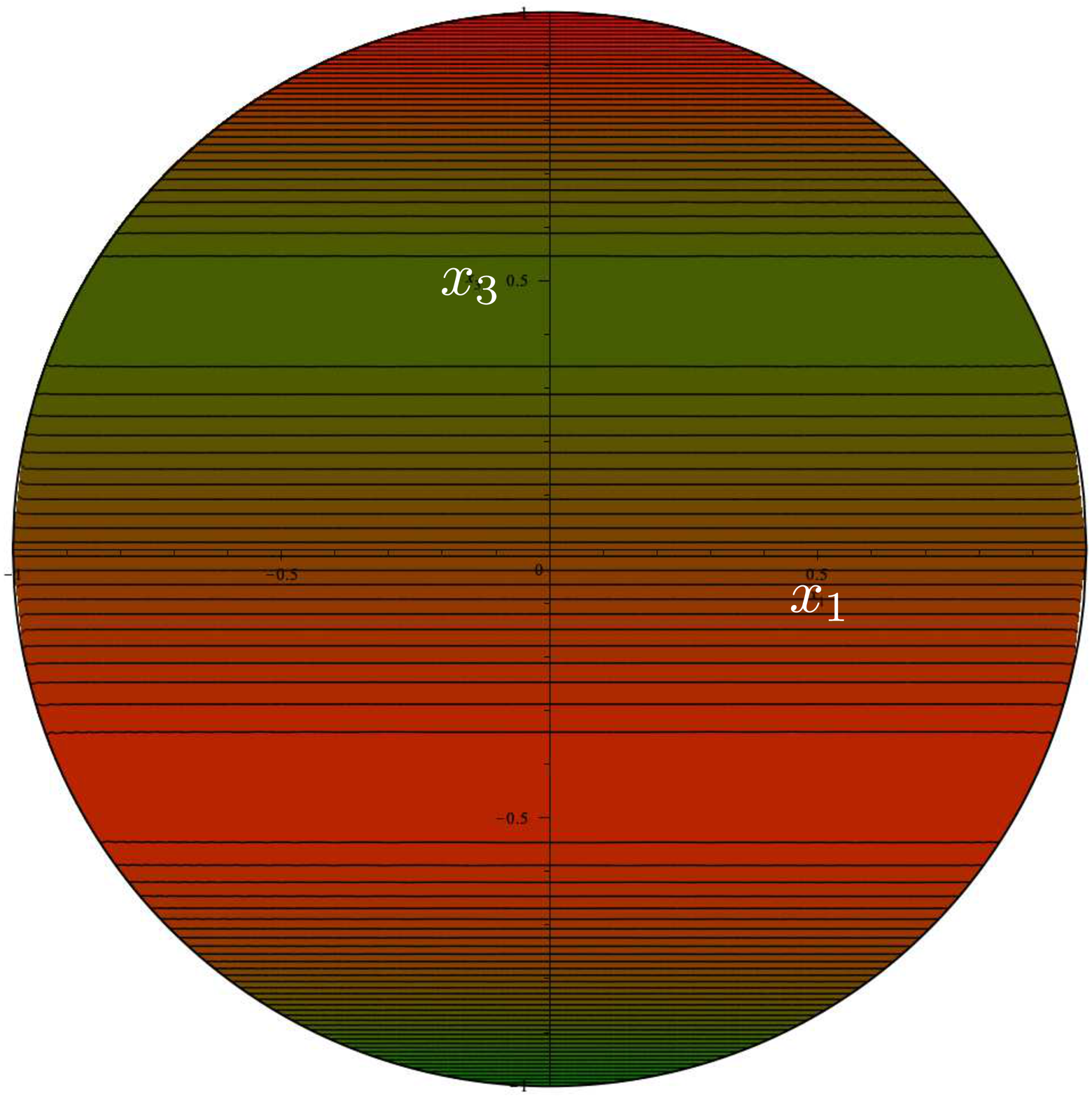}
		\caption{Contour plot on the plane $(x_1,x_3)$.}
		\label{fig:centre_contourplot}
	\end{subfigure}
	\caption{The octupolar potential $\potor$ for $\rho=K=0$.}
	\label{fig:centre}
\end{figure}

\Fref{fig:centre} shows the polar plot and the contour plot for $\potor$ in the centre $\Centre$ in parameter space.
The former (see \fref{fig:centre_polar_plot}) is symmetric about the $x_3$-axis, while the latter (see \fref{fig:centre_polar_plot}) exhibits the same $D_{\infty h}$ symmetry, but seen from a different perspective: the level sets of $\potor$ are parallels and their colour, ranging from green to red, spans the range of values taken by $\potor$, from its minimum (green) to its (opposite) maximum (red). In this specific instance, $\potor$ vanishes on the equator. Alongside the maximum at the North pole (accompanied by its minimum twin at the South pole), a full orbit of maxima (with their twin minima) exist on symmetric parallels.

By construction, in our representation the North pole must be red, whereas the South pole must be green, even if our pictures do not always show this very clearly.  

\subsection{$D_{2h}$}\label{sec:D_2_h}
\begin{figure}[h]
	\centering
	\begin{subfigure}[b]{0.5\linewidth}
		\centering
		\includegraphics[width=\linewidth]{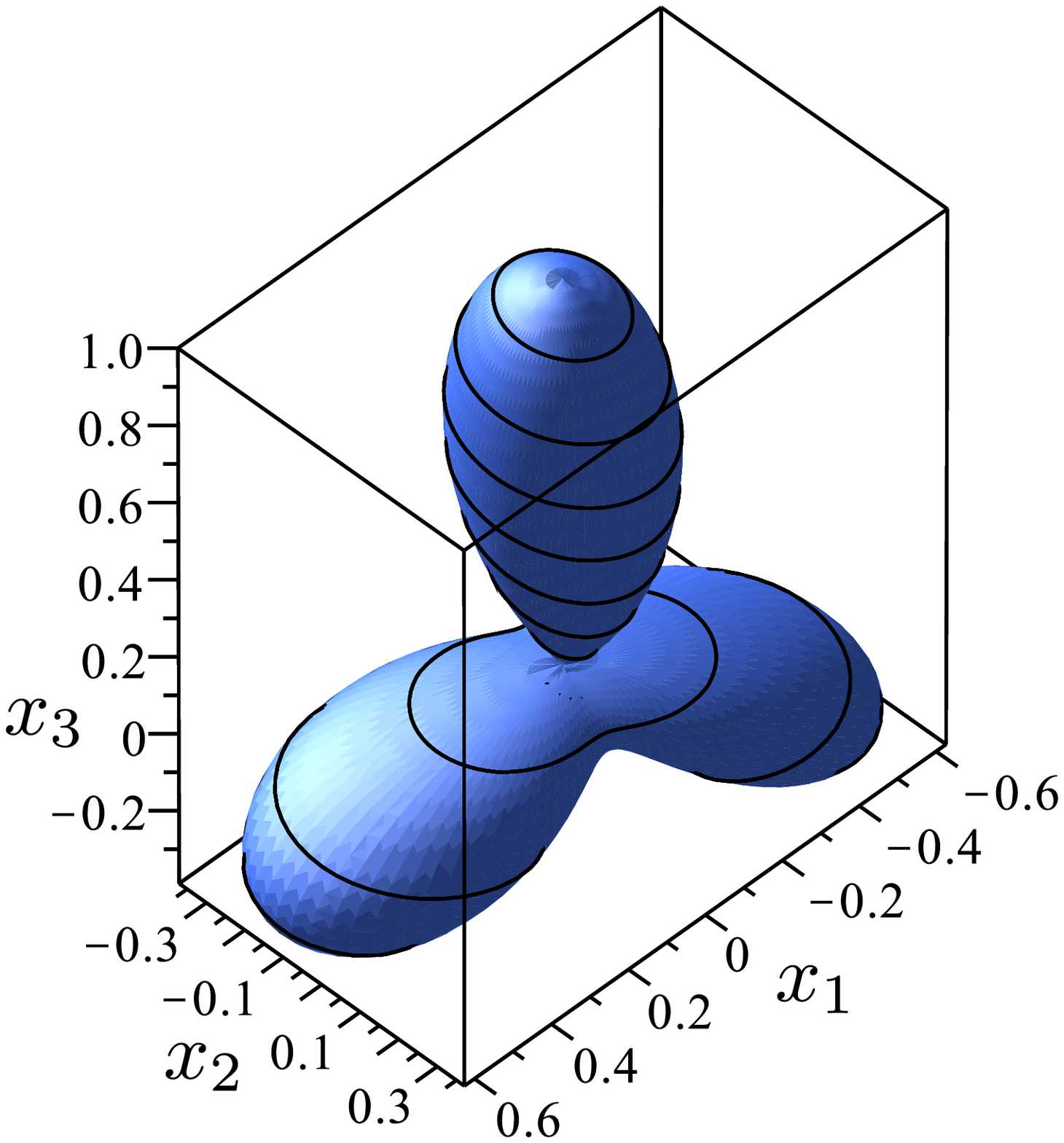}
		\caption{Polar plot.}
		\label{fig:disk_polar_plot}
	\end{subfigure}
	\quad
	\begin{subfigure}[b]{0.4\linewidth}
		\centering
		\includegraphics[width=\linewidth]{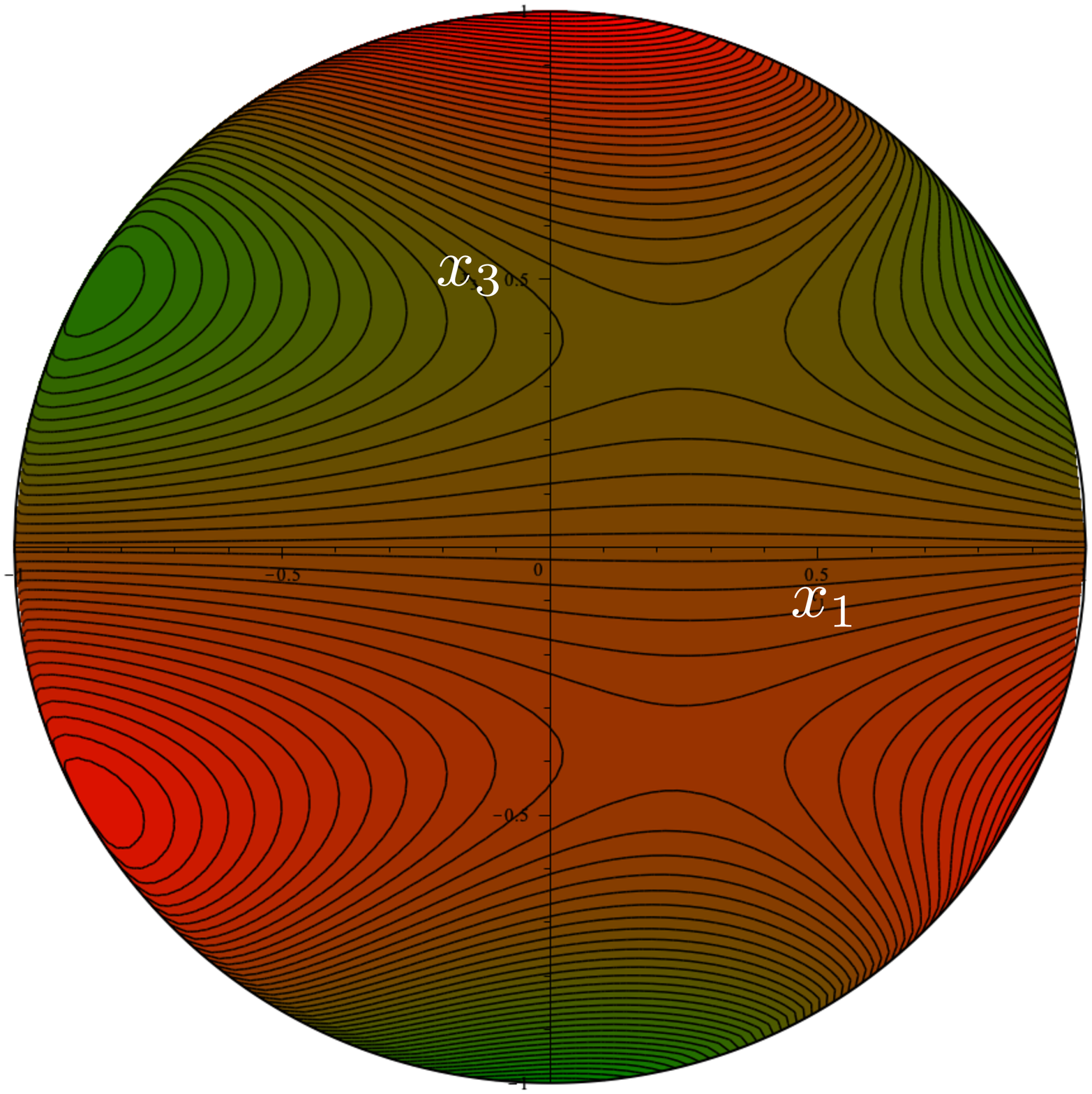}
		\caption{Contour plot on the plane $(x_1,x_3)$.}
		\label{fig:disk_contourplot}
	\end{subfigure}
	\caption{The octupolar potential $\potor$ for $\rho=1/2$, $\chi=-\pi/3$, $K=0$.}
	\label{fig:disk}
\end{figure}

\Fref{fig:disk} shows a case exhibiting the  $D_{2h}$ symmetry  characteristic for the whole disk $\disk$ in parameter space (see  \tref{tab:symmetries}).
The octupolar potential $\potor$ has generically three maxima, three minima, and four saddles, which indices $\iota=+1$, $\iota=+1$, $\iota=-1$, respectively, so that the global constraint \eref{eq:total_index} is satisfied. Two more maxima accompany in the Southern hemisphere the maximum at the North pole (and so do the conjugated minima in the Northern hemisphere). The four (non-degenerate) saddles are two on each hemisphere, for a total of $10$ critical points (see also \sref{sec:case_2}).

\subsection{$D_{3h}$}\label{sec:D_3_h}
\begin{figure}[h]
	\centering
	\begin{subfigure}[b]{0.5\linewidth}
		\centering
		\includegraphics[width=\linewidth]{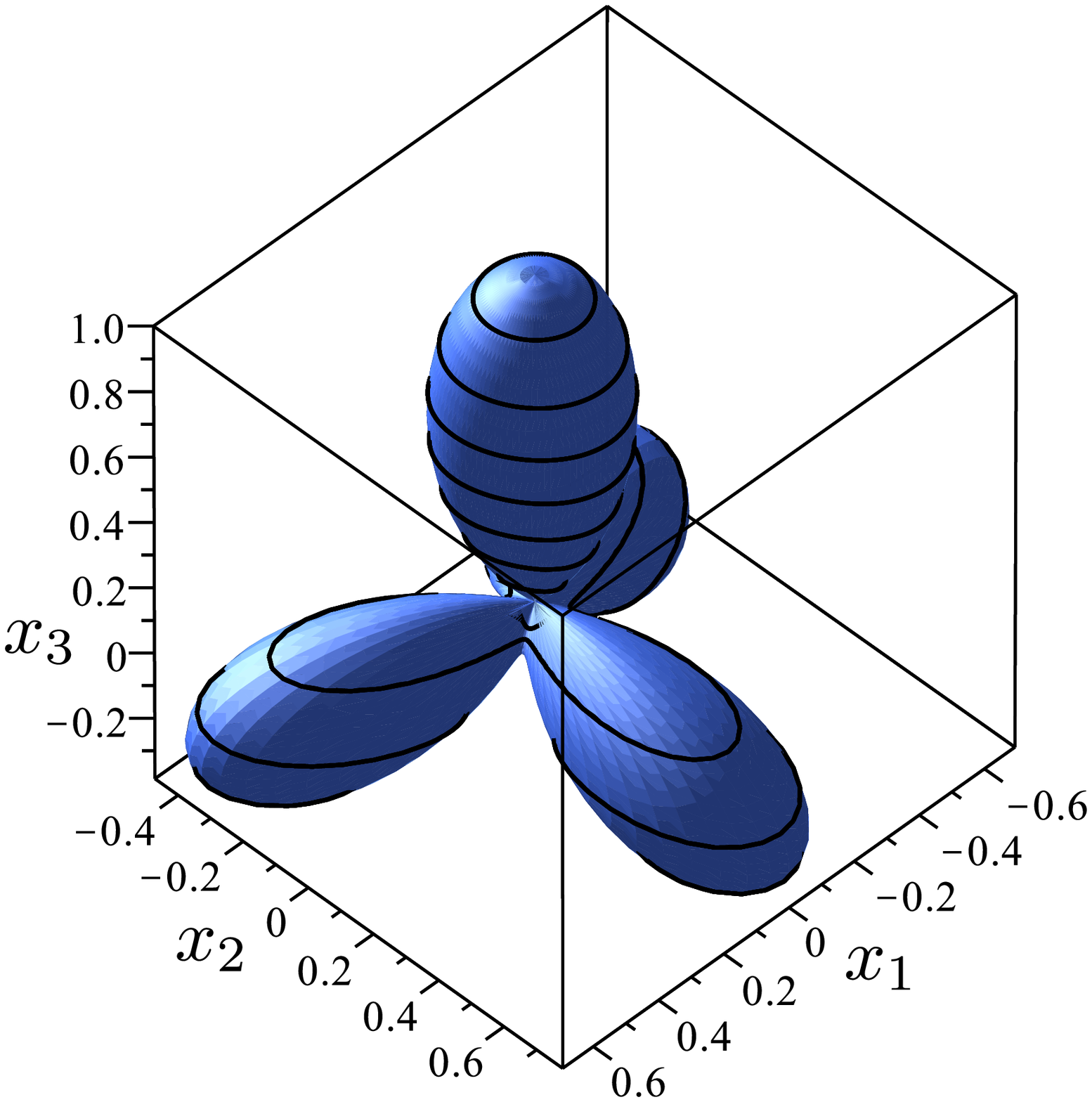}
		\caption{Polar plot.}
		\label{fig:axis_polar_plot}
	\end{subfigure}
	\quad
	\begin{subfigure}[b]{0.4\linewidth}
		\centering
		\includegraphics[width=\linewidth]{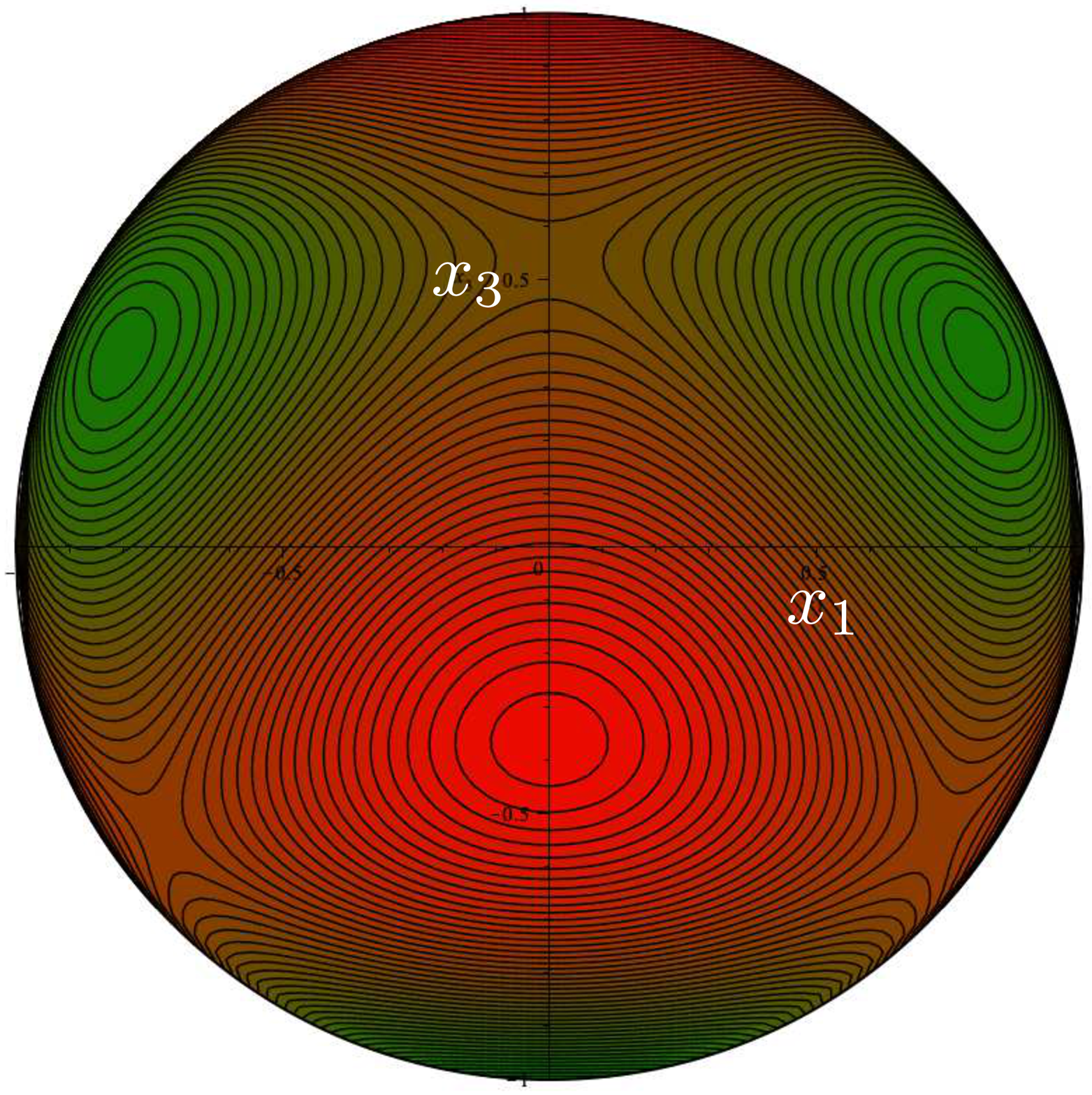}
		\caption{Contour plot on the plane $(x_1,x_3)$.}
		\label{fig:axis_contourplot}
	\end{subfigure}
	\caption{The octupolar potential $\potor$ for $\rho=0$ and $K=1/2$, representing the behaviour on the whole axis $\axis$ in parameter space.}
	\label{fig:axis}
\end{figure}

\Fref{fig:axis} shows the appearance of the octupolar potential $\potor$  on the axis $\axis$ in parameter space.
It enjoys the $D_{3h}$ symmetry and possesses four maxima, four minima, and six saddles, for a total of $14$ isolated critical points.

\subsection{$T_d$}\label{sec:T_d}
\begin{figure}[h]
	\centering
	\begin{subfigure}[b]{0.5\linewidth}
		\centering
		\includegraphics[width=\linewidth]{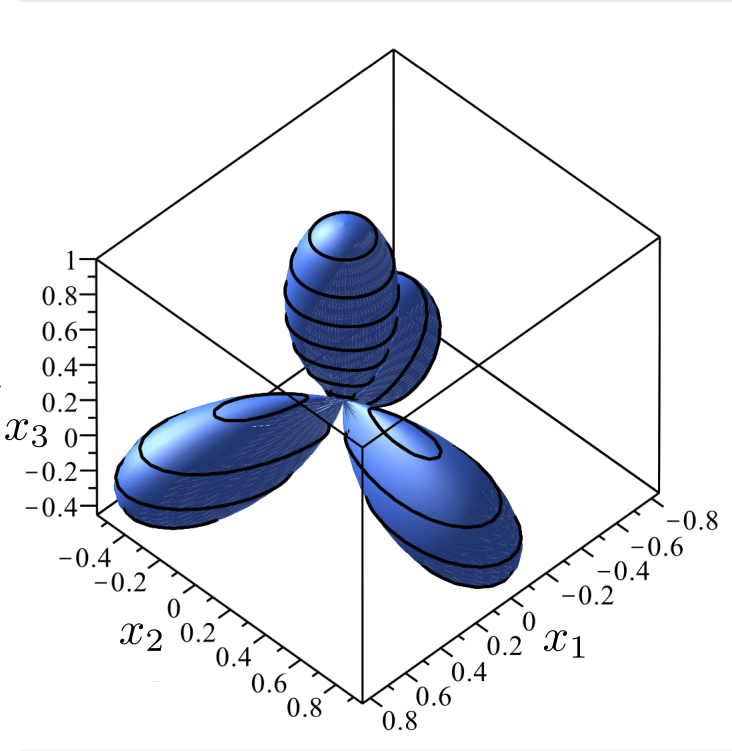}
		\caption{Polar plot.}
		\label{fig:tetra_polar_plot}
	\end{subfigure}
	\quad
	\begin{subfigure}[b]{0.4\linewidth}
		\centering
		\includegraphics[width=\linewidth]{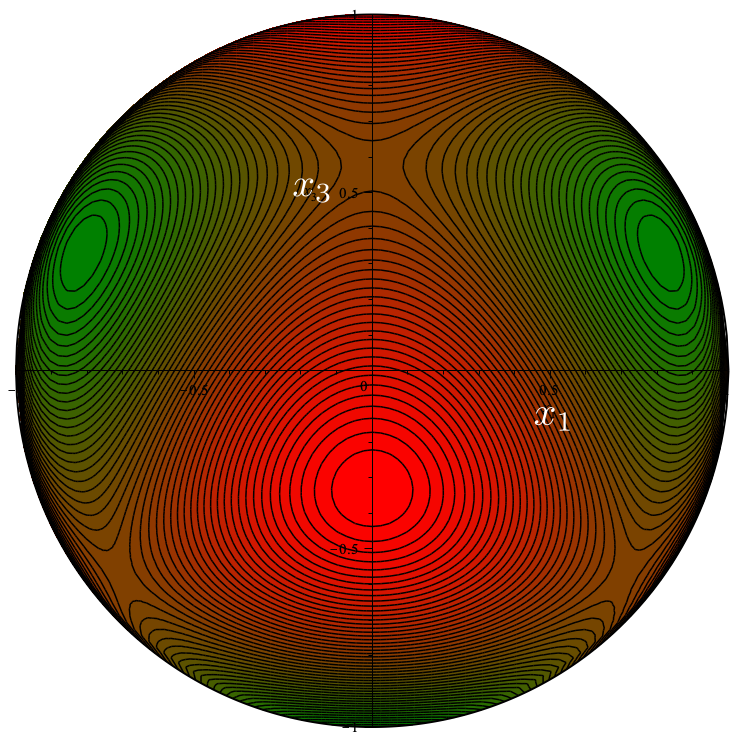}
		\caption{Contour plot on the plane $(x_1,x_3)$.}
		\label{fig:tetra_contourplot}
	\end{subfigure}
	\caption{The octupolar potential $\potor$ for $\rho=0$ and $K=1/\sqrt{2}$, representing one of the two (symmetric) special points $\tetra\in\axis$ in parameter space.}
	\label{fig:tetra}
\end{figure}

Finally, in \fref{fig:tetra} we see $\potor$ at one special tetrahedral point $\tetra$ in parameter space. 
The octupolar potential has four equal maxima at the vertices of a regular tetrahedron (with four antipodal minima) and six saddles with equal values. The total number of critical points is $14$, as in \fref{fig:axis}, but here each maximum, minimum, or saddle cannot be distinguished from all others; $\potor$ enjoys the $T_d$ symmetry.

\subsection{Summary} This survey of the symmetric  cases has shown that $\potor$ can have either $10$ or $14$ critical points, apart from the highly degenerate case corresponding to the singular point $\Centre$ in parameter space, where it has infinitely many. We shall see from the following analysis that cases with either $8$ or $12$ critical points are also possible, thus revealing a more intricate landscape, which we shall also endeavour to illustrate geometrically. 

\section{Algebraic Approach}\label{sec:algebraic}
The (normalized) eigenvectors of the octupolar tensor $\oct$ are identified with the critical points of the octupolar potential $\potor$ on the unit sphere $\sphere$, and the real eigenvalues of $\oct$ are the corresponding critical values. These latter are the only eigenvalues of $\oct$ to bear a physical meaning; as for their number, the general result of Cartwright and Sturmfels~\cite{cartwright:number} only provides an upper bound, which is $14$ in the case of interest here. The number of real eigenvalues  of $\oct$ depends on the parameters $(\rho,\chi,K)$
in a rather complicated and intriguing way, which is explored and fully documented below.

In this pursuit, we found especially expedient a method applied by Walcher~\cite{walcher:eigenvectors}; this reduces the critical points of $\potor$ on $\sphere$ to the roots of an appropriate polynomial in \emph{one} real variable. Here we adapt Walcher's idea to our formalism and draw all our conclusions from the polynomial he introduced. 
The fundamental algebraic tool at the basis of Walcher's method is 
Bezout’s theorem in projective spaces,
for  a full account on which we defer the reader to Chapt.~IV of Shafarevich's book \cite{shafarevich:basic} (precursors of this method can also be retraced in the works \cite{rohrl:zeros,rohrl:projections}). 
The outcomes of our previous analysis \cite{gaeta:symmetries} for the critical points of $\potor$ are confirmed, but an important  detail is added.

Our first move is writing the equilibrium equations for $\potor$ on $\sphere$, whose solutions are the critical points we want to classify. We incorporate in $\potor$ the constraint $\x\cdot\x=1$ by defining the \emph{extended} potential $\Phi_\lambda$ as
\begin{equation}
	\label{eq:Phi_lambda}
	\Phi_\lambda:=\potor+\Phi_\mathrm{c},
\end{equation} 
where the constraint term $\Phi_\mathrm{c}$ is defined by
\begin{equation}
	\label{eq:Phi_c}
	\Phi_\mathrm{c}:=-\frac32\lambda(x_1^2+x_2^2+x_3^2),
\end{equation}
and $\lambda$ is a Lagrange multiplier to be determined by requiring that $\x\in\sphere$. As shown in \cite{gaeta:symmetries}, the scaling of $\Phi_\mathrm{c}$ has been chosen so as to ensure that on a critical point $\lambda$ would equal the corresponding critical value of $\potor$, and hence be a real eigenvalue of $\oct$. In light of this, it should also be recalled that whereas $\potor$ changes sign upon central inversion, $\Phi_\mathrm{c}$ (and so $\Phi_\lambda$) does so only under the simultaneous changes $\x\mapsto-\x$ and $\lambda\mapsto-\lambda$.

With the aid of \eref{eq:oriented_potential_new_parameters}, the equilibrium equations for $\Phi_\lambda$ are easily obtained,
\begin{eqnarray}
	\cases{
\rho\cos\chi x_2x_3-2Kx_1x_2+(\rho\sin\chi-1)x_1x_3=\lambda x_1,\\
\rho\cos\chi x_1x_3-K(x_1^2-x_2^2)-(\rho\sin\chi+1)x_2x_3=\lambda x_2,\\
\rho\cos\chi x_1x_2-(x_2^2-x_3^2)+\frac12(\rho\sin\chi-1)(x_1^2-x_2^2)=\lambda x_3,
}\label{eq:equilibrium_equations_1-3}
\end{eqnarray}	
subject to
\begin{equation}\label{eq:constraint}
	x_1^2+x_2^2+x_3^2=1,
\end{equation}
where the parameters $(\rho,\chi,K)$ are chosen as specified in \eref{eq:cylindrical_sector}.	

Here we split the quest for solutions of \eref{eq:equilibrium_equations_1-3} and \eref{eq:constraint} in two steps. First, we seek solutions with $x_2=0$, and then all others. For the role they will play, the former are called the \emph{background} solutions, for lack of a better name. Clearly, both poles $(0,0,\pm1)$  are solutions of \eref{eq:equilibrium_equations_1-3} and \eref{eq:constraint} by the way the potential has been \emph{oriented}. To avoid double counting, these solutions will be excluded from the background; they should always be added to the ones we are seeking here.

\subsection{Background Solutions}\label{sec:background}
By setting $x_2=0$ in \eref{eq:equilibrium_equations_1-3} and \eref{eq:constraint} and assuming that $x_1\neq0$, so as to exclude both poles, we see that these equations reduce to
\begin{eqnarray}
(\rho\sin\chi-1)x_3=\lambda,\label{eq:x_2=0_star_1}\\
\rho\cos\chi x_3=Kx_2,\label{eq:x_2=0_star_2}\\
x_3^2+\frac12(\rho\sin\chi-1)x_1^2=\lambda x_3,\label{eq:x_2=0_star_3}\\
x_1^2+x_3^2=1.\label{eq:x_2=0_star_4}	
\end{eqnarray}
A number of simple cases arise, which are conveniently described separately, for clarity.
\subsubsection{Case $\rho=K=0$.}\label{sec:case-1}
In this case, the background solutions are $x_1=\pm2\sqrt{5}$ and $x_3=-\lambda=\pm1/\sqrt{5}$, where all choices of sign are possible, so that these roots amount to $4$ critical points of $\potor$ on $\sphere$.

\subsubsection{Case $\rho>0$, $\chi\neq-\pi/2$, $K=0$.}\label{sec:case-2}
This case is the easiest, as \eref{eq:x_2=0_star_2} requires $x_1=0$, which is incompatible with \eref{eq:x_2=0_star_3} in the sector \eref{eq:cylindrical_sector} we selected in parameter space. Thus, no background solution exists. As we shall see below, they do exist for $\chi=-\pi/2$. 

\subsubsection{Case $\rho=0$, $K\neq0$.}\label{sec:case-3} This is another trivial case, as \eref{eq:x_2=0_star_2} again implies $x_1=0$, which is disallowed. Thus, once again no background solution exists for this choice of parameters. 

\subsubsection{Case $\rho>0$, $\chi=-\pi/2$, $K>0$.}\label{sec:case-4} This is another case of non-existence, as \eref{eq:x_2=0_star_2} implies once more that $x_1=0$. Finally, we see now two cases where background solutions do actually exist.

\subsubsection{Case $\rho>0$, $\chi=-\pi/2$, $K=0$.} \label{sec:case-5} For this choice of parameters, equation \eref{eq:x_2=0_star_2} is identically satisfied, while the remaining equations possess the solutions
\begin{equation}
	\label{eq:background_4}
	\fl\qquad x_1=\pm\sqrt{\frac{2(\rho+2)}{5+3\rho}},\quad x_2=0,\quad x_3=\pm\sqrt{\frac{\rho+1}{5+3\rho}},\quad\lambda=\mp\sqrt{\frac{(\rho+1)^3}{5+3\rho}},
\end{equation}
where signs can be chosen independently, provided that $\lambda x_2<0$. Thus, these roots correspond to $4$ critical points of $\potor$, all lying on a great circle of $\sphere$.

\subsubsection{Case $\rho>0$, $\chi\neq-\pi/2$, $K>0$.}\label{sec:case-generic}
This is the generic case for the existence of background solutions. It easily follows from \eref{eq:x_2=0_star_1}-\eref{eq:x_2=0_star_4} that in the selected  sector of parameter space \eref{eq:cylindrical_sector} the background solutions must satisfy the inequalities $x_1x_3>0$ and $\lambda x_3<0$. Elementary calculations deliver
\begin{eqnarray}
	x_1=\mp\sqrt{\frac{2(2-\rho\sin\chi)}{5-3\rho\sin\chi}},\quad x_2=0,\quad x_3=\mp\sqrt{\frac{1-\rho\sin\chi}{5-3\rho\sin\chi}},	\label{eq:x_2=0_non_polar_solutions}\\ \lambda=\pm\sqrt{\frac{(1-\rho\sin\chi)^3}{5-3\rho\sin\chi}},	\label{eq:x_2=0_non_polar_solutions_lambda}
\end{eqnarray}
where signs must be chosen so as to satisfy the above inequalities. However, these solutions do not exist for all values of $K>0$, but only for
\begin{equation}
	\label{eq:x_2=0_K_admissible}
	K=\kappa(\rho,\chi):=\sqrt{\frac{1-\rho\sin\chi}{2(2-\rho\sin\chi)}}\rho\cos\chi.
\end{equation}
Whenever the latter is satisfied, the background solutions correspond to $2$ critical points of $\potor$ on $\sphere$.

\subsection{All other solutions}\label{sec:others}
We now assume that $x_2\neq0$ and set
\begin{equation}
	\label{eq:s_t_mu}
	s:=\frac{x_1}{x_2},\quad t:=\frac{x_3}{x_2},\quad\mu:=\frac{\lambda}{x_2}.
\end{equation}
With the aid of these definitions, equations \eref{eq:equilibrium_equations_1-3} become
\begin{eqnarray}
	\rho\cos\chi t-2Ks+(\rho\sin\chi-1)st=\mu s,\label{eq:equilibrium_equations_s_t_1}\\
	\rho\cos\chi st-K(s^2-1)-(\rho\sin\chi+1)t=\mu,\label{eq:equilibrium_equations_s_t_2}\\
	\rho\cos\chi s-(1-t^2)+\frac12(\rho\sin\chi-1)(s^2-1)=\mu t.\label{eq:equilibrium_equations_s_t_3}	
\end{eqnarray}
While \eref{eq:equilibrium_equations_s_t_2} is by itself an explicit expression for $\mu$ in the new variables $(s,t)$, both \eref{eq:equilibrium_equations_s_t_1} and \eref{eq:equilibrium_equations_s_t_3} can be made into polynomials in these latter  upon insertion of \eref{eq:equilibrium_equations_s_t_2}. These are
\begin{equation}\label{eq:equilibrium_equations_a}
\eqalign{Ks^2t-\rho\cos\chi &st^2+\frac12(\rho\sin\chi-1)s^2+(\rho\sin\chi+2)t^2\\
	&+\rho\cos\chi s-Kt-\frac12(\rho\sin\chi+1)=0,}
\end{equation}
and
\begin{equation}\label{eq:equilibrium_equations_b}
	\rho(\cos\chi s^2-2\sin\chi s-\cos\chi)t=Ks(s^2-3),
\end{equation}
the latter of which has the remarkable feature of being \emph{linear} in $t$.

The general strategy here will be to extract $t$ from \eref{eq:equilibrium_equations_b} and transform \eref{eq:equilibrium_equations_a} into a polynomial of degree $6$ in the single variable $s$. However, in a number of selected case this strategy is not viable and the solutions to the system \eref{eq:equilibrium_equations_a} and \eref{eq:equilibrium_equations_b} can be found by finding the roots of polynomials of lower degree. These special cases will be treated first, as they are somehow related to the symmetries studied above.
Progressing further, we note that once a solution $(s,t)$ of \eref{eq:equilibrium_equations_a} and \eref{eq:equilibrium_equations_b} is known, by \eref{eq:s_t_mu} we obtain the solutions $(x_1,x_2,x_3)$ of \eref{eq:equilibrium_equations_1-3} and \eref{eq:constraint} through the equations
\begin{eqnarray}
\fl\qquad	x_1=\pm\frac{s}{\sqrt{1+s^2+t^2}},\quad x_2=\pm\frac{1}{\sqrt{1+s^2+t^2}},\quad x_3=\pm\frac{t}{\sqrt{1+s^2+t^2}},\label{eq:x_solutions}\\
\fl\qquad	\lambda=\pm\frac{\mu}{\sqrt{1+s^2+t^2}},\label{eq:lambda_solution}
\end{eqnarray}
where $\mu$ is given by \eref{eq:equilibrium_equations_s_t_2}. Thus, as expected, each solution $(s,t)$ of \eref{eq:equilibrium_equations_a} and \eref{eq:equilibrium_equations_b} corresponds to a conjugated pair of critical points of $\potor$. 

\subsubsection{Case $\rho=K=0$.}\label{sec:case_1}
This point corresponds to the centre $\Centre$ in parameter space. For this choice of parameters, equation \eref{eq:equilibrium_equations_b} is identically satisfied and \eref{eq:equilibrium_equations_a} delivers $t^2=\frac14(s^2+1)$, which with the aid of \eref{eq:x_solutions} readily implies that
\begin{equation}\label{eq:x_solutions_1}
\fl\qquad\quad x_1=\pm\frac{2s}{\sqrt{5(1+s^2)}},\quad x_2=\pm\frac{2}{\sqrt{5(1+s^2)}},\quad x_3=\pm\frac{1}{\sqrt{5}},\quad\lambda=\mp\frac{1}{\sqrt{5}},
\end{equation}
which for varying $s$ represent the two orbits of critical points shown in \fref{fig:centre}. It is perhaps worth noting that for $\rho=K=0$ solution \eref{eq:x_solutions_1} reproduces the background solution of \sref{sec:case-1} in the limits as $s\to\pm\infty$, and so no other critical point of $\potor$ is present in this case, besides the poles and the orbits \eref{eq:x_solutions_1}.

\subsubsection{Case $\rho>0$, $\chi\neq-\pi/2$, $K=0$.}\label{sec:case_2}
\begin{figure}[h]
	\centering
	\begin{subfigure}[c]{0.3\linewidth}
		\centering
		\includegraphics[width=\linewidth]{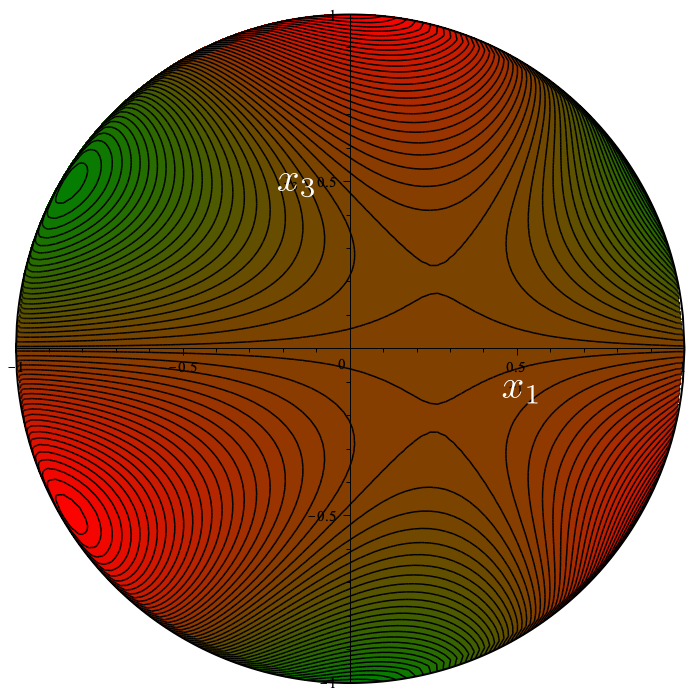}
		\caption{$\rho=0.9$}
		\label{fig:disk_non-generic_a}
	\end{subfigure}
	\quad
	\begin{subfigure}[c]{0.3\linewidth}
		\centering
		\includegraphics[width=\linewidth]{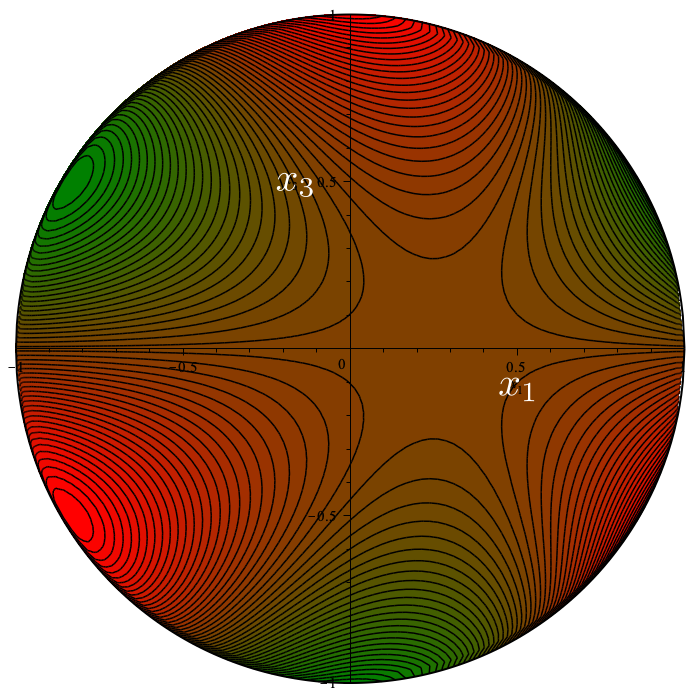}
		\caption{$\rho=1$}
		\label{fig:disk_non-generic_b}
	\end{subfigure}
	\quad
	\begin{subfigure}[c]{0.3\linewidth}
		\centering
		\includegraphics[width=\linewidth]{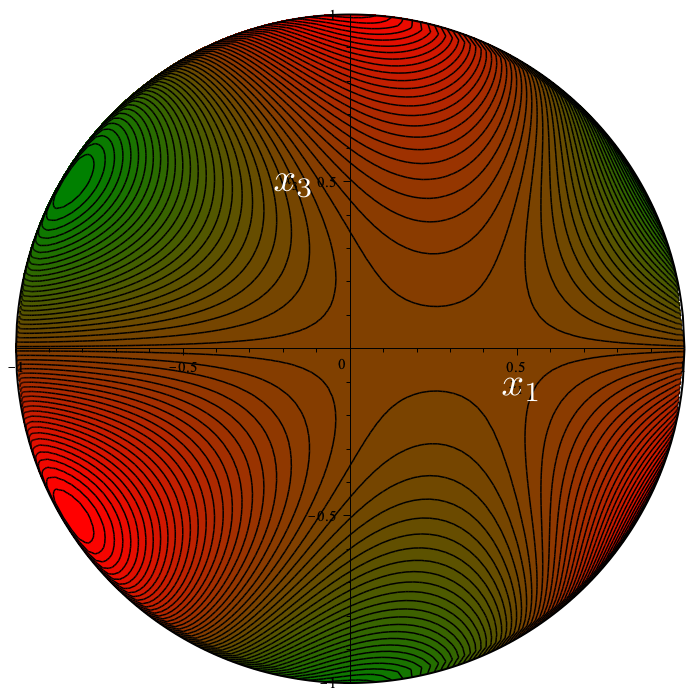}
		\caption{$\rho=1.1$}
		\label{fig:disk_non-generic_c}
	\end{subfigure}
	\caption{Contour plots of $\potor$ on the plane $(x_1,x_3)$ for $\chi-\pi/3$ and $K=0$. For $\rho=1$, $\potor$ has a saddle with index $\iota=-2$ on the equator of $\sphere$. As $\rho$ is either increased or decreased, this saddle splits into two saddles with $\iota=-1$ moving along a meridian or sliding on the equator, respectively.}
	\label{fig:disk_non-generic}
\end{figure}

This is the plane where the disk $\disk$ lies. The case $\chi=-\pi/2$ is again  somewhat special and will be treated separately below. For this choice of parameters, equation \eref{eq:equilibrium_equations_b} requires that either $t=0$ or $s=s_{1,2}=\tan\chi\pm\sqrt{1+\tan^2\chi}$. Inserting the former into \eref{eq:equilibrium_equations_a}, we readily arrive at 
\begin{equation}\label{eq:s_3}
	s=s_3=\frac{-\rho\cos\chi\pm\sqrt{\rho^2-1}}{\rho\sin\chi-1},	
\end{equation}	
which in our admissible sector in parameter space \eref{eq:cylindrical_sector} is valid only for $1\leqq\rho\leqq2$. Upon insertion of $s_1$ and $s_2$, the roots  of the equation \eref{eq:equilibrium_equations_a} transform into 
\begin{equation}
	\label{eq:t_1_2}
	t_1=\pm\sqrt{\frac{1-\rho}{(2-\rho)(1-\sin\chi)}},\quad t_2=\pm\sqrt{\frac{\rho+1}{(2+\rho)(1+\sin\chi)}},
\end{equation}
 respectively, the former of which is valid only for $0<\rho\leqq1$ while the latter is valid for all admissible values of $\rho$. It should be noted that $t_1$ vanishes  for $\rho=1$, while $s_3=s_{2}$, so that these two families of solutions have indeed a member in common. Thus, the total number of critical points $\xc$ of $\potor$ (including the poles) reduces to $8$ from the $10$  shown in \fref{fig:disk}. This latter, special instance is now illustrated in \fref{fig:disk_non-generic}, which shows that for $\rho=1$ a saddle with index $\iota=-2$ lies on the equator of $\sphere$ and it splits into two saddles with $\iota=-1$ as $\rho$ is either increased or decreased.

We close this case by recalling that no extra background solution exists for the present choice of parameters, as shown in \sref{sec:case-2}. 

\subsubsection{Case $\rho=0$, $K\neq 0$.}\label{sec:case_3}
This is the axis $\axis$ in parameter space (deprived of the centre $\Centre$). For this choice of parameters, \eref{eq:equilibrium_equations_b} requires that either $s=s_1=0$ or $s=s_{2,3}=\pm\sqrt{3}$. Inserting the former into \eref{eq:equilibrium_equations_a}, we obtain the roots
\begin{equation}
	\label{eq:t_1_2_bis}
	t_{1,2}=\frac{K \pm\sqrt{K^2+4}}{4},
\end{equation}
resulting in $4$ critical points $\x$. Similarly, the roots of \eref{eq:equilibrium_equations_a} corresponding to $s_{2,3}$ are
\begin{equation}
	\label{eq:t_3_4}
	t_{3,4}=-K\pm\sqrt{K^2+1},
\end{equation}
which together amount to $8$ critical points $\xc$. Adding the poles, also in view of \sref{sec:case-3}, we get the expected total of $14$ critical points for $\potor$ shown in \fref{fig:axis}.

To single out the special case of tetrahedral symmetry depicted in \fref{fig:tetra}, we require that $\potor=1$ at the critical point associated with the roots $s=0$ and negative $t$ in \eref{eq:t_1_2_bis} (as the maxima other than the North pole live in the Southern hemisphere of $\sphere$). Thus, by \eref{eq:lambda_solution}, \eref{eq:equilibrium_equations_s_t_2} and \eref{eq:s_t_mu}, we must have
\begin{equation}
	\label{eq:tetrahedral_equation}
	\lambda=\frac{K-t_2}{\sqrt{1+t_2^2}}=1,
\end{equation}
whose unique root is $K=1/\sqrt{2}$, as expected.

Two more cases deserve a special treatment, as they can be resolved explicitly by finding the roots of lower-degree polynomials. Geometrically, they are related to the special planes delimiting the sector of interest in parameter space \eref{eq:cylindrical_sector}. We treat these cases below, before addressing the generic, more complicated case.

\subsubsection{Case $\rho>0$, $\chi=-\pi/2$, $K>0$.}\label{sec:case_4}
For this choice of parameters, \eref{eq:equilibrium_equations_b} has the trivial solution $s=0$, which inserted in \eref{eq:equilibrium_equations_a} delivers
\begin{equation}
	\label{eq:t_1_2_ter}
	t=t_{1,2}=\frac{K\pm\sqrt{K^2-2(2-\rho)(\rho-1))}}{2(2-\rho)}.
\end{equation}
These are real for all $K>0$, if $0<\rho\leqq1$, but require $K\geqq K_2:=\sqrt{2(2-\rho)(\rho-1))}$ if $1\leqq\rho<2$. The corresponding critical points of $\potor$ are in general $4$, but for $K=K_2$ and $1\leqq\rho<2$, where they reduce to $2$.

The case $\rho=2$ deserves a special notice, as for $s=0$ there is a single root $t=1/2K$ and this branch of solutions only brings in $2$ critical points of $\potor$ (instead of $4$).

\begin{figure}[h]
	\centering
		\includegraphics[width=.5\linewidth]{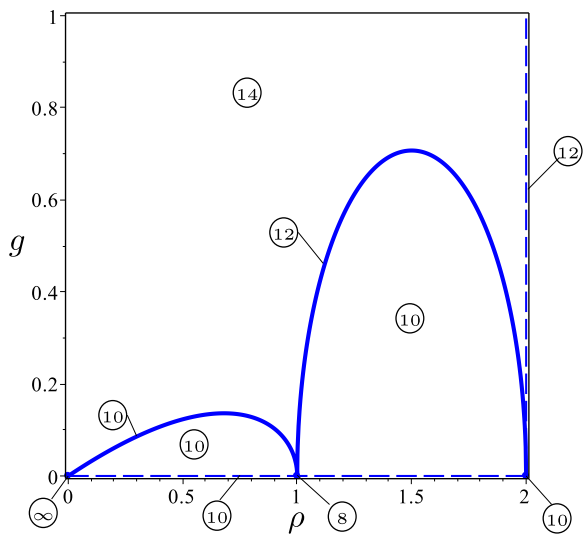}
	\caption{The graph of the function $g$ in \eref{eq:g_definition} against $\rho$. The numbers in different regions of the plane $\chi=-\pi/2$ indicate how many critical points $\potor$ possesses there. For this role, the graph of $g$ is a separating curve, or more shortly, a \emph{separatrix}.}
	\label{fig:g}
\end{figure}

For $s\neq0$, \eref{eq:equilibrium_equations_b} is also solved by
\begin{equation}
	\label{eq:t}
	t=\frac{K}{2\rho}(s^2-3),
\end{equation}
which transforms \eref{eq:equilibrium_equations_a} into a quadratic equation in $\sigma:=s^2$,
\begin{equation}
	\label{eq:biquadratic}
\fl\qquad	K^2(\rho+2)\sigma^2-[K^2(6+\rho)+2\rho^2(1+\rho)]\sigma+3K^2(6-\rho)+2\rho^2(\rho-1)=0,
\end{equation}
whose roots we shall denote $\sigma_1$ and $\sigma_2$. Elementary analysis shows that for $1\leqq\rho\leqq2$ both $\sigma_1$ and $\sigma_2$ are positive, and so they correspond to $8$ critical points of $\potor$. For $0<\rho\leqq1$, the picture is more articulated and changes as $K$ crosses the value
\begin{equation}
	\label{eq:K_1}
	K_1:=\sqrt{\frac{2\rho^2(1-\rho)}{3(6-\rho)}}.
\end{equation} 
For $0<K<K_1$, $\sigma_1$ is negative whereas $\sigma_2$ is positive; the number of corresponding critical points is $4$. For $K=K_1$, $\sigma_1$ vanishes, and so it is not an acceptable root, for $s\neq0$ on this branch; it does not bring any extra critical point, whereas the root $\sigma_2>0$ does brings in $4$, for a total of $10$ (including the poles). Finally, for $K>K_1$, both $\sigma_1$ and $\sigma_2$ are positive and the scene we see is the same as for $1<\rho<2$, with $14$ critical points in total. For $\rho=2$, putting together all roots, we obtain instead $12$ critical points for $\potor$ (see \fref{fig:g}).

The situation is more effectively summarized with the aid of the continuous function
\begin{equation}\label{eq:g_definition}
	g(\rho):=
	\cases{
		\sqrt{\frac{2\rho^2(1-\rho)}{3(6-\rho)}} & for $0\le\rho\le1$,\\
		\sqrt{2(2-\rho)(\rho-1)} & for $1\le\rho\le2$,}
\end{equation}  
whose graph is depicted in \fref{fig:g},
which also shows the total number of critical points of $\potor$ on $\sphere$ associated with different regions in the $\chi=-\pi/2$ plane in parameter space. For its role in separating regions with different numbers of critical points, the curve that represents the graph of $g$ is called a \emph{separatrix}. We shall see below how it extends to a \emph{surface} in parameter space.

\begin{figure}[h]
	\centering
	\begin{subfigure}[c]{0.3\linewidth}
		\centering
		\includegraphics[width=\linewidth]{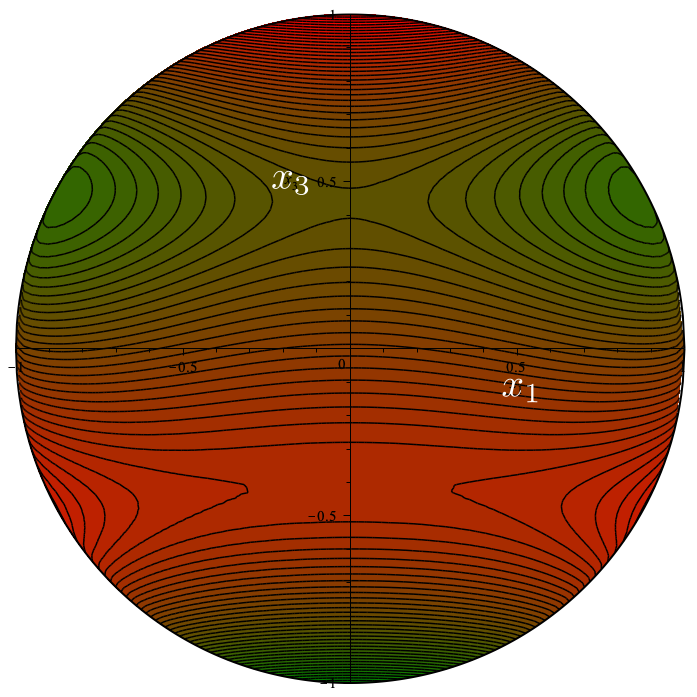}
		\caption{$\rho=1/4$}
		\label{fig:summary_a}
	\end{subfigure}
	\quad
	\begin{subfigure}[c]{0.3\linewidth}
		\centering
		\includegraphics[width=\linewidth]{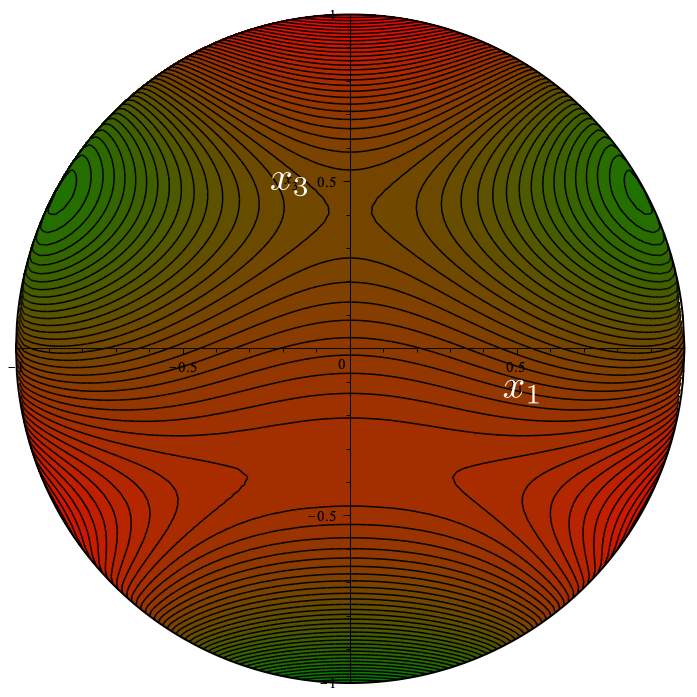}
		\caption{$\rho=2/4$}
		\label{fig:summary_b}
	\end{subfigure}
    \quad
	\begin{subfigure}[c]{0.3\linewidth}
	\centering
	\includegraphics[width=\linewidth]{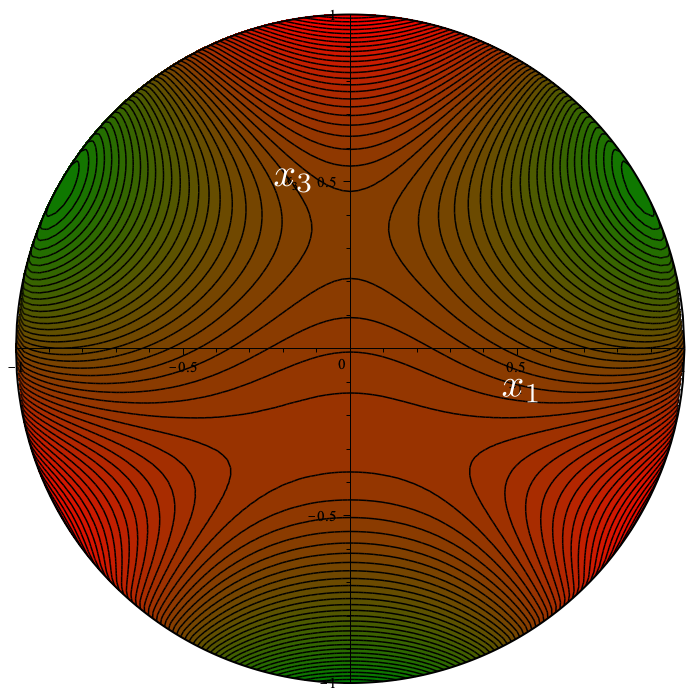}
	\caption{$\rho=3/4$}
	\label{fig:summary_c}
\end{subfigure}
    \quad
\begin{subfigure}[c]{0.3\linewidth}
	\centering
	\includegraphics[width=\linewidth]{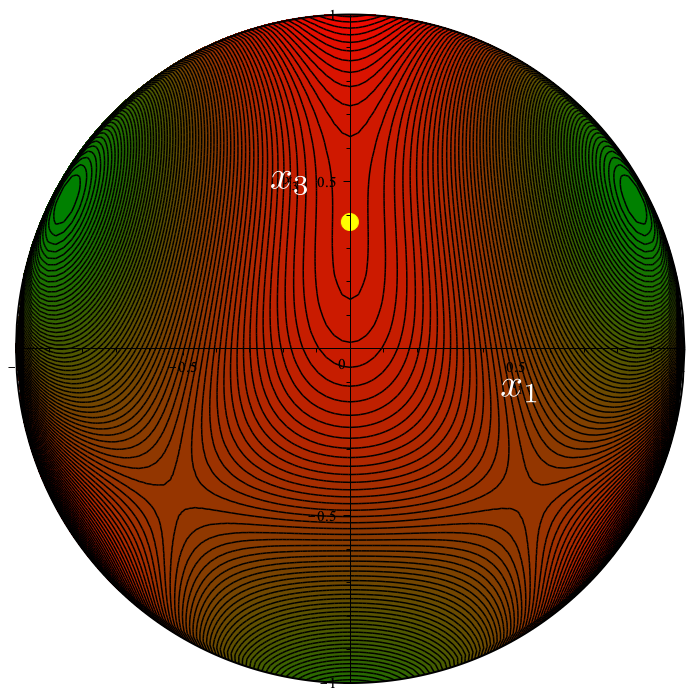}
	\caption{$\rho=5/4$}
	\label{fig:summary_d}
\end{subfigure}
    \quad
\begin{subfigure}[c]{0.3\linewidth}
	\centering
	\includegraphics[width=\linewidth]{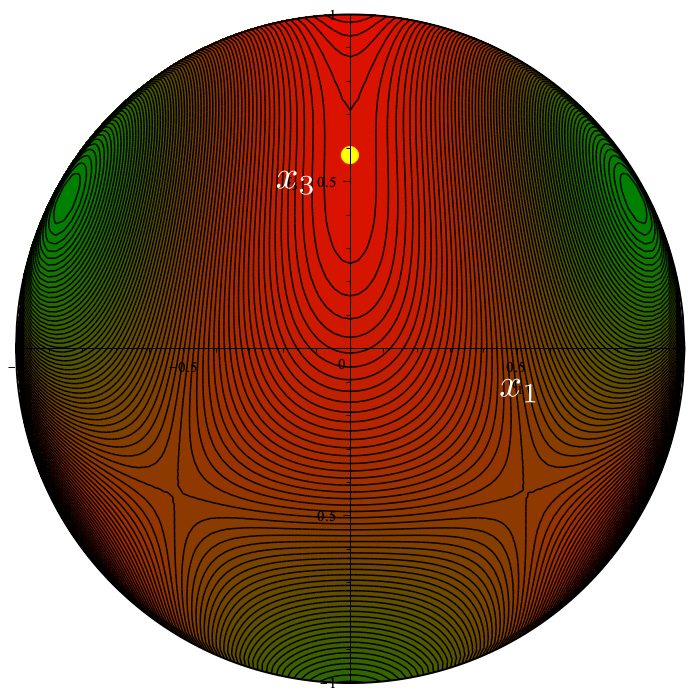}
	\caption{$\rho=6/4$}
	\label{fig:summary_e}
\end{subfigure}
    \quad
    \begin{subfigure}[c]{0.3\linewidth}
    	\centering
    	\includegraphics[width=\linewidth]{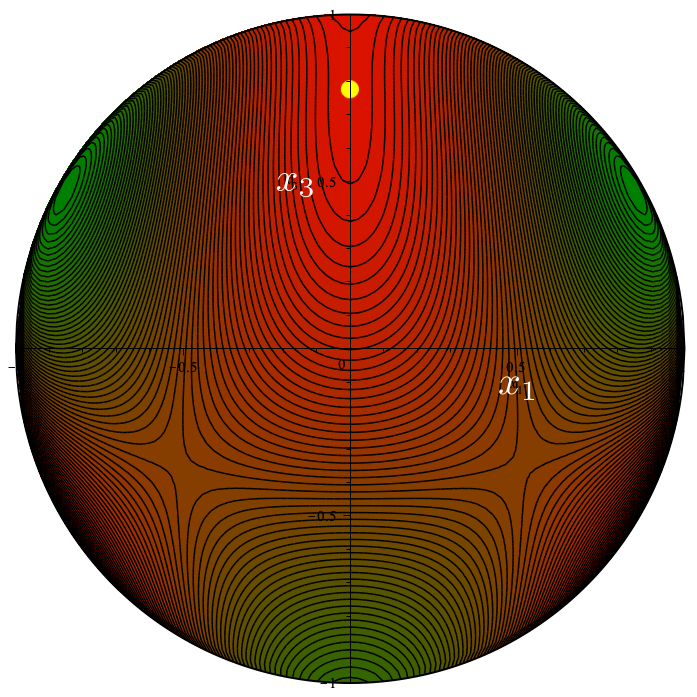}
    	\caption{$\rho=7/4$}
    	\label{fig:summary_f}
    \end{subfigure}
	\caption{Contour plots of $\potor$ on the plane $(x_1,x_3)$ illustrating the critical points of $\potor$ along the line $K=g(\rho)$ on the plane $\chi=-\pi/2$ in parameter space. The centre of the yellow circle in the last three panels designates the position of the degenerate saddle with index $\iota=0$.}
	\label{fig:summary}
\end{figure}

Here we are mainly interested in the algebraic avenue opened by Walcher~\cite{walcher:eigenvectors}, which may readily deliver the total number of critical points of $\potor$, and hence the real eigenvalues and eigenvectors of $\oct$. The following summary supplements the algebraic approach; it relies on stability and bifurcation analyses expounded in \cite{gaeta:symmetries}, to which the reader is referred for further details. 
\begin{enumerate}
	\item For $K > g(\rho)$ there are eight generic critical points beside the two at the poles and  four on the special great circle $x_1 = 0$, for a total of \emph{fourteen} critical points. Four are maxima, four minima, and the remaining six are saddles.  
	\item For $K < g(\rho)$, there are a total   of \emph{ten} critical points, of which three are maxima, three minima, and the remaining four are saddles.
	\item For $K=g(\rho)$, two different scenarios present themselves, according to whether $0<\rho<1$ or $1<\rho<2$. In the former case, the critical points are ten, whereas in the latter case they are \emph{twelve}. In both cases, the total number of maxima is three, as many as the minima; only the number of saddles differs: there are four for $0<\rho<1$ and six for $1<\rho<2$. In the former case, two saddles are degenerate, but all four have index $\iota=-1$. In the latter case, two out of the six saddles are degenerate and have index $\iota=0$ (marked by a yellow circle in \fref{fig:summary}), while the remaining four are not degenerate and have the usual index $\iota=-1$.
	\item The degenerate saddles with $\iota=0$ for $1<\rho<2$ migrate towards the poles as $\rho$ approaches $2$ along the line $K=g(\rho)$ and towards the equator as $\rho$ approaches $1$. Correspondingly, the North pole becomes a degenerate maximum (while the South pole becomes a degenerate minimum) and the equator hosts two symmetric ``monkey saddles'' \cite{gaeta:octupolar}.
\end{enumerate}   

\begin{figure}[h]
	\centering
	\includegraphics[width=.4\linewidth]{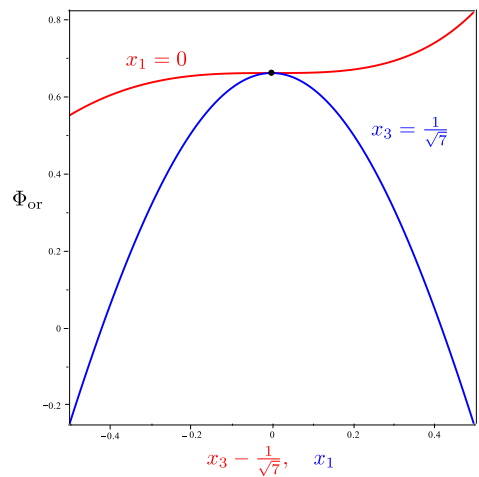}
	\caption{Sections of the graph of $\potor$ for the choice of parameters in \fref{fig:summary_d} on two orthogonal planes through the point depicted as a yellow circle in \fref{fig:summary_d}. The two planes of section have equations $x_1=0$ and $x_3=1/\sqrt{7}$, respectively.}
	\label{fig:iota_zero}
\end{figure}

Degenerate saddles with $\iota=0$ may be elusive and now we show why. \Fref{fig:iota_zero} illustrates the sections of the graph of $\potor$ with two orthogonal planes passing through such a saddle (that marked with a yellow circle in \fref{fig:summary_d}). While on one section the graph of $\potor$ has an inflection point, it has a maximum on the other section.

\begin{figure}[h]
	\centering
	\begin{subfigure}[c]{0.3\linewidth}
		\centering
		\includegraphics[width=\linewidth]{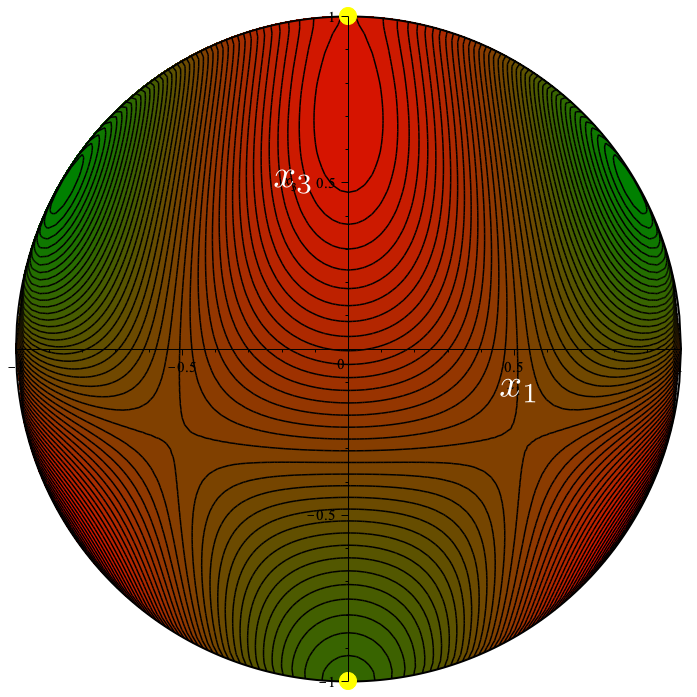}
		\caption{$K=1/2$}
		\label{fig:rho=2_a}
	\end{subfigure}
	\quad
	\begin{subfigure}[c]{0.3\linewidth}
		\centering
		\includegraphics[width=\linewidth]{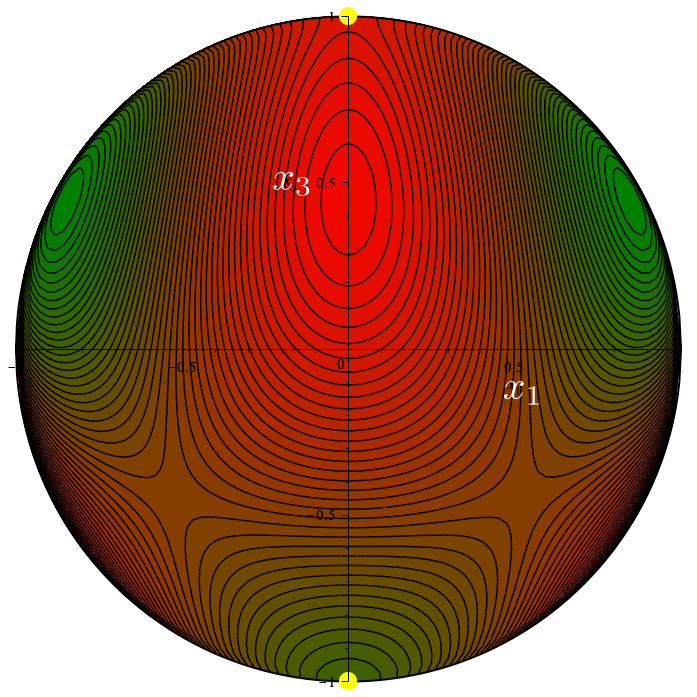}
		\caption{$K=1$}
		\label{fig:rho=2_b}
	\end{subfigure}
	\quad
	\begin{subfigure}[c]{0.3\linewidth}
		\centering
		\includegraphics[width=\linewidth]{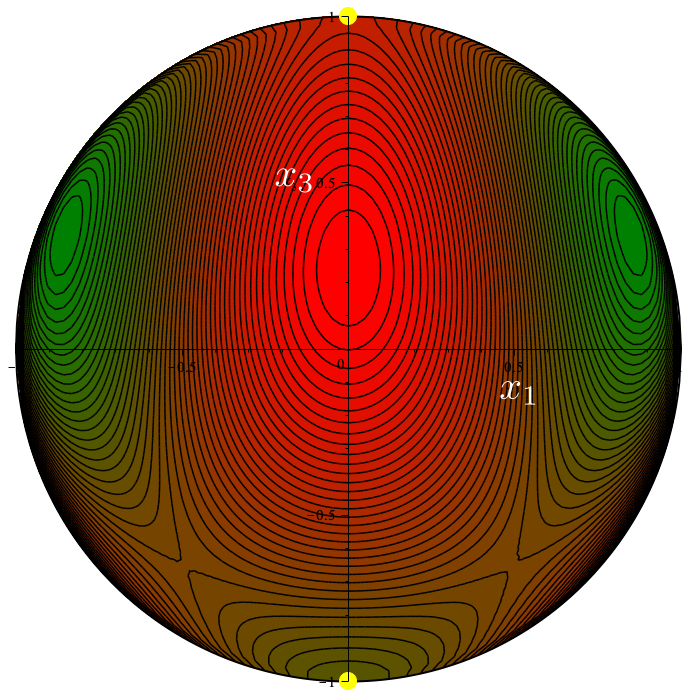}
		\caption{$K=2$}
		\label{fig:rho=2_c}
	\end{subfigure}
	\caption{Contour plots of $\potor$ on the plane $(x_1,x_3)$ for $\rho=2$, $\chi=-\pi/2$, and increasing values of $K$. The yellow circles designate the poles as degenerate saddles with $\iota=0$.}
	\label{fig:rho=2}
\end{figure}

The singular case $\rho=2$, which had escaped our analysis in \cite{gaeta:symmetries}, is further illuminated in \fref{fig:rho=2}, which shows how the $12$ critical points of $\potor$ move on $\sphere$ with increasing $K$. 

As seen in \sref{sec:case-4}, for $\chi=-\pi/2$ and $K>0$, $\potor$ has no critical point with $x_2=0$. Those critical points will however play a role in the case that we now study.

\subsubsection{Case $\rho>0$, $\chi=-\pi/2$, $K=0$.}\label{sec:case_5}
In this case, \eref{eq:equilibrium_equations_b} reduces to $st=0$, and correspondingly the solutions of \eref{eq:equilibrium_equations_a} are
\begin{equation}
	\label{eq:s_t_1}
	s=0,\qquad t=\pm\sqrt{\frac{1-\rho}{2(2-\rho)}},
\end{equation}
valid for $0<\rho\leqq1$, and
\begin{equation}
	\label{eq:s_t_2}s=\pm\sqrt{\frac{\rho-1}{\rho+1}},\qquad t=0,
\end{equation}
valid for $1\leqq\rho\leqq2$. They provide $4$ critical points $\xc$ of $\potor$ for $0<\rho<1$ and $1<\rho\leqq2$, to which we must now add the $4$ corresponding to the background solutions originated from the case studied in \sref{sec:case-5}. Thus, the total number of critical points is generically $10$ (once both poles are added). As made clear by comparing \eref{eq:s_t_1} and \eref{eq:s_t_2}, the case $\rho=1$ is singular, as there the four critical points identified by \eref{eq:s_t_1} and \eref{eq:s_t_2} collapse to $(0,\pm1,0)$, and so the total number of critical points reduces to $8$.

\begin{figure}[h]
	\centering
	\begin{subfigure}[c]{0.3\linewidth}
		\centering
		\includegraphics[width=\linewidth]{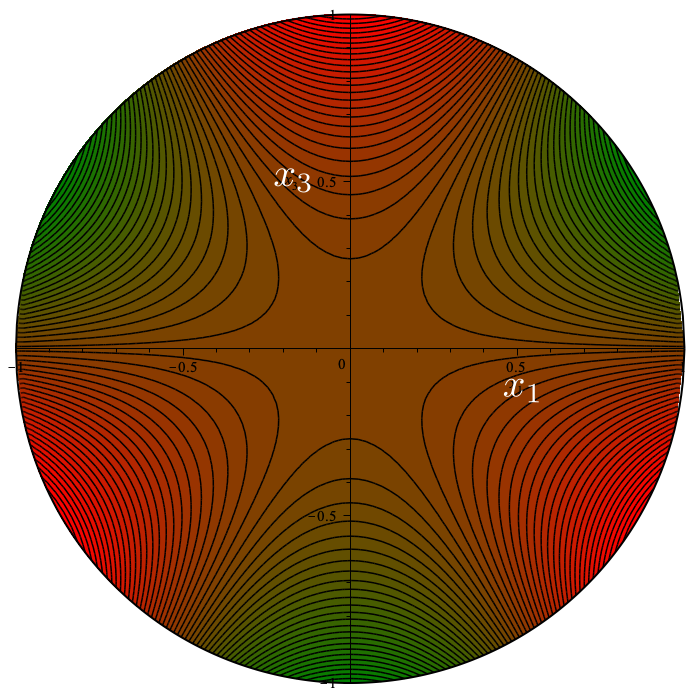}
		\caption{$\rho=1$}
		\label{fig:comparison_a}
	\end{subfigure}
	\quad
	\begin{subfigure}[c]{0.3\linewidth}
		\centering
		\includegraphics[width=\linewidth]{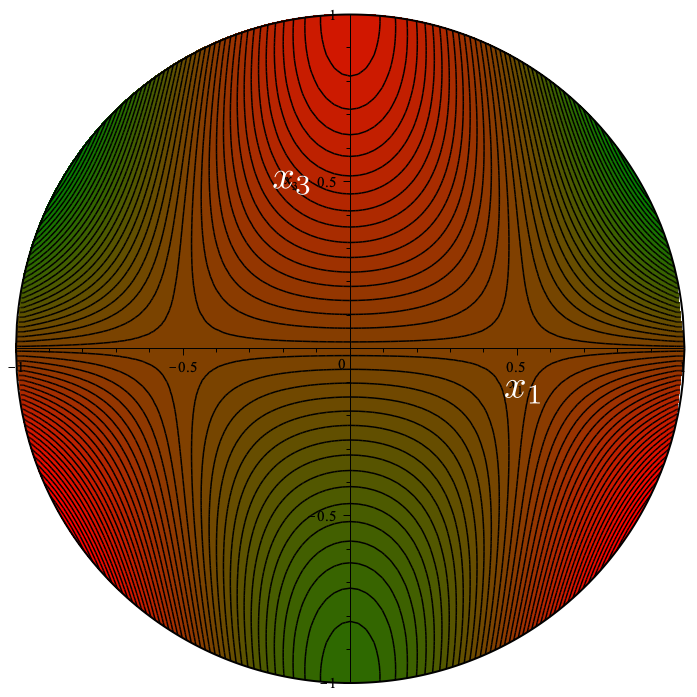}
		\caption{$\rho=2$}
		\label{fig:comparison_b}
	\end{subfigure}
	\caption{Contour plots of $\potor$ on the plane $(x_1,x_3)$ in two special cases for $\chi=-\pi/2$ and $K=0$, with $8$ and $10$ critical points, respectively.}
	\label{fig:comparison}
\end{figure}

For $\rho=1$, three maxima and three minima are accompanied by two degenerate saddles, each with index $\iota=-2$. By contrast, For $\rho=2$, the same number of maxima and minima are accompanied by four degenerate saddles, each  with index $\iota=-1$, for a total of ten critical points (see \fref{fig:comparison}).

\subsubsection{Case $\rho>0$, $\chi=-\pi/6$, $K>0$.}\label{sec:case_6}
As remarked above, by the $2\pi/3$ covariance enjoyed by $\potor$, the plane in parameter space where $\chi=-\pi/6$ can be identified with the the plane where $\chi=\pi/2$. Clearly, the graph of $\potor$ would rotate around the $x_3$- axis as a consequence of the change in $\chi$, but neither the number nor the nature of its critical points  would change.

A glance at equation \eref{eq:x_2=0_star_2} for $\chi=\pi/2$ and $K>0$ suffices to show that there is no background solution in this case. As for the critical points of $\potor$ with $x_2\neq0$, they are determined by the roots $(s,t)$ of \eref{eq:equilibrium_equations_a} and \eref{eq:equilibrium_equations_b}, which now read as
\begin{eqnarray}
(2+\rho)t^2+K(s^2-1)t+\frac12(\rho-1)s^2-\frac12(\rho+1)=0,\label{eq:equilibrium_equations_a_1}\\
2\rho st=Ks(3-s^2),\label{eq:equilibrium_equations_b_2} 
\end{eqnarray}	
respectively.

\begin{figure}[h]
	\centering
		\includegraphics[width=.5\linewidth]{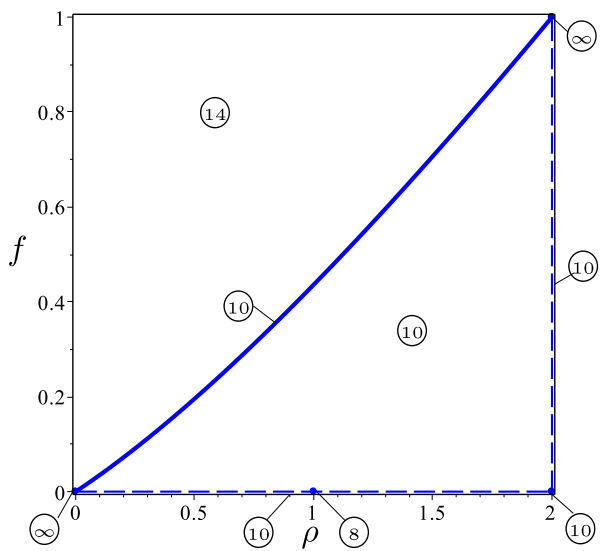}
	\caption{The graph of the function $f$ defined in \eref{eq:f} is plotted against $\rho$ in the interval $0\leqq\rho\leqq2$. It divides the plane $\chi=-\pi/6$ in a number of regions with a different total number of critical points of $\potor$ on $\sphere$ (including the poles).}
	\label{fig:f}
\end{figure}

The latter is solved for either $s=0$ or
\begin{equation}
	\label{eq:t_1}
	t=t_1=K\frac{3-s^2}{2\rho}\quad\mathrm{if}\quad s\neq0.
\end{equation}
Letting $s=0$ in \eref{eq:equilibrium_equations_a_1}, we obtain a quadratic equation for $t$ with roots
\begin{equation}
	\label{eq:t_2_3}
	t_{2,3}=\frac{K\pm\sqrt{K^2+2(\rho+1)(\rho+2)}}{2(\rho+2)},
\end{equation}
which amount to $4$ critical points $\xc$ of $\potor$ on $\sphere$. Setting $t=t_1$ in \eref{eq:equilibrium_equations_a_1} reduces the latter to a quadratic equation in $\sigma:=s^2$,
\begin{equation}
	\label{eq:biquadratic_bis}
\fl	K^2(\rho-2)\sigma^2+2[K^2(6-\rho)+\rho^2(1-\rho)]\sigma-3K^2(6+\rho)+2\rho^2(1+\rho)=0.
\end{equation}
An elementary analysis shows that for $0<\rho<2$ the roots $\sigma_{1,2}$ of this equation are both positives if $K>f(\rho)$, where
\begin{equation}
	\label{eq:f}
	f:=\sqrt{\frac{2\rho^2(1+\rho)}{3(6+\rho)}},
\end{equation}
whereas $\sigma_1=0$ and $\sigma_2>0$ if $K=f(\rho)$, and $\sigma_1<0$ and $\sigma_2>0$ if $K<f(\rho)$. Correspondingly, in complete analogy to our discussion in \sref{sec:case_6}, in the interval $0<\rho<2$, the critical points $\xc$ of $\potor$ on $\sphere$ are $10$ for $K\leqq f(\rho)$ and $14$ for $K>f(\rho)$ (see \fref{fig:f}).

The case $\rho=2$ is once again exceptional, as equation \eref{eq:biquadratic_bis} reduces to 
\begin{equation}\label{eq:biquadratic_bis_reduced}
	(K^2-1)(\sigma-3)=0.
\end{equation}
This shows that, for $K\neq1$, $s=\pm3$ and $t=0$ are the only solutions in one branch (to be accompanied by $s=0$ and $t=t_{2,3}$ in the other branch), which amounts to a total of $10$ critical points for $\potor$. Furthermore, if $K=1$, \eref{eq:biquadratic_bis_reduced} is identically satisfied, and so \eref{eq:t_1} delivers a whole orbit of solutions in this branch, to be again supplemented by $s=0$ and $t=t_{2,3}$ in the accompanying branch. It is easily seen that here $t_2=-1/2$ and $t_3=3/4$; the latter is subsumed in the orbit of the first branch (for $s=0$, of course), whereas the former is not. This special case, where $\potor$ has infinitely many critical points, is nothing but the one considered in  \sref{sec:case_1} above, corresponding to the centre $\Centre$ in parameter space; only, the graph of $\potor$ is rotated in space.

\Fref{fig:f} shows the graph of $f$ in \eref{eq:f} marked with the total number of critical points of $\potor$ in different regions of the plane $\chi=-\pi/6$.

\begin{figure}[h]
	\centering
	\begin{subfigure}[c]{0.3\linewidth}
		\centering
		\includegraphics[width=\linewidth]{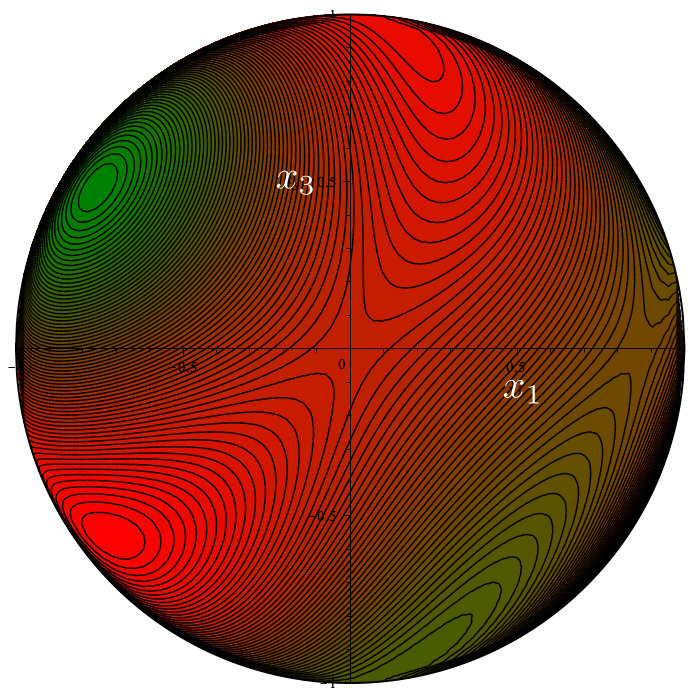}
		\caption{$K=1/2$}
		\label{fig:rho=2_bis_a}
	\end{subfigure}
	\quad
	\begin{subfigure}[c]{0.3\linewidth}
		\centering
		\includegraphics[width=\linewidth]{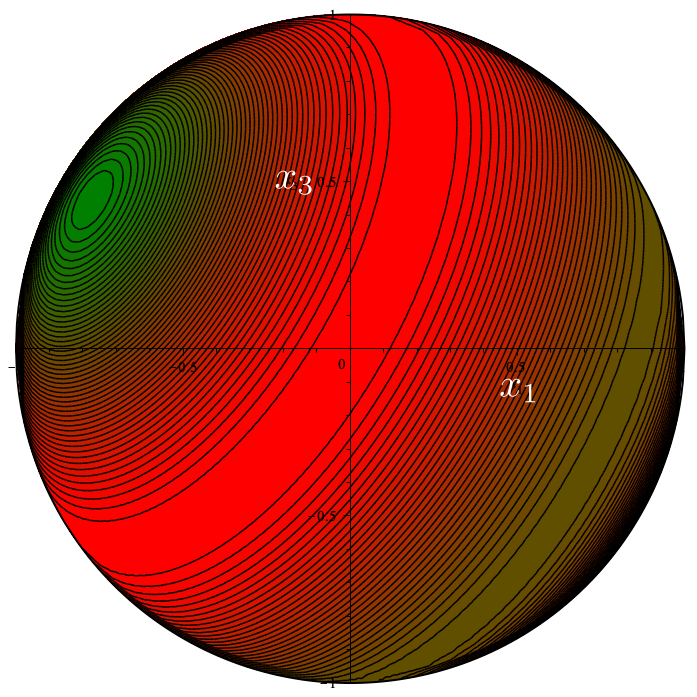}
		\caption{$K=1$}
		\label{fig:rho=2_bis_b}
	\end{subfigure}
\begin{subfigure}[c]{0.3\linewidth}
	\centering
	\includegraphics[width=\linewidth]{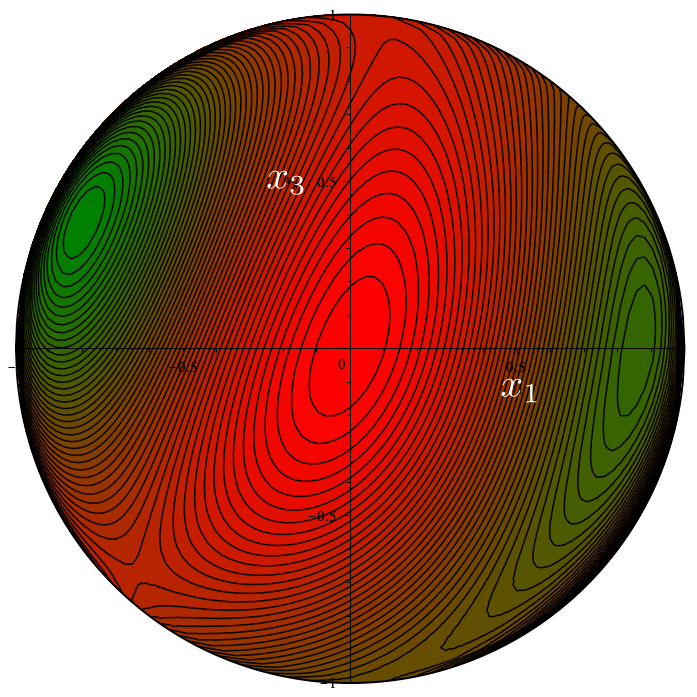}
	\caption{$K=2$}
	\label{fig:rho=2_bis_c}
\end{subfigure}
	\caption{Contour plots of $\potor$ on the plane $(x_1,x_3)$ for $\rho=2$, $\chi=-\pi/6$, and values of $K$ on both sides of the singular value $K=1$. The scenery is quite different in the two adjoining cases, but the total number of critical points is still $10$ for both.}
	\label{fig:rho=2_bis}
\end{figure}

The special case $\rho=2$ is further illuminated in \fref{fig:rho=2_bis}, which suggests a radical change of scenery in the arrangement of the critical points of $\potor$  as $K$ crosses the singular value $K=1$.

Having completed the survey of all special cases where the critical points of $\potor$ are decided by the roots of a low-degree polynomial, we are in a position to address the generic case, which will require handling a polynomial of degree $6$.

\subsubsection{Generic case.}\label{sec:case_generic}
This is the case where $0<\rho\leqq2$, $\pi/2<\chi<-\pi/6$, and $K>0$.  \Eref{eq:equilibrium_equations_b} can be solved for $t$, provided that $s\neq s_\pm$, where
\begin{equation}
	\label{eq:s+-}
	s_\pm:=\tan\chi\pm\sqrt{1+\tan^2\chi}
\end{equation}	 
are the roots of the quadratic polynomial in $s$ on the left hand side of \eref{eq:equilibrium_equations_b}. Whit $t$ thus given by
\begin{equation}
	\label{eq:t_last}
	t=\frac{Ks(s^2-3)}{\rho[(s^2-1)\cos\chi-2s\sin\chi]},
\end{equation}
equation \eref{eq:equilibrium_equations_a} reduces to the polynomial
\begin{equation}
	\label{eq:Walcher_polynomial}
	W(s):=\sum_{i=0}^6S_is^i=0,
\end{equation}
whose coefficients  $S_i$ are given by
\begin{eqnarray}\label{eq:S_0-6}
	\fl\qquad\cases{
S_0:=-\rho^2\cos^2\chi(1+\rho\sin\chi),\\
S_1:=-6K^2\rho\cos\chi+2\rho^2\cos\chi(3\rho\cos^2\chi-2\sin\chi-2\rho),\\
S_2:=6K^2(\rho\sin\chi+6)+5\rho^2\cos^2\chi(3\rho\sin\chi+1)-4\rho^2(1+\rho\sin\chi),\\
S_3:=4\rho\cos\chi[\rho^2(4-5\cos^2\chi)-K^2],\\
S_4:=4K^2(\rho\sin\chi-6)+5\rho^2\cos^2\chi(1-\rho\sin\chi)+4\rho^2(\rho\sin\chi-1),\\
S_5:=2\rho\cos\chi[K^2+\rho(3\rho\cos^2\chi+2\sin\chi-2\rho)],\\
S_6:=2K^2(2-\rho\sin\chi)+\rho^2\cos^2\chi(\rho\sin\chi-1).
}
\end{eqnarray}
Every real root $s\neq s_\pm$ of $W$, once combined with $t$ as in \eref{eq:t_last}, corresponds to two (antipodal) critical points of $\potor$ on $\sphere$. So, if all roots of $W$ are real and not coincident with either $s_\pm$, and if the case in \sref{sec:case-generic} for the existence of background solutions does not apply, then $\potor$ possesses $14$ critical points (two of which are at the poles), thus reaching the allowed maximum number, according to the theorem of \cite{cartwright:number} applied to tensor $\oct$.

We see now that only $s_+$ can be a spurious root of $W$ (and must be suppressed) in the selected sector of parameter space \eref{eq:cylindrical_sector} where our analysis is confined. This follows from a direct inspection, which yields
\begin{equation}
	\label{eq:Walcher_inspected}
	W(s_\pm)=4K^2(2\mp\rho)\frac{(\sin\chi\pm1)(2\sin\chi\mp1)^2}{\cos^2\chi(\sin\chi\mp3)\pm4(1\mp\sin\chi)},
\end{equation}
so that, for $0<\rho\leqq2$ and $-\pi/2<\chi<-\pi/6$, only $W(s_+)$ vanishes, for $\rho=2$. This shows that on the lateral boundary of the selected sector one root of $W$ is inadmissible and two critical points of $\potor$ are lost. In particular, for $\rho=0$, whenever  $W$ has $6$ real roots, $\potor$ possesses only $12$ critical points, not $14$.

Next we prove that $W$ has indeed $6$ real roots asymptotically for large $K$. It follows from \eref{eq:Walcher_polynomial} and \eref{eq:S_0-6} that for $K\gg1$
\begin{equation}
	\label{eq:W_asymptotic}
	W(s)=-K^2s(s^2-3)w(s)+O(1),
\end{equation}
where
\begin{equation}
	\label{eq:w_polynomial}
	w(s):=(\rho\sin\chi-2)s^3-\rho\cos\chi s^2+(\rho\sin\chi+6)s-\rho\cos\chi.
\end{equation}
The algebraic discriminant $\Delta(w)$ of $w$ is readily computed,
\begin{equation}
	\label{eq:discriminant_w}
	\Delta(w)=4[432-\rho^4+16\sin\chi(4\cos^2\chi-1)\rho^3-72\rho^2]
\end{equation}
and can be shown to be positive for $0<\rho\leqq2$ and $\pi/2<\chi<-\pi/6$ (it vanishes only along a line in our selected sector, where $\rho=2$ and $\chi=-\pi/6$). Thus, for $K$ sufficiently large, all roots of $W$ are real, and since the function $\kappa(\rho,\chi)$ in \eref{eq:x_2=0_K_admissible} is bounded, we easily conclude that $\potor$ possesses $14$ critical points on $\sphere$.

It should be noted that $S_6$ vanishes precisely for $K=\kappa(\rho,\chi)$. This means that when $W$ becomes a polynomial of degree $5$,  loosing at least one root (and $\potor$, correspondingly,  $2$ critical points), the $2$ critical points connected with the background solutions studied in \sref{sec:case-generic} come into the picture, replacing the lost ones. Actually, it follows from \eref{eq:x_solutions} that whenever a root of $W$ flies to $\pm\infty$, the corresponding critical points of $\potor$ approach the great circle of $\sphere$ where $x_2=0$, so that crossing the surface $K=\kappa(\rho,\chi)$ in parameter space does not result in a discontinuity of the critical points of $\potor$, neither for their number nor for their position. This suggests that the background solutions classified in \sref{sec:background} could never play a role for $K>0$. There is, however, one singular instance where they can, if $W$ loses more than one root. This is the case where $S_6$ vanishes alongside with $S_0$, $S_1$, and $S_2$, along the curve in the space $(\rho,\chi,K)$ parameterized by
\begin{equation}
	\label{eq:line_cusps}
	\chi=-\arcsin\left(\frac1\rho\right),\quad K=h(\rho):=\sqrt{\frac{\rho^2-1}{3}},\quad\mathrm{for}\quad1<\rho\leqq2,
\end{equation} 
where $W$ reduces to
\begin{equation}
	\label{eq:W_0}
	W_0(s):=\frac83(\rho^2-4)s^3(\sqrt{\rho^2-1}s^2+s-2\sqrt{\rho^2-1}).
\end{equation}   
The polynomial $W_0$ has clearly $3$ distinct real roots that generate $6$ critical points of $\potor$, to which we must add the $2$ associated with the background solutions and the $2$ poles (as usual), for a total of $10$ critical points.

We know from our analysis for $K=0$ that the total number of real roots of $W$ must decrease upon decreasing $K$. Since all coefficients of $W$ are real, this can only happen through the coalescence of two real roots. To identify the critical values of $K$, for given $\rho$ and $\chi$, where this takes place, we need to find a common root for $W(s)$ and its derivative $W'(s)$. The conventional way is to compute the algebraic discriminant  $\Delta(W)$ of $W$ and look for its roots. Unfortunately, $\Delta(W)$ turns out to possess a very complicated expression (involving a polynomial of degree $20$ in $\rho$).

Our strategy will be different. The system requiring that both $W$ and $W'$ vanish has the following general structure
\begin{eqnarray}
K^2a_{11}+a_{12}=0,	\label{eq:system_structure_1}\\
K^2a_{21}+a_{22}=0,	\label{eq:system_structure_2}
	\end{eqnarray} 
where $a_{ij}=a_{ij}(\rho,\chi)$ are the entries of a matrix $A$. This system is compatible only if $\det A=0$, which turns out to be a polynomial in $s$ of degree $10$, whose complex roots can easily be computed numerically. Among these, we are only interested in the real roots $s_\ast$ that deliver a positive $K^2$ through either \eref{eq:system_structure_1} or \eref{eq:system_structure_2}; these are as well all possible double roots of $W$. We systematically found a single root $s_\ast$ for all $0<\rho\leqq2$ and $-\pi/2<\chi<-\pi/6$.

\begin{figure}[h]
	\centering
	\begin{subfigure}[t]{0.4\linewidth}
		\centering
		\includegraphics[width=\linewidth]{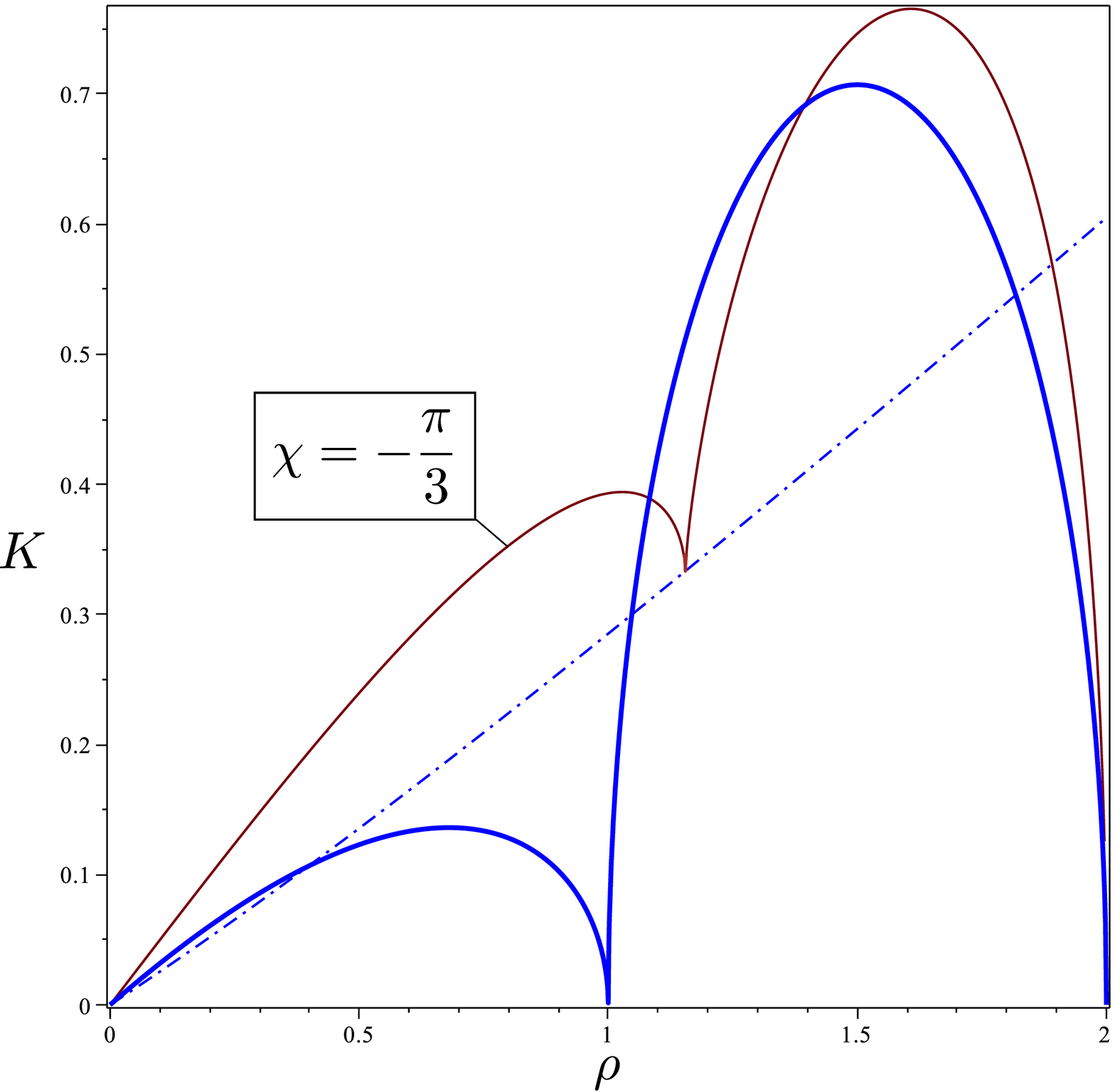}
		\caption{The graphs of $K_\ast$ (brown solid line) and $\kappa$ (blue dash-dotted line) against $\rho$ for $\chi=-\pi/3$ . The graph of $g$ (blue solid line), which is a separatrix for $\chi=-\pi/2$,  is drawn for reference (see \fref{fig:g}).}
		\label{fig:P5_plot}
\end{subfigure}
	\qquad\quad
	\begin{subfigure}[t]{0.4\linewidth}
	\centering
	\includegraphics[width=\linewidth]{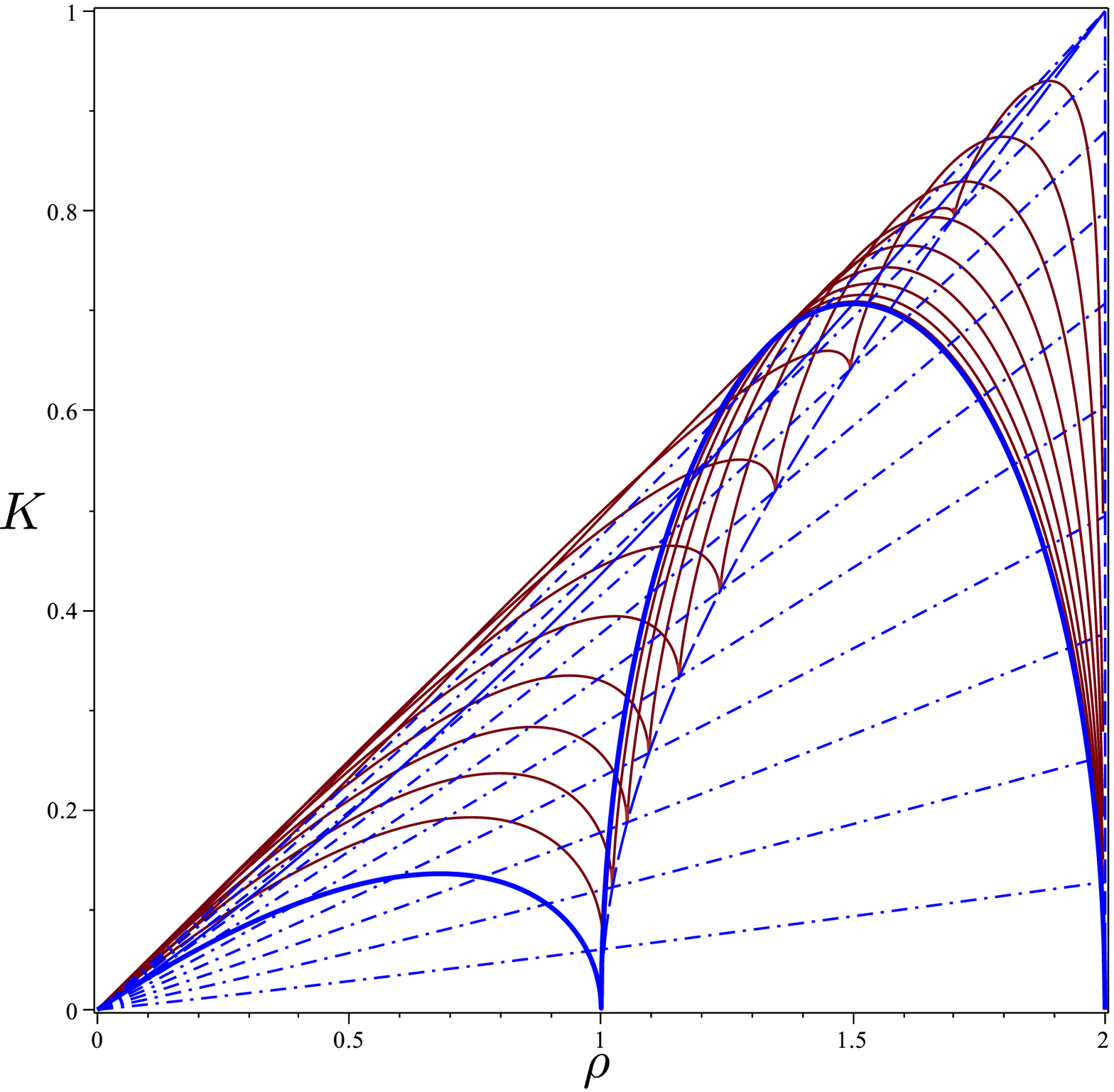}
	\caption{The graphs of $K_\ast$ (brown solid lines) and $\kappa$ (blue dash-dotted lines) against $\rho$ for $\chi=-\pi/2+j\pi/30$, $j=\{1,\dots,9\}$. The cusps on the separatrix lie all on the graph (blue long-dashed line) of the function $h$  defined in \eref{eq:line_cusps}. Here, barely visible in the background, is also the graph of the function $f$, which is a separatrix for $\chi=-\pi/6$ (see \fref{fig:f}).}
	\label{fig:P_plot_all}
\end{subfigure}
\caption{Representations of the separatrix $K=K_\ast(\rho,\chi)$ and the surface $K=\kappa(\rho,\chi)$ described by \eref{eq:x_2=0_K_admissible}.}
	\label{fig:P_plots}
\end{figure}

\Fref{fig:P5_plot} shows the critical value $K_\ast$ of $K$ corresponding to $s_\ast$ for $\chi=-\pi/3$ and $0\leqq\rho\leqq2$, along with the graph of the function $\kappa$ defined in \eref{eq:x_2=0_K_admissible}.
The graph of $K_\ast$ has two branches connected by a cusp at $\rho=\rho_\mathrm{c}$; along the branch with $\rho<\rho_\mathrm{c}$, $s_\ast<0$, whereas $s_\ast>0$ for $\rho>\rho_\mathrm{c}$; for $\rho=\rho_\mathrm{c}$, where $K_\ast$ and $\kappa$ cross, $s_\ast=0$. Thus, on both branches of $K_\ast$ the $5$ distinct real roots of $W$ correspond to $12$ critical points of $\potor$ (including the poles), whereas on the cusp the $3$ distinct real roots of $W$ and the $2$ background solutions amount to $10$ critical points of $\potor$. \Fref{fig:P_plot_all} illustrates the branches of $K_\ast$ and their cusp for a sequence of values $\chi$ in our selected sector \eref{eq:cylindrical_sector}. They play the same separating role that $g$ and $f$ play on the planes $\chi=-\pi/2$ and $\chi=-\pi/6$, respectively. Together they form a two-vaulted surface, which we call the \emph{separatrix}, traversed by a \emph{groin} represented by the line of cusps described by \eref{eq:line_cusps}.

\begin{figure}[h]
	\centering
	\includegraphics[width=.5\linewidth]{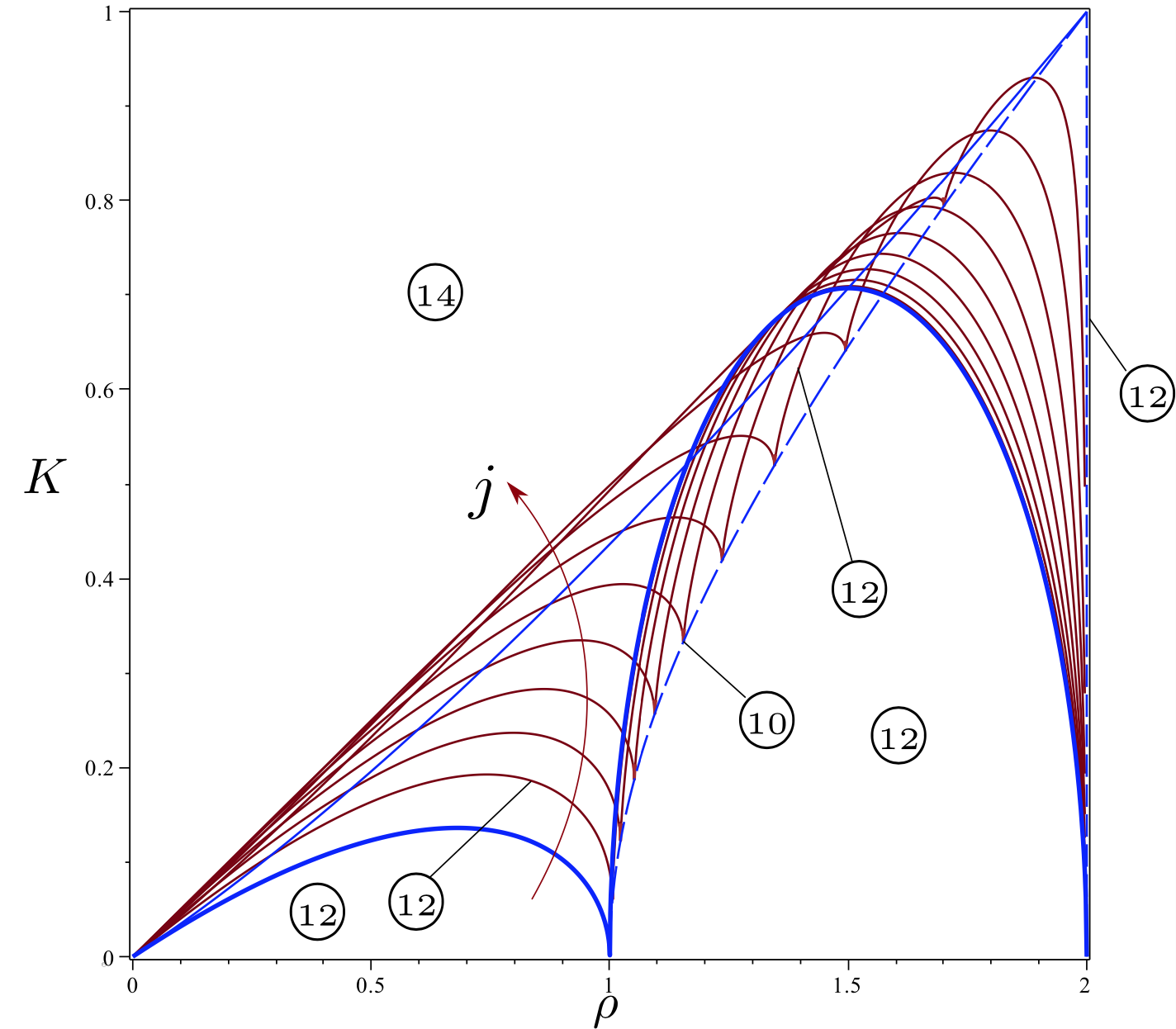}
	\caption{Overview on the critical points of $\potor$ for $0\leqq\rho\leqq2$ and $\chi=-\pi/2+j\pi/30$, $j=\{1,\dots,9\}$ as in \fref{fig:P_plot_all}. The graph of $g$ (thick blue solid line) and $f$ (thin blue solid line) are shown for comparison. The (blue dashed) line at $\rho=2$ is where $2$ critical points are lost as $W$ acquires the spurious root $s_+$. The whole landscape of critical points is obtained by combining this picture with figures~\ref{fig:g} and \ref{fig:f}.}
	\label{fig:critical_point_count}
\end{figure}

Above the separatrix, $\potor$ has $14$ critical points, below and on each vault of the separatrix $\potor$ has $12$ critical points, whereas it has only $10$ on the groin (see \fref{fig:critical_point_count}).

\begin{figure}
	\centering
	\begin{subfigure}[t]{0.3\linewidth}
		\centering
		\includegraphics[width=\linewidth]{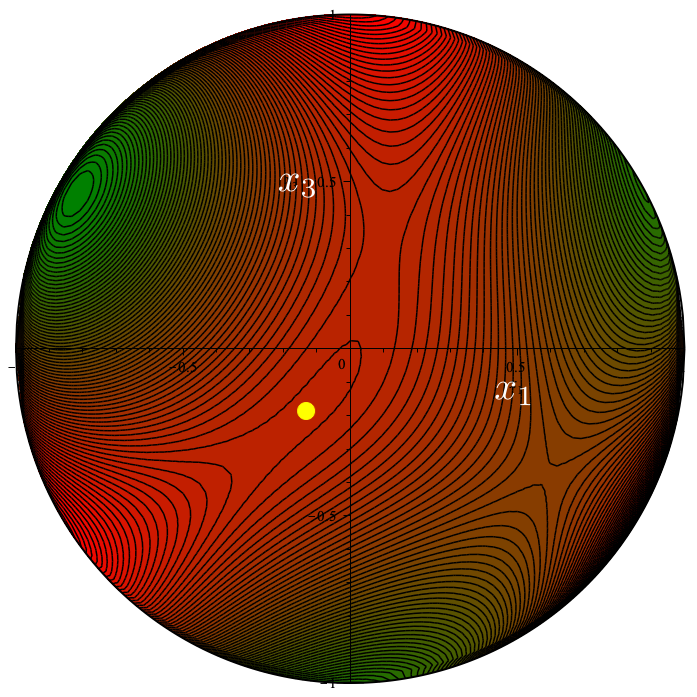}
		\caption{$\rho=1.10$}
		\label{fig:merge_cusp_a}
	\end{subfigure}
	\quad
	\begin{subfigure}[t]{0.3\linewidth}
		\centering
		\includegraphics[width=\linewidth]{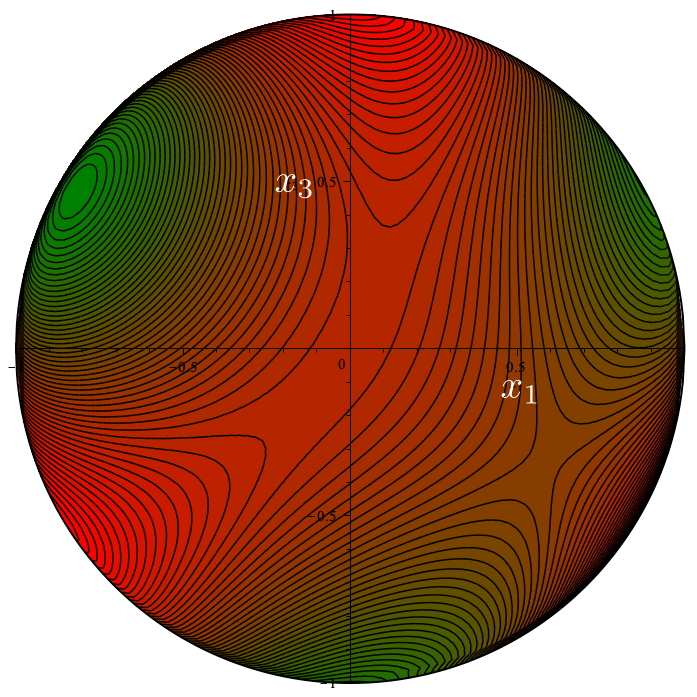}
		\caption{$\rho\doteq1.15$}
		\label{fig:merge_cusp_b}
	\end{subfigure}
	\quad
	\begin{subfigure}[t]{0.3\linewidth}
		\centering
		\includegraphics[width=\linewidth]{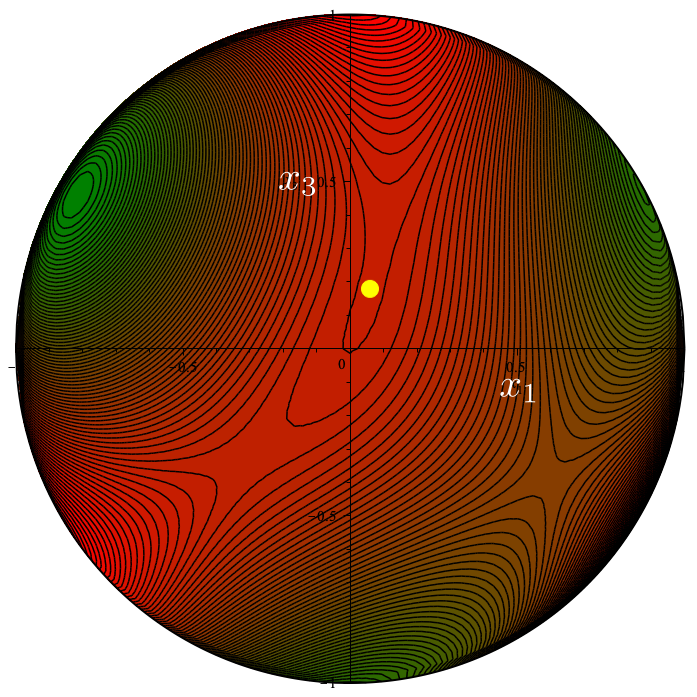}
		\caption{$\rho=1.20$}
		\label{fig:merge_cusp_c}
	\end{subfigure}
	\caption{For $\chi=-\pi/3$, the cusp in the separatrix is hit at $\rho\doteq1.15$ (see \fref{fig:P5_plot}). Here we show the contour plots of $\potor$ on the plane $(x_1,x_3)$ on the two branches of the separatrix (panels (a) and (c)) and on the cusp (panel (b)). $\potor$ has $12$ critical points in (a) and (c), two of which are degenerate saddles with $\iota=0$ (marked by yellow circles). $\potor$ has $10$ critical points in panel (b), none of which has index $\iota=0$.}
	\label{fig:merge_cusp}
\end{figure}

The behaviour of $K_\ast$ around a cusp can be obtained from a standard asymptotic analysis. For given $\chi$, the value of $K_\ast$ at the cusp is delivered by setting $\rho=-1/\sin\chi$ in $h(\rho)$  as defined by \eref{eq:line_cusps}. For $\rho$ close to this value, $K_\ast$ is expressed by
\begin{equation}
	\label{eq:K_near_cusp}
\fl\qquad	K_\ast=-\frac{1}{\sqrt{3}}\frac{\cos\chi}{\sin\chi}+\frac{3^{1/6}}{2^{4/3}}\left(\frac{3-4\cos^2\chi}{\cos\chi\sin^2\chi}\right)\left(\rho+\frac{1}{\sin\chi}\right)^{2/3}
	+O\left(\rho+\frac{1}{\sin\chi}\right).
\end{equation} 
\Fref{fig:merge_cusp} shows how two critical points of $\potor$ merge upon approaching the cusp from both branches of the separatrix for a given value of $\chi$: a degenerate saddle with $\iota=0$ and a standard saddle with $\iota=-1$ coalesce into a standard saddle.

\begin{figure}
	\centering
	\begin{subfigure}[t]{0.3\linewidth}
		\centering
		\includegraphics[width=\linewidth]{CP_rho1o2_chi-Pio2_Kg_deco.png}
		\caption{$\rho=1/2$, $\chi=-\pi/2$}
		\label{fig:approaching_1_a}
	\end{subfigure}
	\quad
	\begin{subfigure}[t]{0.3\linewidth}
		\centering
		\includegraphics[width=\linewidth]{CP_rho1_chi-Pio2_Kg_deco.png}
		\caption{$\rho=1$, $\chi=-\pi/2$}
		\label{fig:approaching_1_b}
	\end{subfigure}
	\quad
	\begin{subfigure}[t]{0.3\linewidth}
		\centering
		\includegraphics[width=\linewidth]{CP_rho3o2_chi-Pio2_Kg_deco.png}
		\caption{$\rho=3/2$, $\chi=-\pi/2$}
		\label{fig:approaching_1_c}
	\end{subfigure}\\
\begin{subfigure}[t]{0.3\linewidth}
	\centering
	\includegraphics[width=\linewidth]{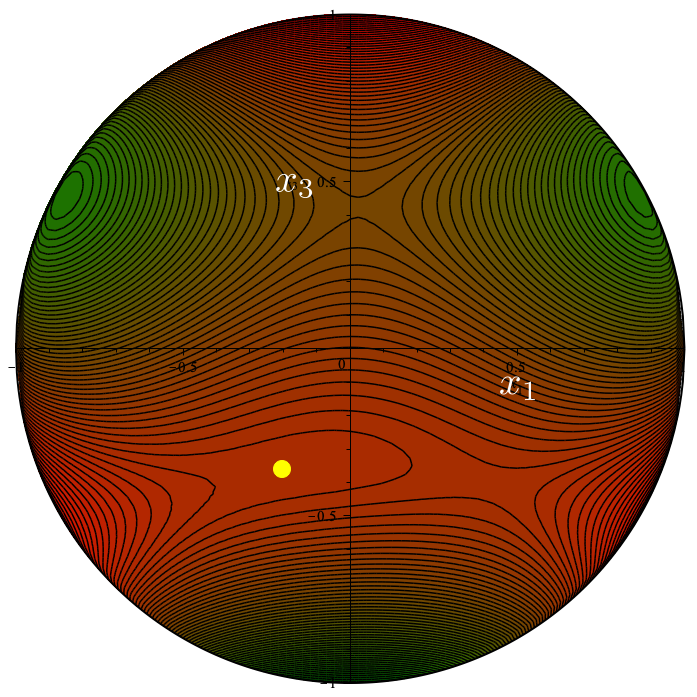}
	\caption{$\rho=1/2$, $\chi=-7\pi/15$}
	\label{fig:approaching_1_d}
\end{subfigure}
\quad
\begin{subfigure}[t]{0.3\linewidth}
	\centering
    \includegraphics[width=\linewidth]{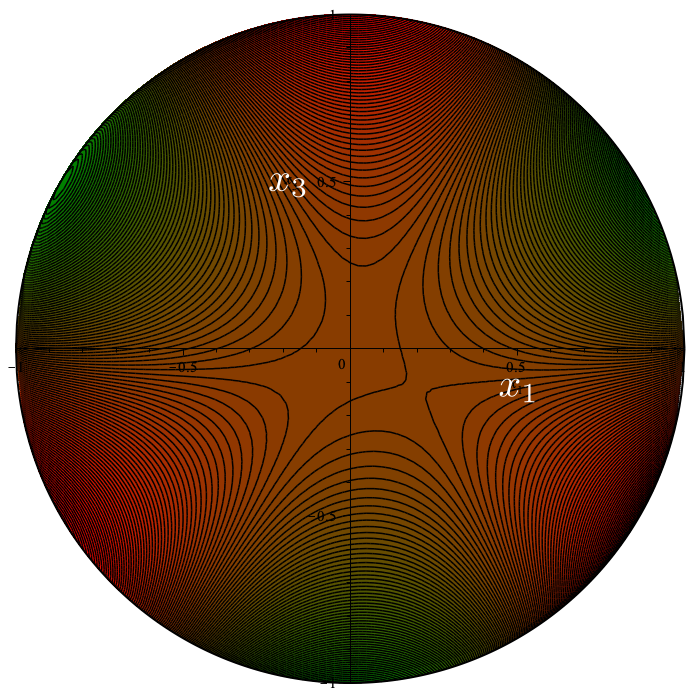}
	\caption{$\rho\doteq1.006$, $\chi=-7\pi/15$}
	\label{fig:approaching_1_e}
\end{subfigure}
\quad
\begin{subfigure}[t]{0.3\linewidth}
	\centering
	\includegraphics[width=\linewidth]{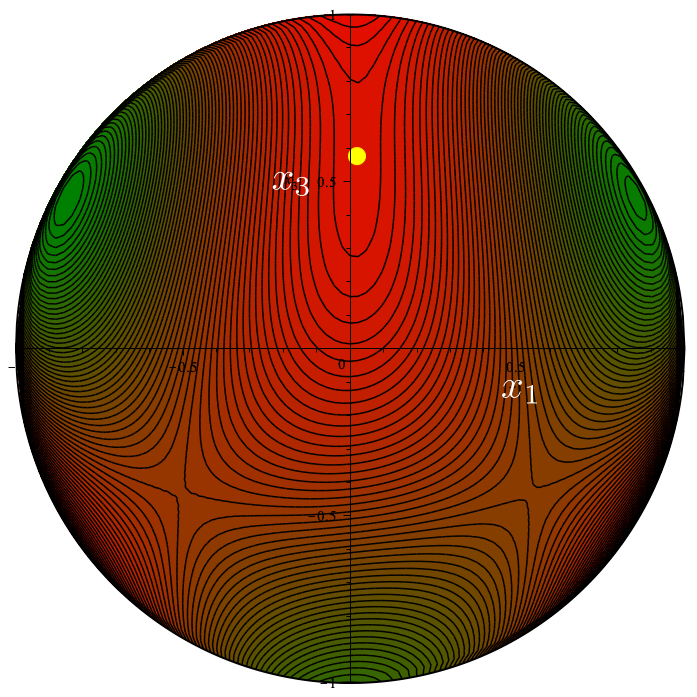}
	\caption{$\rho=\frac32$, $\chi=-7\pi/15$}
	\label{fig:approaching_1_f}
\end{subfigure}
	\caption{Comparison between the contour plots of $\potor$ on the plane $(x_1,x_3)$ for different points on the separatrix in parameter space, taken for the two values of $\chi$ corresponding to the graph of $g$ in \fref{fig:g} and to the graph with $j=1$ in \fref{fig:critical_point_count}, respectively. Panels (b) and (e) refer to the two cusps involved. The number of critical points of $\potor$ changes as follows: from $10$ to $12$ going from (a) to (d), from $8$ to $10$ going from (b) to (e); it is the same in  (c) and (f). Yellow circles mark degenerate saddles with index $\iota=0$.}
	\label{fig:approaching_1}
\end{figure}

\begin{figure}
	\centering
	\begin{subfigure}[t]{0.3\linewidth}
		\centering
		\includegraphics[width=\linewidth]{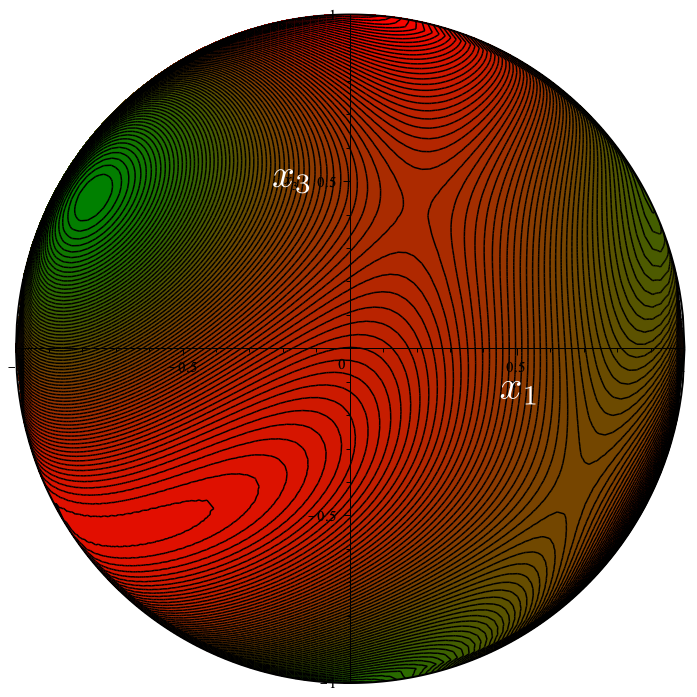}
		\caption{$\rho=1$, $\chi=-\pi/6$}
		\label{fig:approaching_2_a}
	\end{subfigure}
	\quad
	\begin{subfigure}[t]{0.3\linewidth}
		\centering
		\includegraphics[width=\linewidth]{CP_rho2_chi-Pio6_K1_deco.png}
		\caption{$\rho=2$, $\chi=-\pi/6$}
		\label{fig:approaching_2_b}
	\end{subfigure}
	\quad
	\begin{subfigure}[t]{0.3\linewidth}
		\centering
		\includegraphics[width=\linewidth]{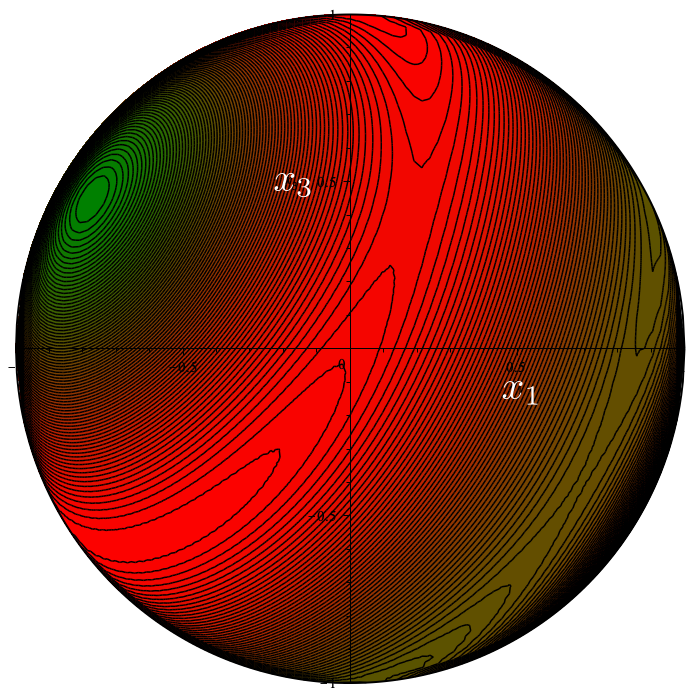}
		\caption{$\rho=1.8$, $\chi=-\pi/6$}
		\label{fig:approaching_2_c}
	\end{subfigure}\\
	\begin{subfigure}[t]{0.3\linewidth}
		\centering
		\includegraphics[width=\linewidth]{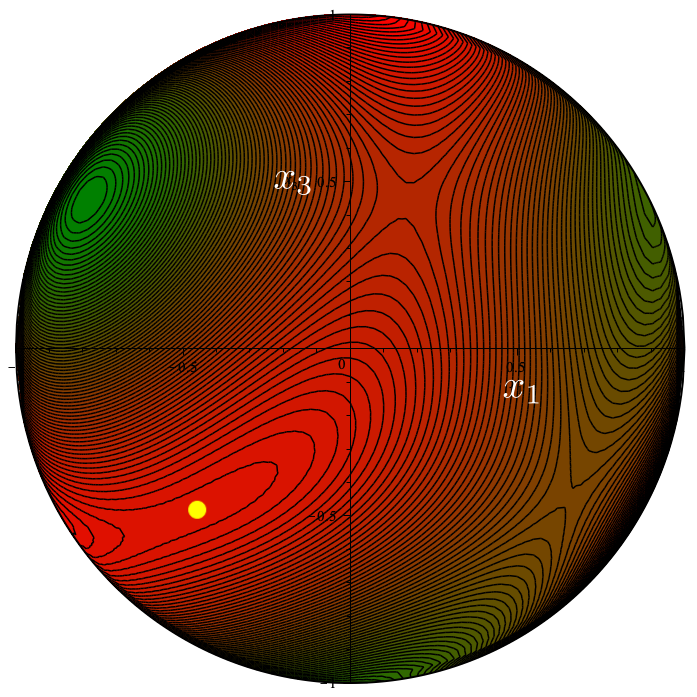}
		\caption{$\rho=1$, $\chi=-\pi/5$}
		\label{fig:approaching_2_d}
	\end{subfigure}
	\quad
	\begin{subfigure}[t]{0.3\linewidth}
		\centering
		\includegraphics[width=\linewidth]{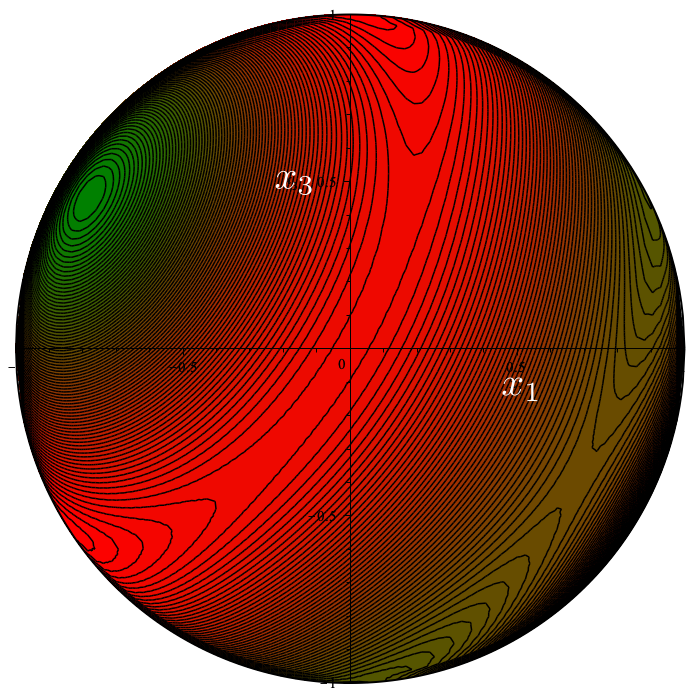}
		\caption{$\rho\doteq1.70$, $\chi=-\pi/5$}
		\label{fig:approaching_2_e}
	\end{subfigure}
	\quad
	\begin{subfigure}[t]{0.3\linewidth}
		\centering
		\includegraphics[width=\linewidth]{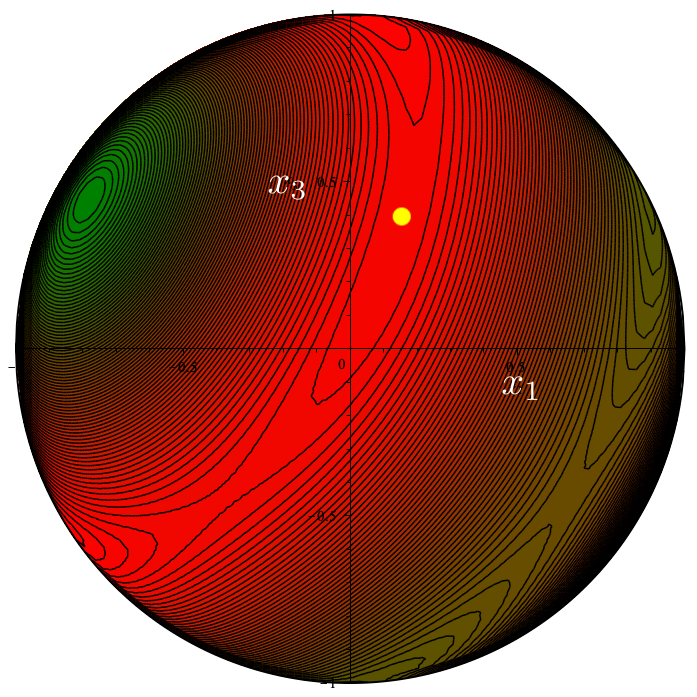}
		\caption{$\rho=1.8$, $\chi=-\pi/5$}
		\label{fig:approaching_2_f}
	\end{subfigure}
	\caption{Comparison between the contour plots of $\potor$ on the plane $(x_1,x_3)$ for different points on the separatrix in parameter space, taken for the two values of $\chi$ corresponding to the graph of $f$ in \fref{fig:f} and to the graph with $j=9$ in \fref{fig:critical_point_count}, respectively. Panels (e) and (b) refer to the cusp involved in one separatrix and to its cusp-free limit in the other, respectively. The number of critical points of $\potor$ changes as follows: from $10$ to $12$ going both from (a) to (d) and from  (c) to (f), from $\infty$ to $10$ going from in  (b) and (e). Yellow circles mark degenerate saddles with index $\iota=0$.}
	\label{fig:approaching_2}
\end{figure}

As clearly emerges from combining figures~\ref{fig:g}, \ref{fig:f}, and \ref{fig:critical_point_count}, the number of critical points of $\potor$ suffers discontinuities on the planes $\chi=-\pi/2$ and $\chi=-\pi/6$ that delimit the selected sector. Figures~\ref{fig:approaching_1} and \ref{fig:approaching_2} illustrate these transitions.

\begin{figure}[h!]
	\centering
	\begin{subfigure}[t]{0.3\linewidth}
		\centering
		\includegraphics[width=\linewidth]{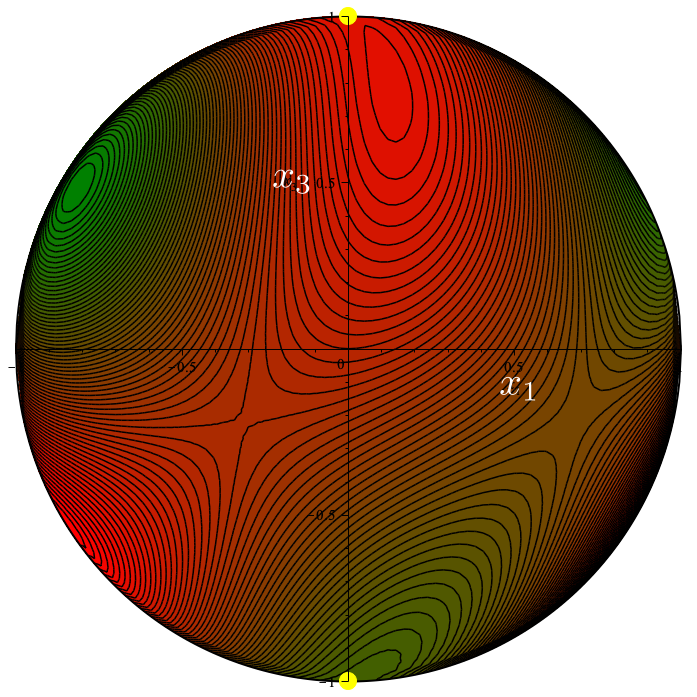}
		\caption{$\chi=-\pi/3$}
		\label{fig:approaching_3_a}
	\end{subfigure}
	\quad
	\begin{subfigure}[t]{0.3\linewidth}
		\centering
		\includegraphics[width=\linewidth]{CP_rho2_chi-Pio6_K0p5_deco.png}
		\caption{$\chi=-\pi/6$}
		\label{fig:approaching_3_b}
	\end{subfigure}
\caption{Contour plots of $\potor$ on the plane $(x_1,x_3)$ for $\rho=2$, $K=1/2$. and two different values of $\chi$. The number of critical points of $\potor$ is $12$ in (a) and $10$ in (b). In going from (a) to (b), a maximum (minimum) lands on the North (South) pole changing its singular  nature of degenerate saddle with index $\iota=0$.}
\label{fig:approaching_3}
\end{figure}

Finally, we show in \fref{fig:approaching_3} how the number of critical points of $\potor$ changes on the lateral boundary of the selected sector, where $\rho=2$, upon approaching the plane $\chi=-\pi/6$. Here the poles are degenerate saddles with $\iota=0$ for all $-\pagebreak/2\leqq\chi<-\pi/6$; their nature changes as a maximum (minimum) lands on the North (South) pole at $\chi=-\pi/6$. 
 	
\subsection{Comparison with previous studies}\label{sec:comparison}
In our previous studies \cite{gaeta:octupolar,gaeta:symmetries}, we have taken a combined geometric-analytic approach for the determination of the critical points of $\potor$ (and the corresponding eigenvalues and eigenvectors of the octupolar tensor $\oct$). Symmetry was at the basis of our geometric considerations, and path continuation was at the basis of our analytic ones. In that approach, the special cases for $\chi=-\pi/2$ and $\chi=-\pi/6$ were suggested by symmetry; they were handled directly by solving explicitly the equilibrium equations  \eref{eq:equilibrium_equations_1-3} and \eref{eq:constraint} for $\potor$.

In the algebraic approach put forward by Walcher~\cite{walcher:eigenvectors} and fully adopted here, the determination of the critical points of $\potor$ on the symmetry planes in parameter space stems from the study of the roots of low-degree polynomials, for which resolvent formulas are available. The outcomes of this analysis, which has been detailed above, confirmed our previous findings and are summarized in figures~\ref{fig:g} and \ref{fig:f}.

Things were different in the interior of the selected sector in parameter space, representative of its whole. The algebraic approach, albeit perhaps more pedantic  (as testified by the detailed case distinction we had to work out not to loose solutions), revealed itself more accurate. The major differences with our previous findings are summarized below.
\begin{enumerate}
	\item We found an explicit, analytic expression \eref{eq:line_cusps} for the line of cusps that traverses the separatrix, acting as a groin joining two vaults.
	\item We showed that the total number of critical points of the octupolar potential is $10$ along the line of cusps, instead of the $8$ we had found in \cite{gaeta:symmetries}.
	\item  We showed that in the interior of the representative sector in parameter space the whole separatrix (away from the line of cusps) bears $12$ critical points for the octupolar potential, instead of the $10$ we had found in \cite{gaeta:symmetries} on one component bordering on the line of cusps.
	\item We found another singular case with only $8$ critical points for the octupolar potential, a whole circle in parameter space (corresponding to $\rho=1$, $K=0$ in our representation).
\end{enumerate}

What was predominantly responsible for the incompleteness of our previous analyses is a type of potentially baffling critical point of the octupolar potential, which we had partly missed. This is a singular point of the index field $\field_{\potor}$ in \eref{eq:index_field} that can be lifted by a local surgery of $\potor$. More particularly, a degenerate saddle with index $\iota=0$, easily missed in a standard topological analysis of the index field $\field_{\potor}$ (usually  visually associated with the features of a countour plot). The algebraic method, on the other hand, clearly identifies these elusive critical points with the real roots of even multiplicity of the polynomials involved. These are indeed the roots that a slight, surgical perturbation of the polynomial may make either disappear or unfold in a number of simple roots (with vanishing total topological index). Contrariwise, a non-simple root with odd multiplicity cannot be associated with a critical point with index $\iota=0$, as perturbations of the polynomial cannot remove it.

We have seen both these mechanisms at work here: a critical point with index $\iota=0$ suddenly appearing, disappearing, or splitting; three critical points coming together in a single one with $\iota\neq0$. We have seen the first instance on the separatrix and the second on the line of cusps and the circle with the least number of critical points (eight).

This also explains, for what is worth, why critical points were missed in \cite{gaeta:symmetries}. These were the degenerate saddles with $\iota=0$ on the fold of the separatrix that borders the plane $\chi=-\pi/2$ for $0\leqq\rho\leqq1$. No critical point with $\iota=0$ lives on this border, and so it could not be propagated to the rest of the separatrix, as was instead the one that lives on the adjoining border for $1<\rho\leqq2$.  
 
\section{Trace Extensions}\label{sec:extensions}
Our analysis so far has been confined to fully symmetric octupolar tensors $\oct$ with vanishing traces. Here, we broaden the scope of our study by allowing $\oct$ to have non-vanishing traces, while still retaining full symmetry. This will add $3$ more parameters to an already crowded scene. However, the octupolar potential will again prove a useful tool to describe this larger class of tensors.

\subsection{General symmetric and trace type tensors}
Let us consider tensors which are fully symmetric, but not necessarily traceless.
The most general potential associated to a fully symmetric  tensor is written in \eref{eq:A_components_full_symmetry}; there we now make use of definitions \eref{eq:parametrization_alpha_0_beta} and
\begin{equation}
	\label{eq:gamma_definitions}
	\ga_1:=A_{133},\quad  \ga_2:=A_{112},\quad \ \ga_3:=A_{223}.
\end{equation}
Traceless tensors are characterized by having
\begin{equation}
	\label{eq:gamma_traceless}
	\ga_i=- ( \a_i+\b_i),\quad i=1,2,3.
\end{equation}
For later reference, we will write
\beq \ga_i  =  \frac13A_i  - ( \a_i  +  \b_i ); \eeq
thus the coefficients $A_i$ will characterize the \emph{trace type} part of tensors: traceless tensors are characterized by having $A_i = 0$ for $i=1,2,3$.

We will consider a general fully symmetric tensor as being the sum of a traceless tensor and  a \emph{trace type} tensor; the latter are thus identified as having $A_i$ arbitrary real constants, and $\a_i = \b_i = 0$.

The most general octupolar potential associated with a  fully symmetric tensor can  be written more compactly (understanding cyclic permutations in $i$, i.e., $i=4$ means $i=1$ and $i=0$ means $i=3$) as
\beq\label{eq:octupolar_potential_s} 
\pot_\mathrm{s}=6\a_0 x_1x_2x_3 + \sum_{i=1}^3 \a_ix_i^3+3\sum_{i=1}^3 \b_i x_ix_{i+1}^2+3\sum_{i=1}^3 \ga_ix_ix_{i-1}^2. \eeq
In the same formalism,
the most general octupolar potential associated with a \emph{traceless}  fully symmetric tensor in \eref{eq:octupolar_potential} can be rewritten as
\beq\label{eq:tracelesspotential}
	\pot = 6\a_0x_1x_2x_3 +\sum_{i=1}^3 \a_ix_i\( x_i^2 - 3 x_{i-1}^2 \)  +  \sum_{i=1}^3 \b_i  x_i  \( x_{i+1}^2 -  x_{i-1}^2 \). \eeq
The difference between these is the potential associated with \emph{trace type} tensors, and turns out to be
\beq\label{eq:octupolar_potential_t}
\pot_\mathrm{t} := \pot_\mathrm{s} -  \pot =  3  \sum_{i=1}^3 (\a_i +  \b_i  + \ga_i ) x_i x_{i-1}^2=  \sum_{i=1}^3 A_i x_i x_{i-1}^2. \eeq
\begin{remark}
In the original notation introduced in  \eref{eq:A_components_full_symmetry}, $\pot_\mathrm{t}$ can alternatively be written as
\begin{eqnarray}
	\label{eq:auxiliary_writing}
	\pot_\mathrm{t} &=(A_{311}+A_{322}+A_{333}) x_{3}^3+3 
	(A_{111}+A_{122}+A_{133})x_{1} x_{3}^2\nonumber \\
	&+3(A_{211}+A_{222}+A_{233}) x_{2}x_{3}^2. 
\end{eqnarray}
\end{remark}
\begin{remark}
	Reasoning as in \sref{sec:oriented_potential}, we can lower by $4$ the number of independent parameters appearing in $\pot_\mathrm{s}$ by 
\emph{orienting} the potential in \eref{eq:octupolar_potential_s}.
\end{remark}

\subsection{Trace type potential}
Here our attention will be confined to the general trace type potential in \eref{eq:octupolar_potential_t}, which we write in expanded form as
\beq\label{eq:octupolar_potential_t_expanded}
\pot_\mathrm{t} = A_1  x_1  x_3^2 + A_2 x_2 x_1^2 + A_3 x_3  x_2^2.
\eeq
This potential shares several of the remarkable properties of the potential $\pot$ corresponding to traceless tensors studied in sections~\ref{sec:geometric} and \ref{sec:algebraic}. It is covariant under inversion of $\x$,
$ x_i \to  - x_i$ ($i=1,2,3$), and also under inversion of parameters $A_i$, collected in a vector $\bm{A}$, $ A_i  \to -  A_i$ ($i=1,2,3$). Formally, we write these properties as follows
\beq
\pott( - \x , \bm{A} )=  - \pott \( \x , \bm{A} \) = \pott ( \x , - \bm{A} ), \eeq 
which implies that $\pott$ is also 
invariant under a simultaneous inversion of $\x$ and $\bm{A}$,
\beq
\pott( - \x , - \bm{A} ) = \pott ( \x , \bm{A} ).
\eeq
It is likewise invariant under a simultaneous identical permutation of the $x_i$ and of the $A_i$,
\beq
\pott( \pi (\x) , \pi(\bm{A}) ) =  \pott ( \x , \bm{A} ).
\eeq
By the inversion covariance of $\pott$,  we can just study it on a hemisphere (e.g. for $x_3 \geqq 0$) and for non-negative values of one of the control parameters (e.g. for $A_2 \geqq 0$); in the following, we shall explore this possibility.

We  restrict $\pott$ to the unit sphere $\sphere$; the two standard ways of doing this (which we will use alternatively according to convenience) are:
\begin{itemize}
	\item[(a)] Consider the upper (Northern) and the lower (Southern) hemispheres separately; on these we can just set
	\beq
	x_3 = \pm \sqrt{1-x_1^2 - x_2^2} ;
\eeq
	we denote the potential thus obtained as $\pott^\pm$.
	\item[(b)] Pass to spherical coordinates:
	\beq\label{eq:spherical_coordinates}
	x_1 = \ \cos\theta\cos\phi,\quad x_2 = \sin\theta\cos\phi,\quad x_3 = \sin\phi,
	\eeq
	where $\theta \in [- \pi, \pi]$ and $\phi \in [- \pi/2 , \pi/2 ]$. \end{itemize}

\bigskip\noindent
We will mostly consider the restriction to the unit sphere using Cartesian coordinates (that is, (a) above); this will lead us to consider separately the potential in the two hemispheres.

\subsubsection{Oriented potential on hemispheres.}
The potential in the Northern hemisphere is explicitly written as  
\beq
\pott^+ = A_1  x_1  ( 1 - x_1^2 - x_2^2 )  +  A_2  x_1^2 x_2  + A_3  x_2^2  \sqrt{1 - x_1^2 - x_2^2 },
\eeq
and its gradient is immediately computed to be
\beq
\nabla \pott^+ = \pmatrix{ A_1  ( 1 - 3 x_1^2 - x_2^2 ) +  2 A_2 x_1 x_2  -  A_3  \frac{x_1 x_2^2}{\sqrt{1 - x_1^2 - x_2^2 }} \cr - 2 A_1 x_1 x_2  +  A_2 x_1^2  + A_3 \frac{(2 - 2 x_1^2 - 3 x_2^2) x_2 }{\sqrt{1 - x_1^2 - x_2^2}} \cr}.
\eeq

It should be stressed that $\pot_\mathrm{t}^+$ has no special invariance or covariance properties under reflections in the $x_1,x_2$ variables (together or one at a time), while it retains of course the covariance under reflection in the $A_i$ parameters.
On the other hand, $\pot_\mathrm{t}^+$ is invariant under either one of the following transformations:
\begin{eqnarray} \label{eq:flipThetaN}
	( A_1 , A_2 , A_3 ; x_1 , x_2 , x_3 ) & \to & ( - A_1 , A_2 , A_3 ; - x_1 , x_2 , x_3 )  , \nonumber \\
	( A_1 , A_2 , A_3 ; x_1 , x_2 , x_3 ) & \to & ( A_1 , - A_2 , A_3 ; x_1 , -  x_2 , x_3 ) .
\end{eqnarray}

We can orient the potential requiring that it has a critical point in the North pole (and hence also in the South pole); the pole corresponds to $x_1=0$, $x_2=0$, and it is immediately seen from the formula for $\nabla \pott^+$ above that this is a critical point if and only if
\beq\label{eq:A_1=0}
A_1 = 0 .
\eeq
We will assume this to be the case. However, there is no guarantee that the critical points at the poles are either a maximum or a minimum.

In this way we are led to consider the oriented potential
\beq\label{eq:oriented_trace_potential}
\pott^+  = A_2  x_1^2  x_2  +  A_3  x_2^2  \sqrt{1 - x_1^2 - x_2^2 }.
\eeq
Looking at \eref{eq:flipThetaN}, we see that this retains the second of those invariance properties, while the first is now reduced to the statement that the potential is even in $x_1$.

The gradient of the oriented potential in \eref{eq:oriented_trace_potential} is
\beq
\nabla \pott^+ =  \pmatrix{ 2 A_2 x_1 x_2  -  A_3 \frac{x_1 x_2^2}{\sqrt{1 - x_1^2 - x_2^2 }} \cr A_2 x_1^2  + A_3  \frac{(2 - 2 x_1^2 - 3 x_2^2) x_2 }{\sqrt{1 - x_1^2 - x_2^2}}\cr }.
\eeq
Similarly, the potential in the Southern hemisphere is
\beq
\pott^-  = A_1  x_1  ( 1 - x_1^2 - x_2^2 )  +  A_2 x_1^2  x_2  -  A_3  x_2^2  \sqrt{1 - x_1^2 - x_2^2 },
\eeq
and its gradient is immediately computed to be
\beq
\nabla \pott^- =  \pmatrix{ A_1  ( 1 - 3 x_1^2 - x_2^2 )  +  2 A_2 x_1 x_2  -  A_3 \frac{x_1 x_2^2}{\sqrt{1 - x_1^2 - x_2^2 }} \cr - 2 A_1 x_1 x_2 +  A_2 x_1^2  -  A_3  \frac{(2 - 2 x_1^2 - 3 x_2^2) x_2 }{\sqrt{1 - x_1^2 - x_2^2}}\cr} .
\eeq
Again to guarantee having a critical point in the South pole we need $A_1 = 0$; this reduces $\pott^-$ to
\beq
\pott^-  =  A_2  x_1^2 x_2  -  A_3  x_2^2  \sqrt{1 - x_1^2 - x_2^2 } ,
\eeq

We are thus left with the two control parameters, $A_2$ and $A_3$. It is convenient to consider separately the cases with $A_2 = 0$ and with $A_2 \not= 0$.

\subsubsection{The case $A_2 = 0$.}
In this case (assuming $A_3 \not=0$, lest $\pott$ would identically vanish), the potential on the Northern hemisphere reduces to
\beq
\pott^+  =  A_3 x_2^2  \sqrt{1  - x_1^2  - x_2^2}.
\eeq
This has degenerate critical points on the whole set $x_2=0$ (which corresponds to a meridian on the hemisphere, and by symmetry there is a whole circle $\mathbb{S}^1 \subset \sphere$ of degenerate critical points), including the pole, and two isolated critical points at
\begin{equation}
	\label{eq:critical_points}
	\( 0  , \pm \sqrt{2/3} \).
\end{equation}

The meridian $x_2 = 0$  (in the Northern hemisphere) is hyperbolically unstable for $A_3 >0$ and hyperbolically stable for $A_3 < 0$.
As for the two isolated critical points \eref{eq:critical_points}, these are maxima (for $A_3 > 0$).
Finally, analyzing the situation on the equator, we detect two critical points at $(1,0)$ and $(-1,0)$; these are degenerate saddles.

\begin{figure}
	\centering
	\includegraphics[width=.35\linewidth]{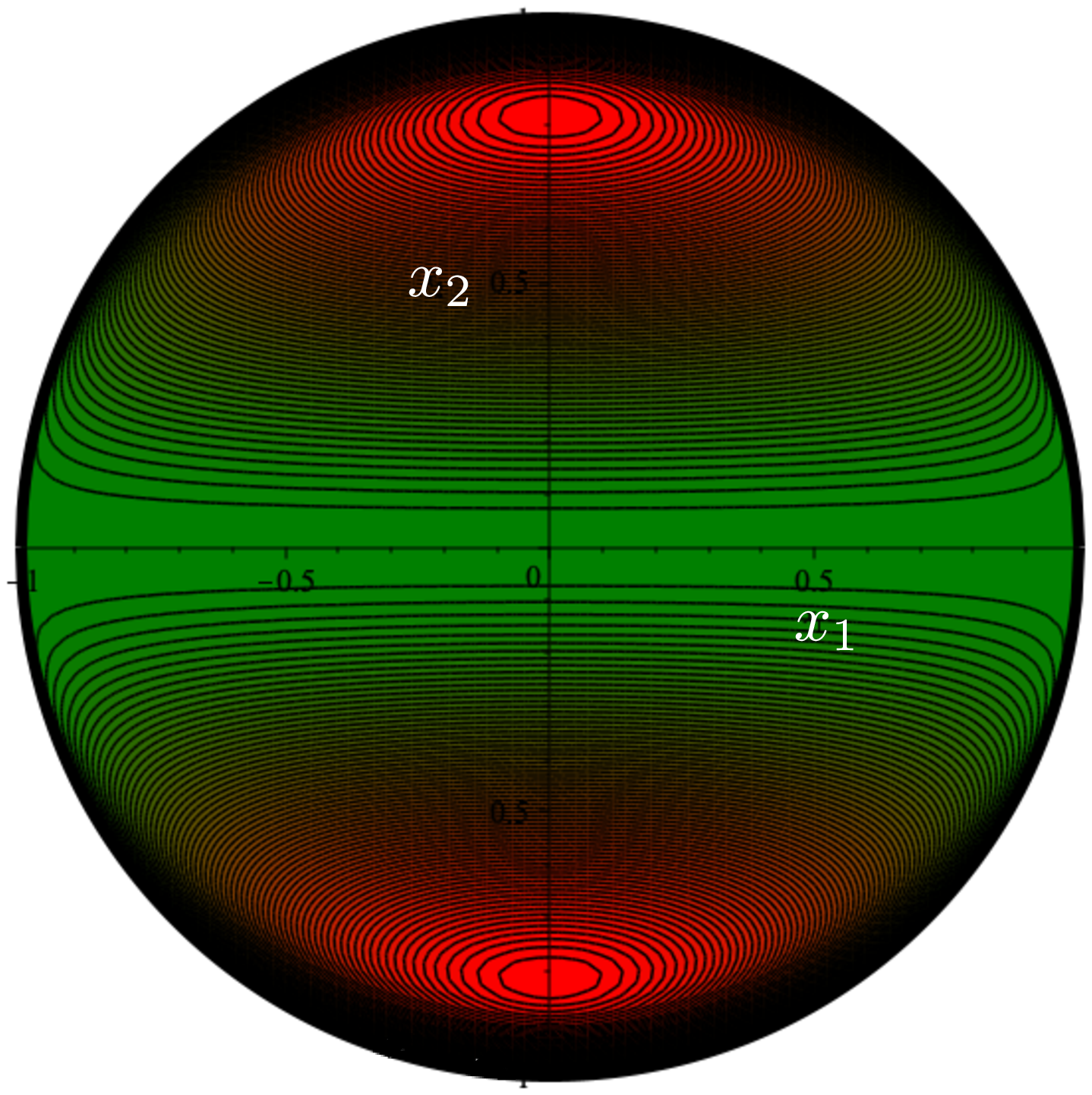}
	\caption{Contour plot on the plane $(x_1,x_2)$ of the potential  $\pott^+$ in \eref{eq:oriented_trace_potential} for $A_2 = 0$, $A_3 = 1$.}
	\label{fig:A20}
\end{figure}

\begin{figure}
	\centering
	\includegraphics[width=.55\linewidth]{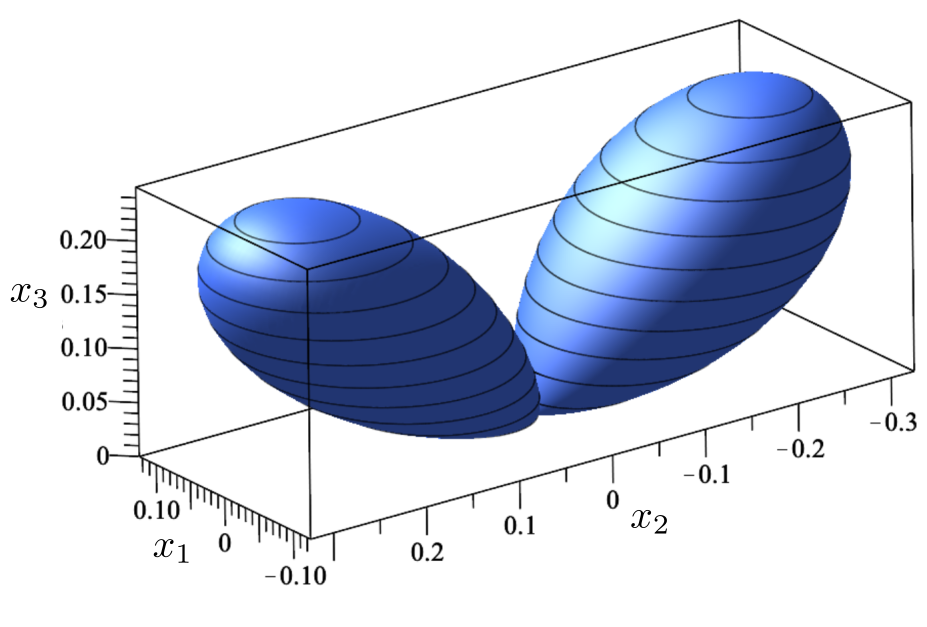}
	\caption{Polar plot of $\pott$ in \eref{eq:octupolar_potential_t_expanded}  for the case $A_1=A_2 = 0$, $A_3 = 1$. As customary here,  minima are invaginated under maxima since $\pott$ is odd under central inversion.}
	\label{fig:A203D}
\end{figure}

Figures~\ref{fig:A20} and \ref{fig:A203D} illustrate and confirm the analysis just performed.

\subsubsection{The case $A_2 \not= 0$.}
In this case it is convenient to write
\beq\label{eq:conveniency}
A_3  =  \mu  A_2  ,
\eeq
so that 
\beq \label{eq:potential_mu}
\pott^+  =  A_2 x_2 \( x_1^2  +  \mu x_2 \sqrt{1 - x_1^2 - x_2^2} \)  ,
\eeq
and
\beq
\nabla \pott^+ = A_2\pmatrix{  x_1 x_2  \( 2  -  \mu\frac{  x_2}{\sqrt{1 -  x_2^2 - x_3^2}} \) \cr  x_1^2 + \mu  \frac{x_2  \( 2 - 2 x_1^2 - 3 x_2^2 \) }{\sqrt{1  -  x_2^2  -  x_3^2}} \cr}.
\eeq
It is clear from \eref{eq:conveniency} that $A_2$ is a multiplicative factor for the potential; thus, rescaling $\pott^+$ we can just consider the cases $A_2 = 1$, with no prejudice for our analysis.

Similar formulas hold for $\pott^-$; in view of the inversion symmetry of $\pott$, we can just work with $\pot_\mathrm{t}^+$, which we will do from now on.

A change of sign in $A_2$ (keeping $\mu$ unchanged, which means changing also the sign of $A_3$) would just flip the potential---in particular, minima would become maxima, and viceversa---so we can as well consider just the case $A_2 = 1$, which we do from now on.\footnote{We stress that this holds as far as we only consider the potential associated with trace type tensors \emph{per se}; if we also consider the potential associated with traceless tensors, the scales of the two potentials cannot be set independently.}

Note that for $A_2 >0$, we always have that
\beq
\pott^+ (x_1,|x_2|) \ge  \pott^+ (x_1 , - | x_2 |);
\eeq
more precisely,
\beq
\pott^+ (x_1,|x_2|) - \pott^+ (x_1 , - | x_2 |)  = 2  A_2  x_1^2  |x_2|  .
\eeq
Moreover, for $x_1 = 0$ the potential is even in $x_2$, i.e.,
\beq
\pott^+ (0,x_2) = -  \pott^+ (0 , -  x_2 ) .
\eeq

Similar formulas hold, with changes of sign, for $A_2 < 0$. (We recall that for $A_2=0$ we have a degenerate situation, the meridian $x_2=0$ being critical, see above; this corresponds to a global bifurcation.)

Summarizing, we are reduced to study
\beq
\pott^+ = x_2 \( x_1^2  +  \mu x_2  \sqrt{1 - x_1^2 - x_2^2} \) ,
\eeq
and
\beq
\nabla \pott^+  = \pmatrix{ x_1 x_2 \( 2  - \mu\frac{  x_2}{\sqrt{1  - x_2^2  -  x_3^2}} \) \cr x_1^2  +  \mu  \frac{x_2  \( 2 - 2 x_1^2 - 3 x_2^2 \) }{\sqrt{1  -  x_2^2  -  x_3^2}}\cr}
\eeq
in the Northern hemisphere; while in the Southern one we have
\beq
\pott^-  =  x_2 \( x_1^2  - \mu  x_2  \sqrt{1 - x_1^2 - x_2^2} \) ,
\eeq
and
\beq
\nabla \pott^- = \pmatrix{ x_1 x_2  \( 2 -  \mu\frac{ x_2}{\sqrt{1  -  x_2^2  -  x_3^2}} \) \cr  x_1^2 - \ \mu \frac{x_2  \( 2 - 2 x_1^2 - 3 x_2^2 \) }{\sqrt{1  -  x_2^2 - x_3^2}} \cr} .
\eeq
Note that the formulas for the Northern and Southern hemispheres are interchanged under a change of sign in $\mu$; that is, in an obvious notation, we have that
\begin{equation}
\pott^- (x_1,x_2;\mu) =  \pott^+ (x_1,x_2;- \mu) . 
\end{equation}

\subsubsection{Critical points.}
We will now look at the critical points for $\pott^+$; first we determine their location, and then we will study their nature.

\subsubsection{The case $\mu = 0$.}\label{sec:mu_0}
It is convenient to single out the case $\mu = 0$; in this case, we simply have that
\beq\label{eq:potential_mu_0}
\pott^+  =  x_1^2  x_2  ,
\eeq
which, being independent of $x_3$, is the same as $\pott^-$ (and $\pott$ itself).

Despite its simplicity, equation \eref{eq:potential_mu_0} is unfit to reveal the critical points of $\pott$ on the equator of $\sphere$ at $x_3=0$. For this purpose, we find it convenient to consider the representation in spherical coordinates introduced in \eref{eq:spherical_coordinates},
\begin{equation}
	\label{eq:potential_speherical_coordinates}
	\pott =  \sin \theta  \cos^2 \theta \cos^3 \phi .
\end{equation}
Hence
\begin{equation}
	\nabla \pott = \pmatrix{- 3 \sin \theta \sin \phi \cos^2 \theta \cos^2 \phi \cr \cos^3 \theta \cos^3 \phi -  2 \cos^3\phi \sin^2 \theta \cos \theta \cr}.
\end{equation}
The first component of the gradient vanishes for $\phi = \pm \pi/2$ (these corresponds to North and South poles respectively), for $\phi = 0$ (the equator), and for $\theta = m \pi/2$.
Looking also at the second component, we get that the critical points on the equator are located at
\begin{equation}
\theta = \pm  \pi/2,\quad\theta =  \pm \arccos \( \pm \sqrt{2/3} \).
\end{equation}
Looking instead at critical points on $\theta = \pm \pi$, these reduce again to the poles; as for $\theta = \pm \pi/2$, the whole curve $\phi \in [-\pi/2,\pi/2]$ is critical; this is just the $x_1=0$ meridian.

The stability of these critical points is also easily analyzed by considering the matrix of second derivatives in the angular coordinates.
It turns out that critical points at the poles ($\phi = \pm \pi/2$) and on the meridian $x_1=0$ ($\theta = \pm \pi/2$) have  degenerate stability; as for the other critical points on the equator ($\phi = 0$), those at
$\theta = \arccos (\sqrt{2/3})$ are maxima, those at $\theta = - \arccos (\sqrt{2/3})$ are minima.

\begin{figure}[h]
	\centering
	\begin{subfigure}[c]{0.35\linewidth}
		\centering
		\includegraphics[width=\linewidth]{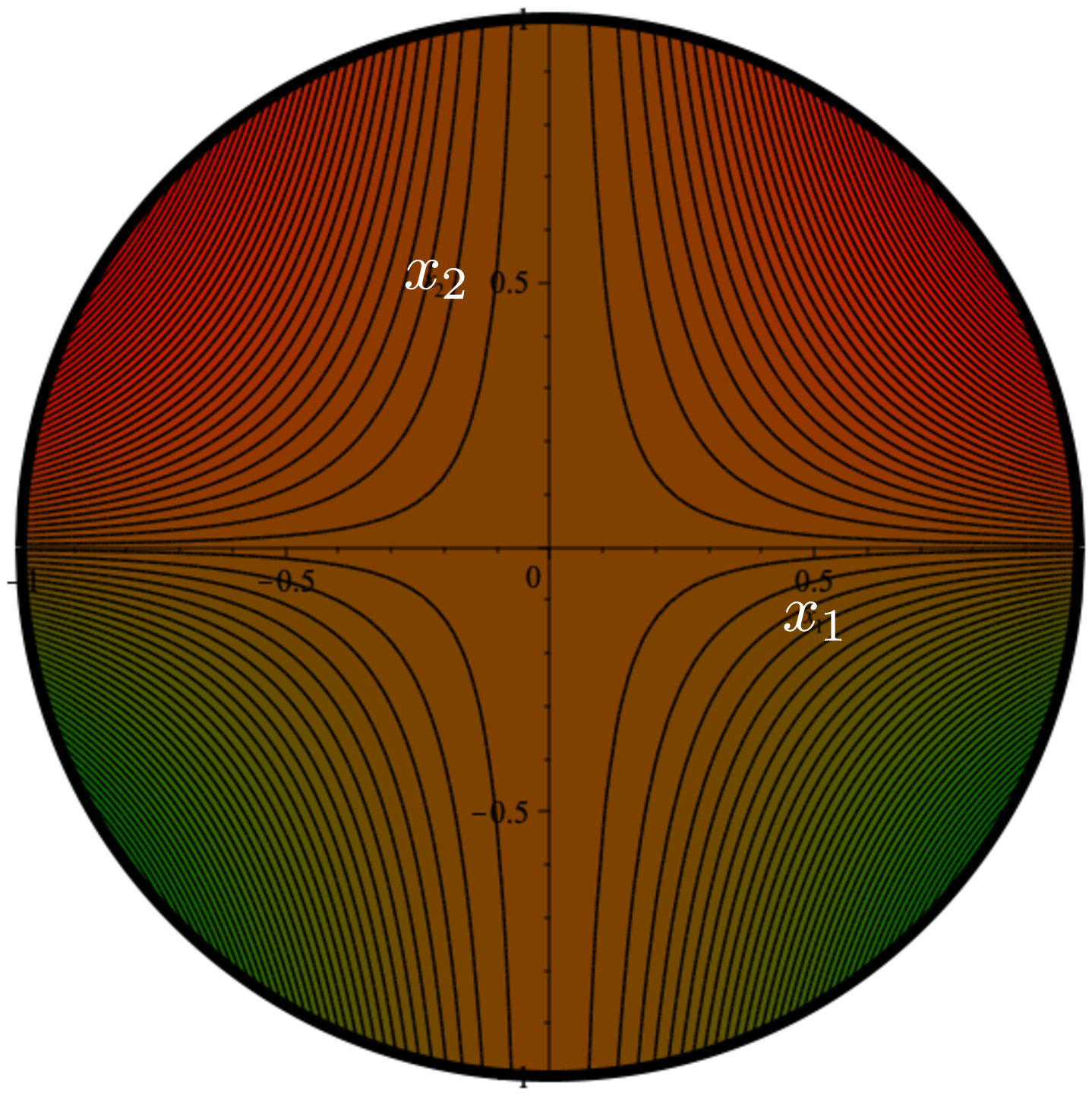}
		\caption{Contour plot on the plane $(x_1,x_2)$.}
		\label{fig:centre_contourplot_bis}
	\end{subfigure}
	\qquad
	\begin{subfigure}[c]{0.55\linewidth}
		\centering
		\includegraphics[width=\linewidth]{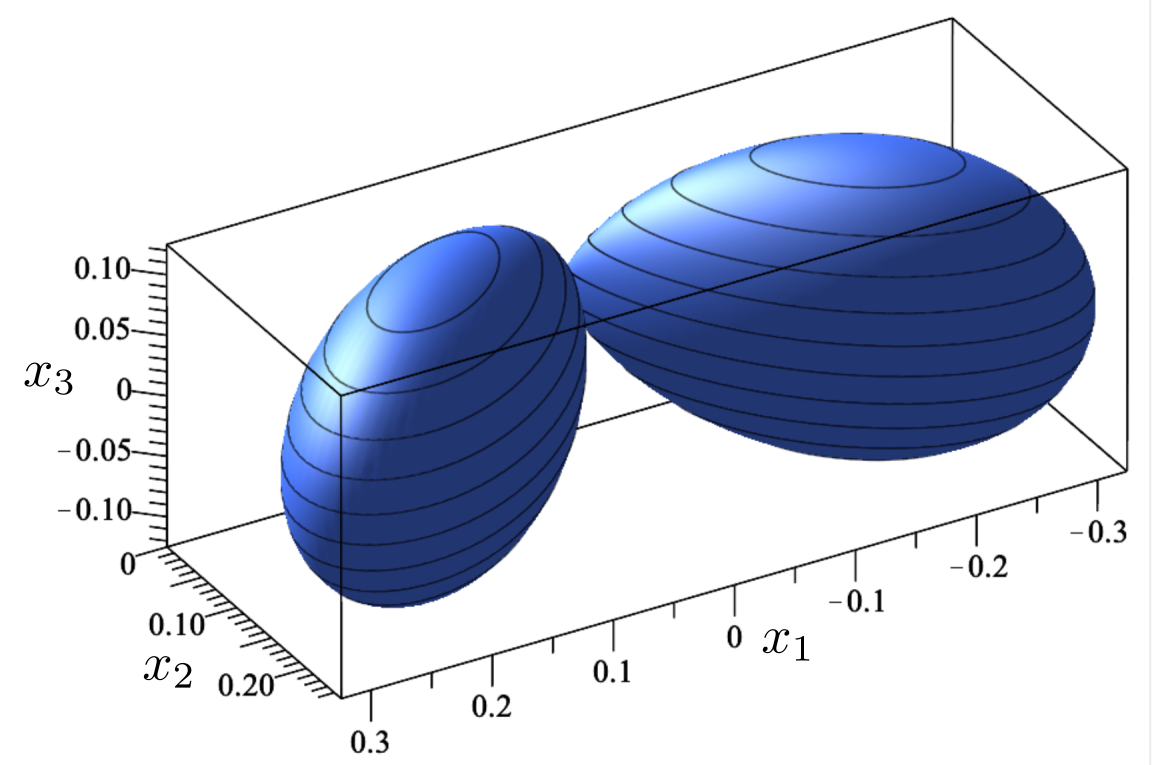}
		\caption{Polar plot (with minima invaginated under maxima).}
		\label{fig:centre_polar_plot_bis}
	\end{subfigure}
	\caption{The potential $\pott^+$ in \eref{eq:potential_mu_0} (which is the same as $\pott^-$ and $\pott$) exhibits five critical points.}
	\label{fig:mu_0}
\end{figure}

This completes the analysis of the $\mu = 0$ case; \fref{fig:mu_0} illustrates it. 

\subsubsection{The case $\mu \not= 0$.}\label{sec:mu_not_0}
As noted above while discussing the case $\mu = 0$, the restriction to Northern (or Southern) hemisphere fails to detect critical points lying on the equator. It is thus convenient to analyze first the equatorial region by considering the spherical coordinates representation \eref{eq:spherical_ccordinates}. This yields (also in view of \eref{eq:A_1=0} and  \eref{eq:conveniency} with $A_2 = 1$)
\beq
\pott  = \cos^2\theta\sin\theta\cos^3\phi  +  \mu\sin\phi\sin^2\theta\cos^2\phi. \eeq
The gradient in the spherical coordinates reads as
\begin{equation}\label{eq:potential_gradient_spherical_ccordinates}
	\fl\qquad
\nabla\pott= \pmatrix{\cos ^2\phi \cos \theta \left(\frac{1}{2} \cos \phi (3
	\cos2 \theta-1)+2 \mu  \sin\phi \sin \theta \right) \cr
	\cos
	\phi \sin \theta \left(\mu  \sin \theta  \cos ^2\phi-3
	\cos ^2\theta  \sin \phi  \cos \phi -2 \mu  \sin ^2\phi 
	\sin \theta\right) \cr} .
\end{equation}
Since at this stage we \emph{only} want  to identify the critical points lying on the equator, we  set $\phi = 0$ in \eref{eq:potential_gradient_spherical_ccordinates}, which becomes
\begin{equation}
\nabla\pott|_{\phi=0}= \pmatrix{\frac{1}{2} \cos \theta  \(3 \cos 2 \theta  -1 \) \cr \mu \sin^2\theta\cr}.
\end{equation}
For $\mu \not= 0$, vanishing of the second component requires $\theta = 0$ or $\theta = \pm \pi$; but at these points the first component does not vanish. We conclude that for $\mu \not= 0$ there are no critical points lying \emph{exactly} on the equator.

This analysis assures us that use of Cartesian coordinates and reduction to hemispheres will be able to detect all critical points in the case $\mu \not= 0$. We shall thus consider $\pot_\mathrm{t}^+$ and its gradient $\nabla \pot_\mathrm{t}^+$ in the coordinates $(x_1,x_2)$.

Some standard algebra shows that the equation $\nabla \pot_\mathrm{t}^+ = 0$ has
three roots independent of $\mu$, namely,
\begin{eqnarray}\label{eq:three_roots}
	\cases{
p_1 :x_1 = 0  ,\quad  x_2 = 0,\\
p_2  : x_1 = 0 ,\quad x_2 = - \sqrt{2/3},\\
p_3  :  x_1 = 0 ,\quad x_2 =  \sqrt{2/3}, 
}
\end{eqnarray}
and two roots depending on $\mu$, which only exist for $0<|\mu|\le\sqrt{2}$, namely,
\begin{eqnarray}
p_4 &:x_1 =-\frac{2}{\sqrt{3}}\cos\xi,\quad x_2=\sqrt{\frac23}\sin\xi\label{eq:root_4}\\ 
p_5 &: x_1 =\frac{2}{\sqrt{3}}\cos\xi,\quad x_2=\sqrt{\frac23}\sin\xi,\label{eq:root_5}
\end{eqnarray}
where $\xi$ is related to $\mu$ through the equation
\begin{equation}
	\label{eq:xi_of_mu}
	\xi=\arcsin\left(\sgn(\mu)\sqrt{\frac{2}{4-\mu^2}}\right)
\end{equation}
and ranges in the interval $-\pi/2\leqq\xi<-\pi/4$ for $-\sqrt{2}\leqq\mu<0$ and in the interval $\pi/4<\xi\leqq\pi/2$ for $0<\mu\leqq\sqrt{2}$.

\begin{figure}[h]
	\centering
	\includegraphics[width=.5\linewidth]{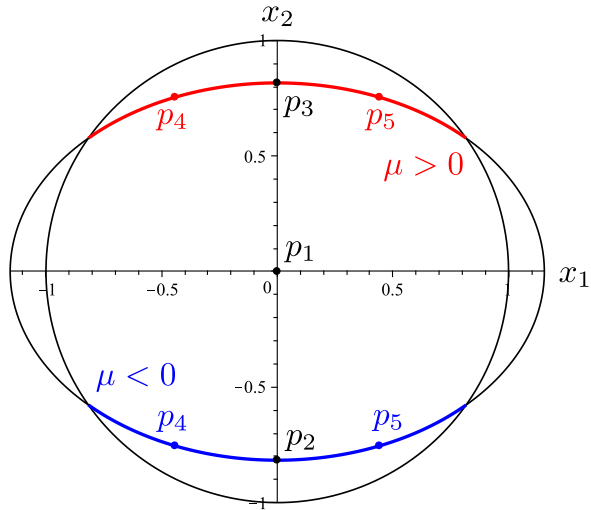}
	\caption{The critical points of $\pott^+$ in \eref{eq:potential_mu} for $A_2=1$. Three of them, namely $p_1$, $p_2$, and $p_3$, are independent of $\mu$; they are marked as black dots. The other critical points, namely $p_4$ and $p_5$, exist only for $0<|\mu|\le\sqrt{2}$ and are located within the unit circle (equator of $\sphere$) either on the lower half of the ellipse \eref{eq:ellipse} for $\mu<0$ (blue dots), or on the upper half for $\mu>0$ (red dots). For $\mu=-\sqrt{2}$, $p_4$ and $p_5$ coalesce on $p_2$, while for $\mu=\sqrt{2}$ they coalesce on $p_3$. At $\mu=0$, they reach the equator and jump from one side to the other of the ellipse.}
	\label{fig:critical_points}
\end{figure}

All critical points $p_1$-$p_5$ are illustrated in \fref{fig:critical_points}.
In particular, according to \eref{eq:root_4} and \eref{eq:root_5}, $p_4$ and $p_5$ run on the ellipse
\begin{equation}
	\label{eq:ellipse}
	\frac34x_1+\frac32x_2^2=1,
\end{equation}
which intersects the unit circle precisely for $\xi=\pm\pi/4$ and $\xi=\pm3\pi/4$. The critical points $p_4$ and $p_5$ are symmetric with respect to the $x_2$-axis; they are located on the lower half of the ellipse \eref{eq:ellipse} for $\mu<0$ and on the upper part for $\mu>0$. For $\mu=-\sqrt{2}$, both $p_4$ and $p_5$ collapse on $p_2$; as $\mu$ increases, they separate and move symmetrically until tending to reach the unit circle (equator of $\sphere$) as $\mu\to0$. There, they jump onto the upper part of the ellipse; as $\mu$ further increases, $p_4$ and $p_5$ move symmetrically towards the $x_2$-axis, which is reached for $\mu=\sqrt{2}$, when $p_4$ and $p_5$ coalesce on $p_3$ (see \fref{fig:critical_points}).
\begin{remark}
As noted in \sref{sec:mu_0}, for $\mu=0$ the potential  $\pott$ has more critical points than those we retrieve from the preceding analysis in the limit as $\mu\to0$.
\end{remark} 
In terms of the original parameters $A_i$, we summarize our conclusions as follows:
\begin{proposition}
Let $A_1$ be zero and $A_2$ be nonzero. For $0<|A_3| < \sqrt{2} |A_2|$ all the critical points listed above are real, and the potential has five critical points in the Northern hemisphere (and, by symmetry, five critical points in the Southern hemisphere), hence \emph{ten} critical points in total. For $|A_3| > \sqrt{2} |A_2|$ the potential has three critical points in the Northern hemisphere (and, by symmetry, three critical points in the Southern hemisphere), hence  \emph{six} critical points in total.
\end{proposition}
In the limiting case where 
$|A_3| = \sqrt{2} |A_2|$, all  critical points reduce to \eref{eq:three_roots};  one of them becomes degenerate, thus hosting a   local bifurcation.
In summary,
\begin{proposition}
Let $A_1$ be zero and $A_2$ be nonzero. At the bifurcation, i.e., for $|A_3| = \sqrt{2} | A_2|$, there are three critical points, $p_1$, $p_2$, and $p_3$, in the Northern hemisphere (and three mirroring critical points in the Southern one), hence a total of \emph{six} critical points; either the point $p_2$ or the point $p_3$ is degenerate, depending on the sign of $A_3/A_2$.
\end{proposition}

\begin{figure}[h]
	\centering
	\includegraphics[width=.5\linewidth]{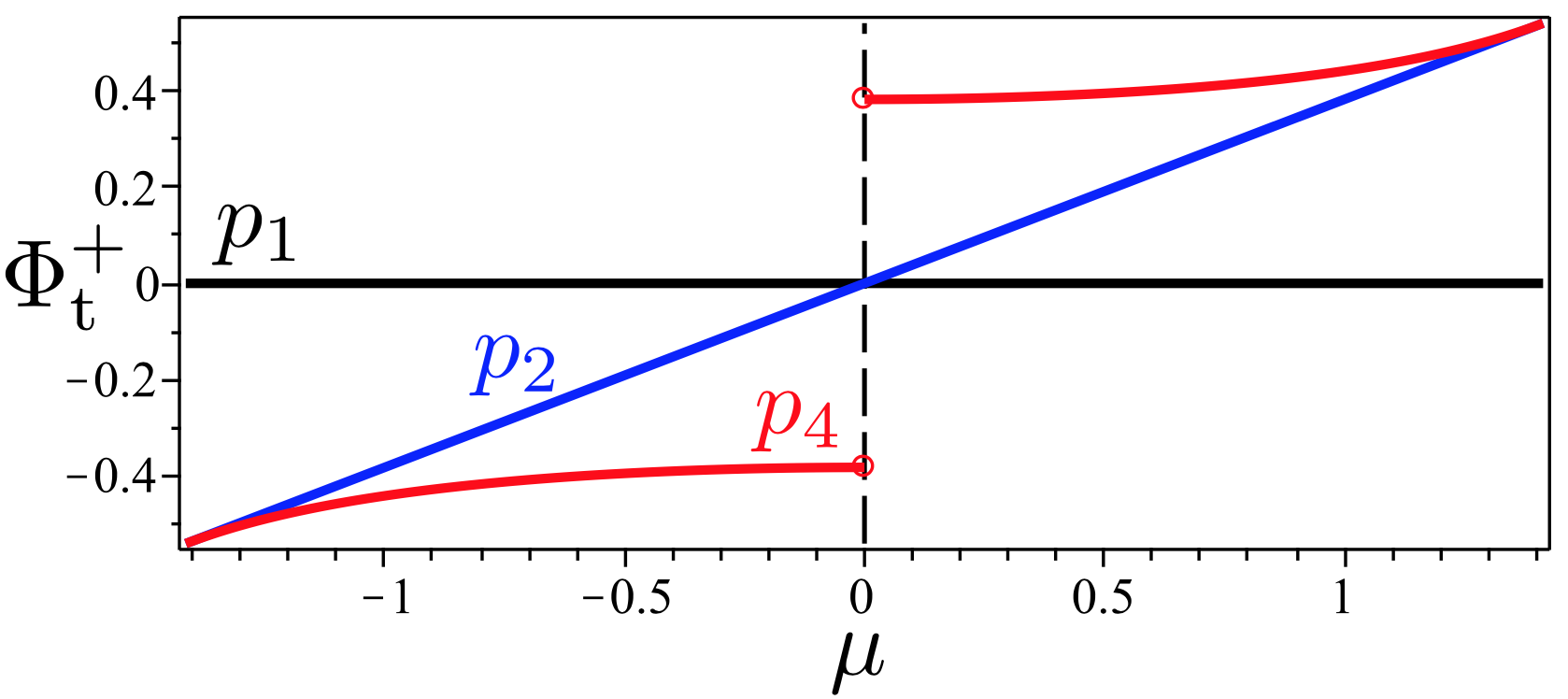}\\
	\caption{The critical values of the potential $\pot_\mathrm{t}^+$  as a functions of $\mu \in [- \sqrt{2},\sqrt{2}]$. In black, $\pot_\mathrm{t}^+ (p_1) = 0$; in blue, $\pot_\mathrm{t}^+ (p_2) = \pot_\mathrm{t}^+ (p_3)$; in red, $\pot_\mathrm{t}^+ (p_4) = \pot_\mathrm{t}^+ (p_5)$.}
	\label{fig:THcp}
\end{figure}

The values taken  by the potential at these critical points are promptly computed: 
\begin{eqnarray}
	\pot_\mathrm{t}^+ (p_1) &=& 0  ,\label{eq:critical_value_1} \\
	\pot_\mathrm{t}^+ (p_2) &=& \pot_\mathrm{t}^+ (p_3)  = \frac{2\mu}{3  \sqrt{3}} , \label{eq:critical_value_2_3}\\
	\pot_\mathrm{t}^+ (p_4) &=& \pot_\mathrm{t}^+ (p_5)  =  \sgn (\mu ) \frac{4}{3  \sqrt{3}  \sqrt{4-\mu^2}}.\label{eq:critical_value_4_5}
\end{eqnarray}
These critical values are plotted in \fref{fig:THcp} as functions of $\mu$.
\begin{remark}
Note that $\pot_\mathrm{t}^+(p_2)=\pot_\mathrm{t}^+(p_3)$, which corresponds to $\pot_\mathrm{t}^+$ being even in $x_2$ on the $x_1 = 0$ line; and $\pot_\mathrm{t}^+ (p_4) = \pot_\mathrm{t}^+ (p_5)$, which corresponds to the invariance of $\pot_\mathrm{t}^+$ under inversion in $x_1$.
\end{remark}

\subsubsection{Nature of critical points.}\label{sec:nature_critical_points}
Having identified the critical points, we should enquire if these are maxima, minima, or saddles.
In order to ascertain the nature of the critical points, we should consider the matrix of second derivatives
\beq\label{eq:hessian_matrix}
\mathsf{H}:=\nabla^2\pott^+ = \pmatrix{ \frac{\pa^2 \pot_\mathrm{t}^+}{\pa x_1^2 } & \frac{\pa^2 \pot_\mathrm{t}^+}{\pa x_1 \pa x_2} \cr \frac{\pa^2 \pot_\mathrm{t}^+}{\pa x_1 \pa x_2} & \frac{\pa^2 \pot_\mathrm{t}^+}{\pa x_2^2} \cr} ,
\eeq
and compute its eigenvalues, or at least their sign, at the critical points $p_j$ identified in \sref{sec:mu_not_0} above.

We will stick to our assumption that $A_1 = 0$ and $A_2 = + 1$ (with $A_3 = \mu$); the case of negative $A_2$ can be recovered recalling that a change of sign in $A_2$ corresponds to a change of sign in the potential, and that $\mu = A_3/A_2$ (so a change of sign in $A_2$ leaving $A_3$ unchanged corresponds to a change of sign in $\mu$, while a change of sign in both $A_2$ and $A_3$ leaves $\mu$ unchanged).

For the first three critical points the eigenvalues are easily computed and provide simple formulas:
\begin{eqnarray}
	\cases{
	p_1  :  \la_1  =  0,\quad \la_2 = 2  \mu, \\
	p_2 :  \la_1 = - 4 \sqrt{3} \mu,\quad  \la_2 = - \sqrt{2/3}  \( 2  + \sqrt{2} \mu \) , \\
	p_3 :  \la_1 = - 4 \sqrt{3}  \mu,\quad  \la_2 = + \sqrt{2/3}  \( 2  - \sqrt{2} \mu \). 
}
\end{eqnarray}
Note that in the degenerate case where $\mu = 0$ the first eigenvalue of all these critical point vanishes, while for $\mu=\pm\sqrt{2}$ the second eigenvalue vanishes in one.
\begin{remark}
This simple analysis is not conclusive for the critical point at the North pole. Actually, the series expansion along $x_2=0$ is  flat to all orders, as clear from the explicit expression of $\pot_\mathrm{t}$.
\end{remark}
The stability of points $p_2$ and $p_3$ is promptly analyzed:
\begin{itemize}
	\item The point $p_2$ is a minimum for $\mu > 0$, a saddle for $- \sqrt{2} < \mu < 0$, and a maximum for $\mu < - \sqrt{2}$;
	\item The point $p_3$ is a minimum for $\mu < 0$, a saddle for $0 < \mu < \sqrt{2}$, and a maximum for $\mu > \sqrt{2}$.
\end{itemize}

\bigskip\noindent
For the other critical points, $p_4$ and $p_5$, we find it more convenient to characterize their nature by computing the trace and determinant of the Hessian matrix $\mathsf{H}$ in \eref{eq:hessian_matrix} in terms of the parameter $\xi$ introduced in \eref{eq:xi_of_mu},
\begin{equation}
	\label{eq:trace_determinant_H}
	\tr\mathsf{H}=\frac{2\sqrt{6}(4\cos^4\xi-11\cos^2\xi+12)}{3\sin\xi(2\cos^2\xi-1)},\quad\det\mathsf{H}=\frac{16\cos^2\xi}{1-2\cos^2\xi},
\end{equation}
where use has also been made of the inverse of the function in \eref{eq:xi_of_mu},
\begin{equation}
	\label{eq:mu_of_xi}
	\mu=\frac{\sqrt{2}}{\sin\xi}\sqrt{2\sin^2\xi-1}.
\end{equation}
It is a simple matter to conclude from the study of the signs of $\tr\mathsf{H}$ and $\det\mathsf{H}$ that 
\begin{itemize}
	\item For $- \sqrt{2} < \mu < 0$ the points $p_4$ and $p_5$ are local minima;
	\item For $0 < \mu < \sqrt{2}$ the points $p_4$ and $p_5$ are local maxima.
\end{itemize}

\begin{figure}[h!]
	\centering
	\begin{subfigure}[c]{0.29\linewidth}
		\centering
		\includegraphics[width=\linewidth]{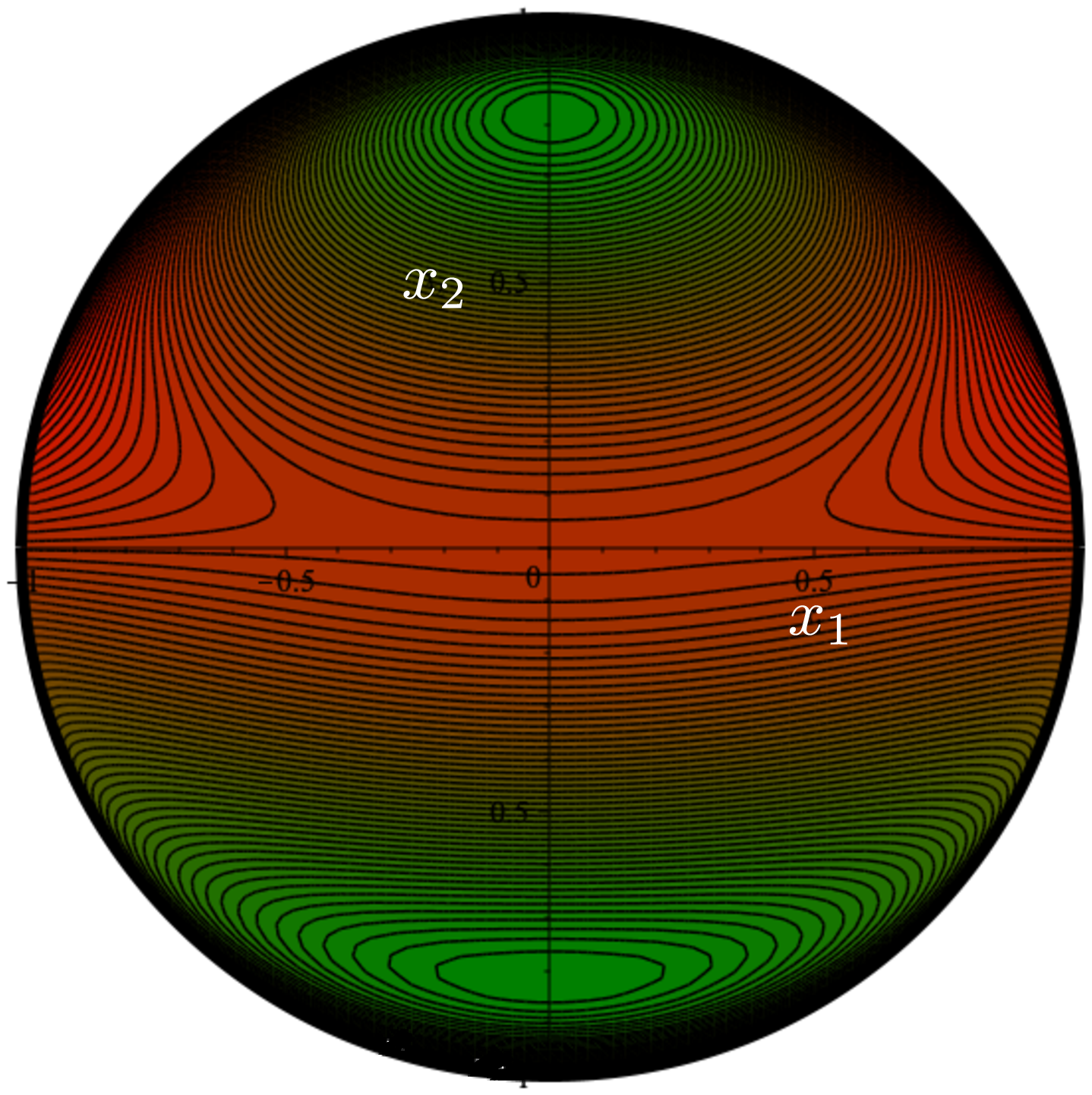}
		\caption{$\mu=-2$}
	\end{subfigure}
	\quad
	\begin{subfigure}[c]{0.29\linewidth}
		\centering
		\includegraphics[width=\linewidth]{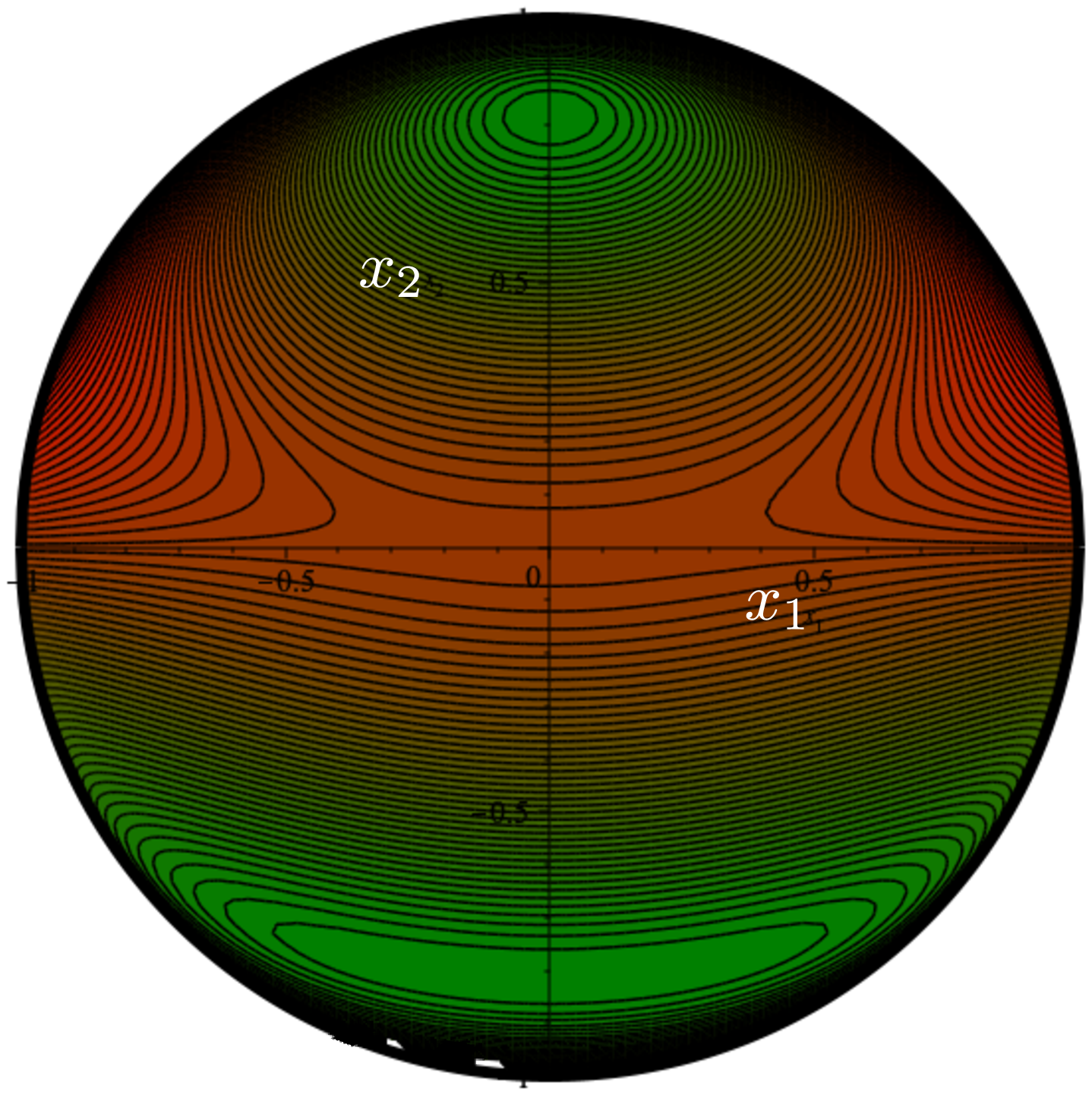}
		\caption{$\mu=-\sqrt{2}$}
	\end{subfigure}\\
	\begin{subfigure}[c]{0.29\linewidth}
	\centering
	\includegraphics[width=\linewidth]{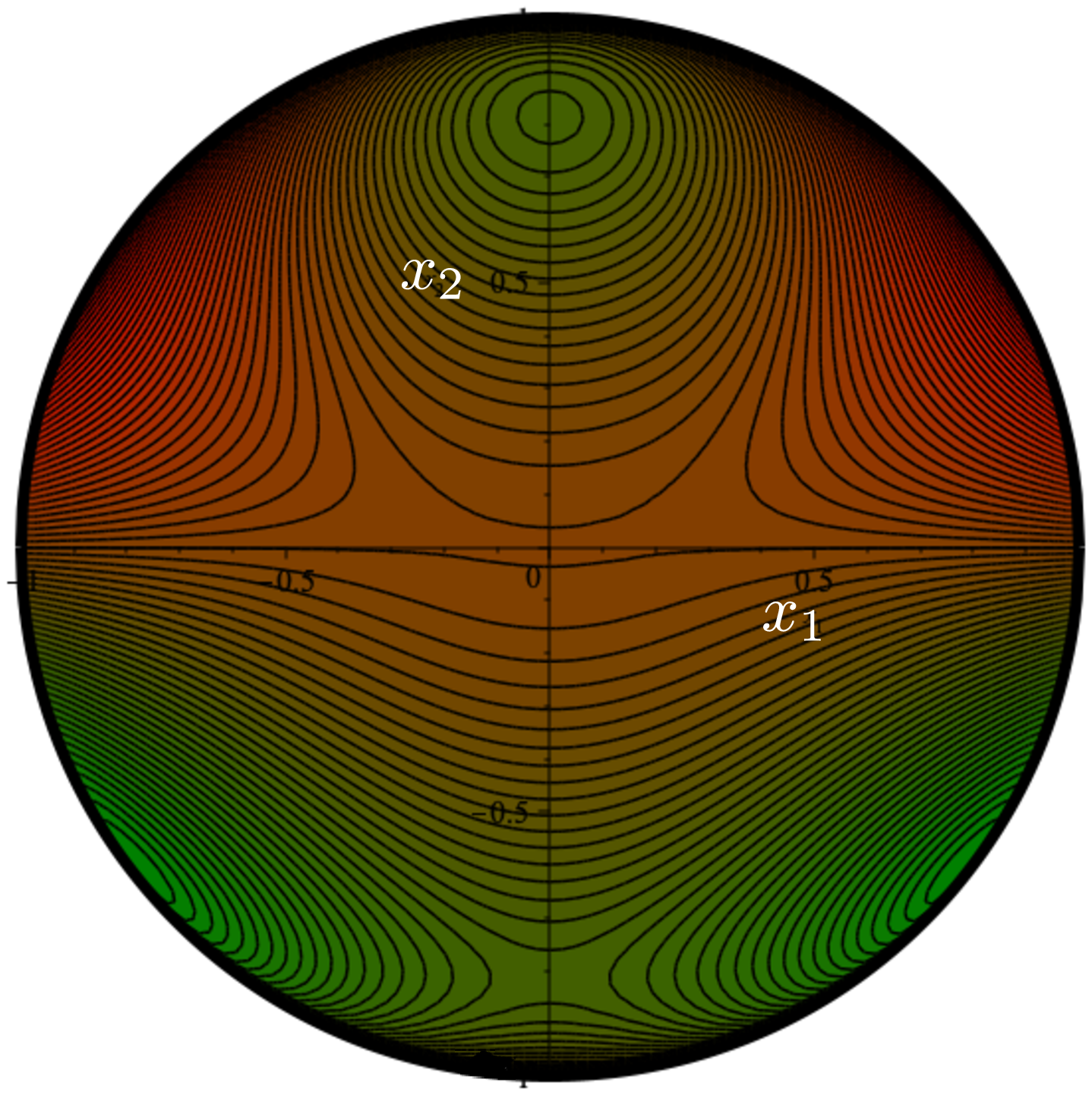}
	\caption{$\mu=-1/2$}
\end{subfigure}
\quad
\begin{subfigure}[c]{0.29\linewidth}
	\centering
	\includegraphics[width=\linewidth]{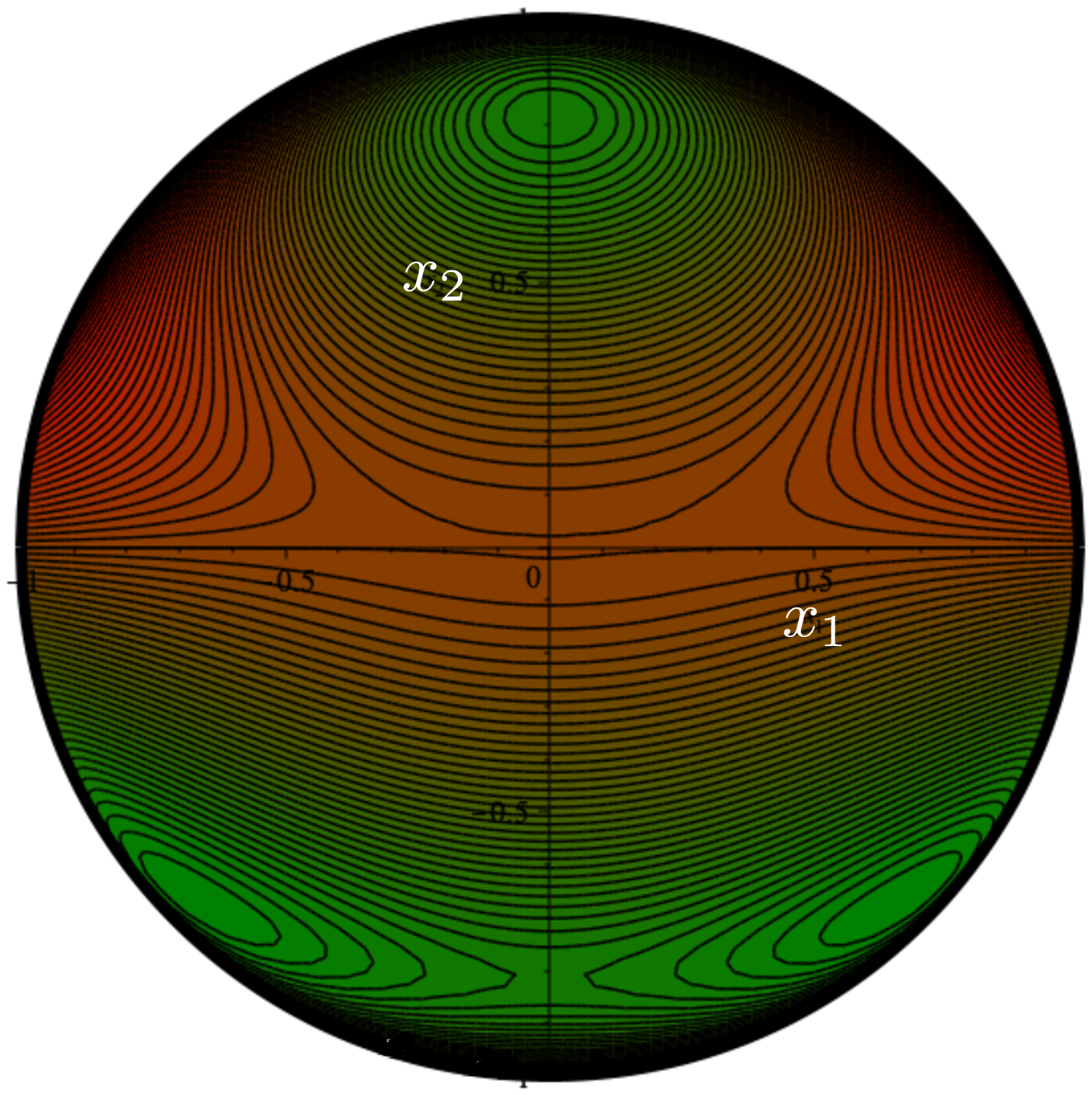}
	\caption{$\mu=-1$}
\end{subfigure}\\
\begin{subfigure}[c]{0.29\linewidth}
\centering
\includegraphics[width=\linewidth]{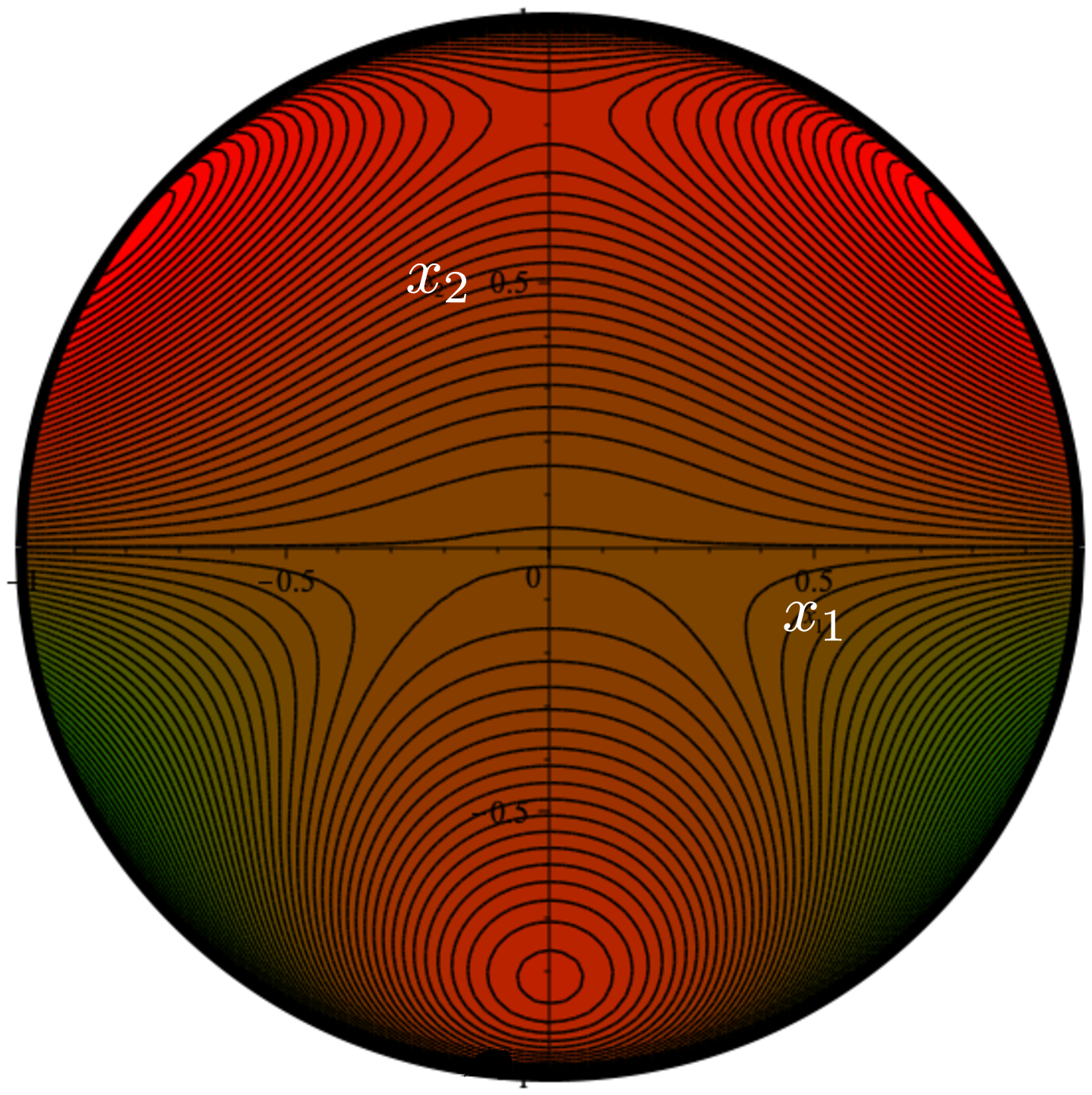}
\caption{$\mu=1/2$}
\end{subfigure}
\quad
\begin{subfigure}[c]{0.29\linewidth}
\centering
\includegraphics[width=\linewidth]{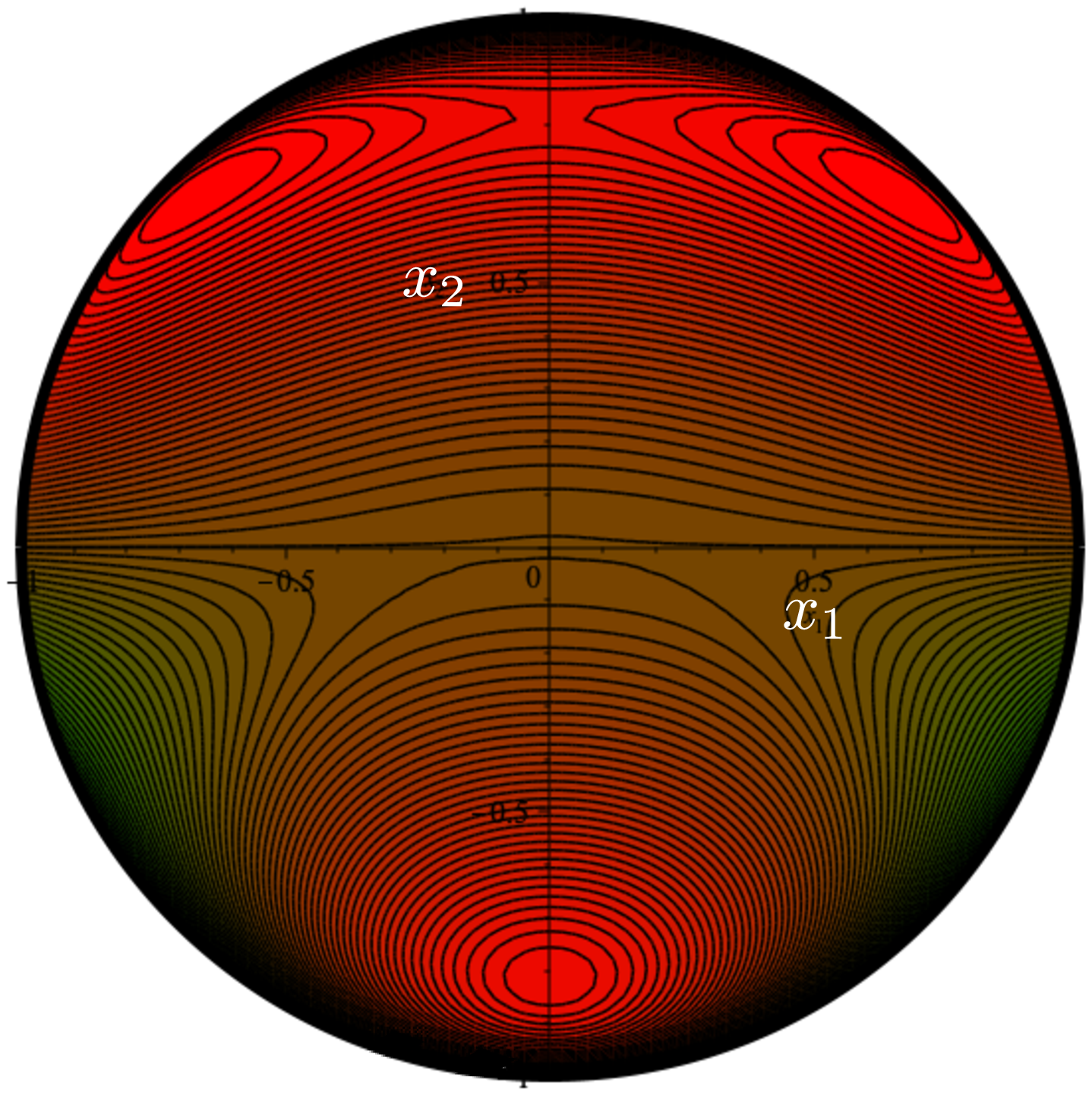}
\caption{$\mu=1$}
\end{subfigure}\\
\begin{subfigure}[c]{0.29\linewidth}
\centering
\includegraphics[width=\linewidth]{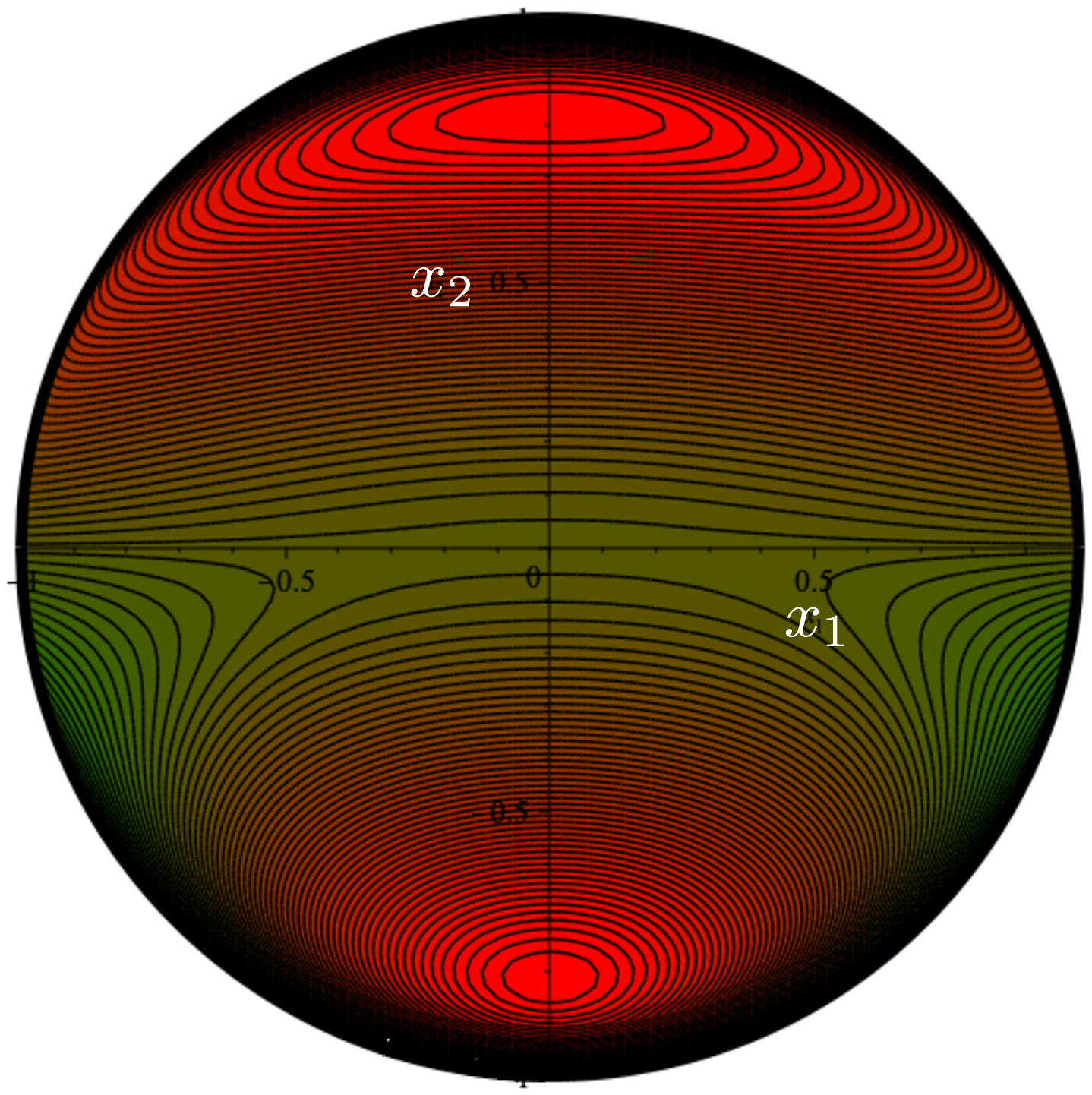}
\caption{$\mu=2$}
\end{subfigure}
\quad
\begin{subfigure}[c]{0.29\linewidth}
		\centering
	\includegraphics[width=\linewidth]{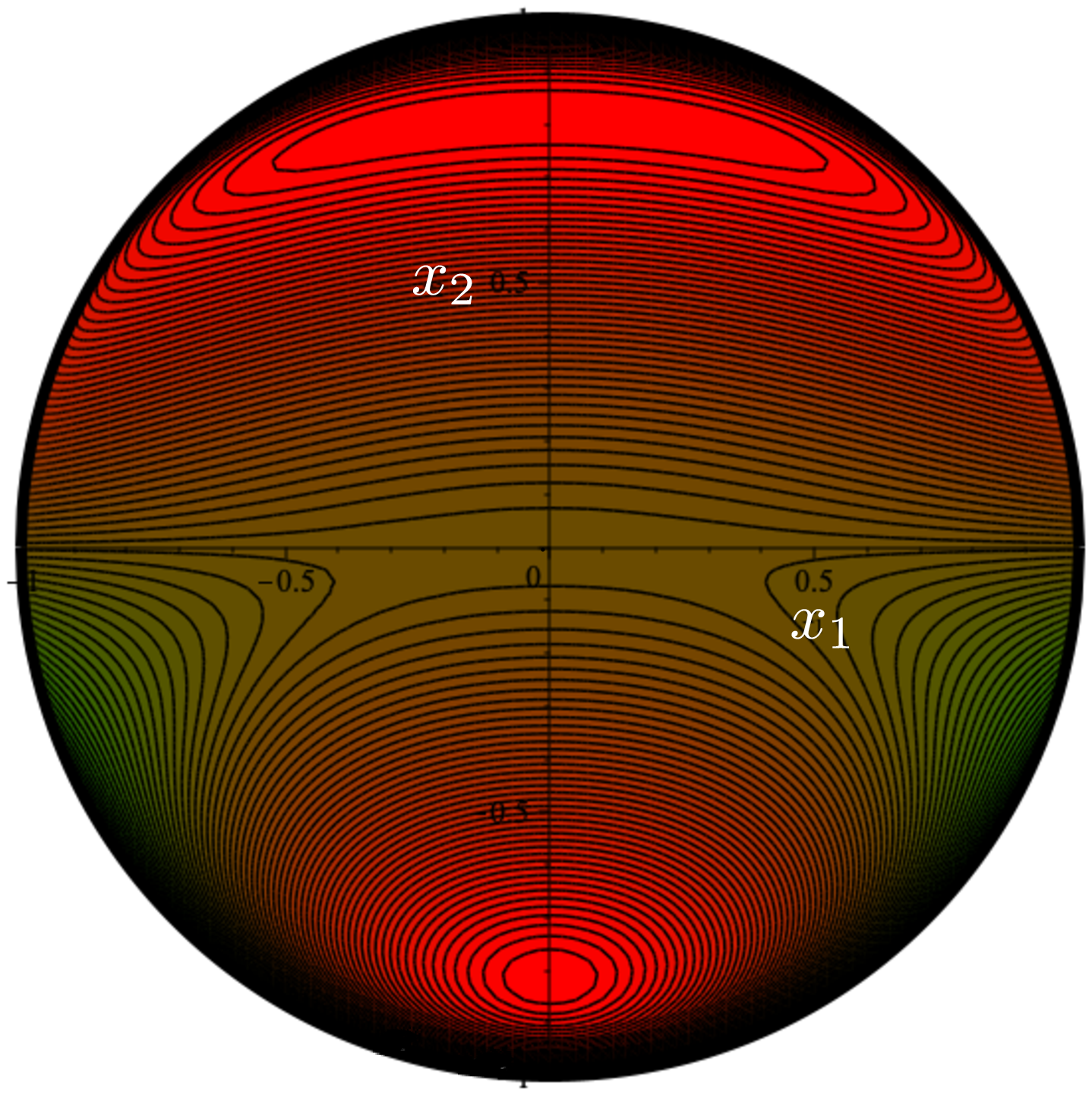}
	\caption{$\mu=\sqrt{2}$}
\end{subfigure}
	\caption{Contour plots of the potential $\pot_\mathrm{t}^+$ for $A_3 = \mu A_2$ and $A_2=1$ on the $(x_1,x_2)$ plane, for different values of $\mu$.}
	\label{fig:T_contour_plots}
\end{figure}

\begin{figure}[h!]
	\centering
	\begin{subfigure}[c]{0.34\linewidth}
		\centering
		\includegraphics[width=\linewidth]{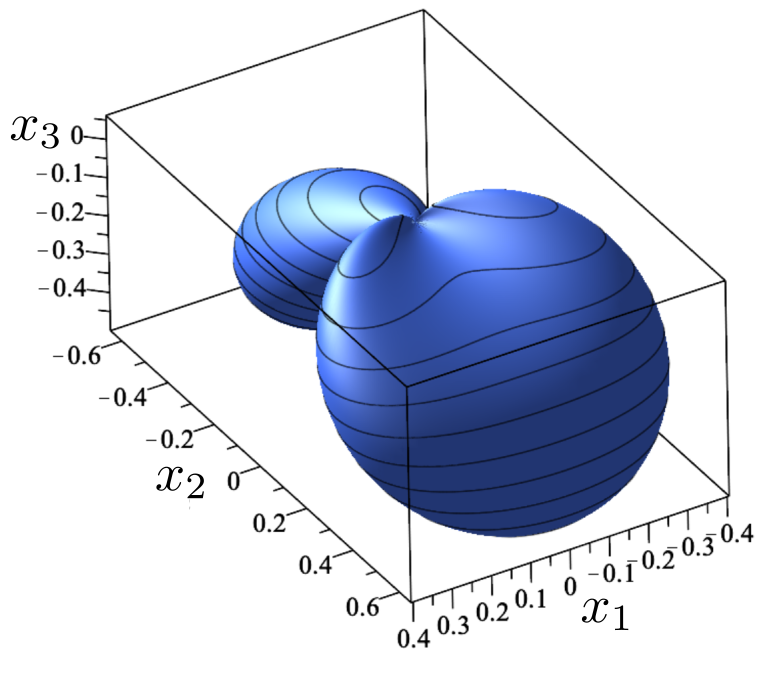}
		\caption{$\mu=-2$}
	\end{subfigure}
	\quad
	\begin{subfigure}[c]{0.34\linewidth}
		\centering
		\includegraphics[width=\linewidth]{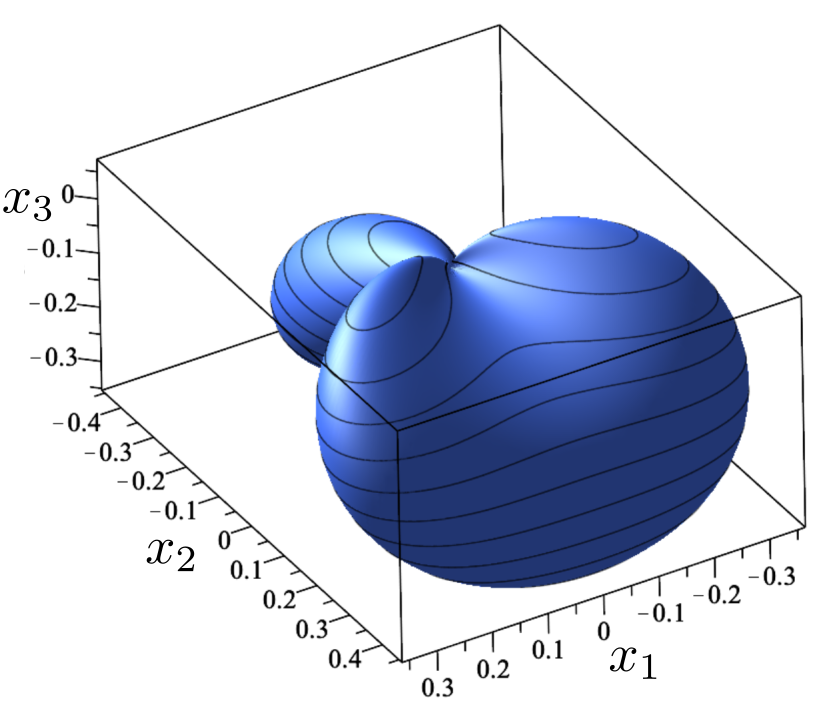}
		\caption{$\mu=-\sqrt{2}$}
	\end{subfigure}\\
	\begin{subfigure}[c]{0.34\linewidth}
		\centering
		\includegraphics[width=\linewidth]{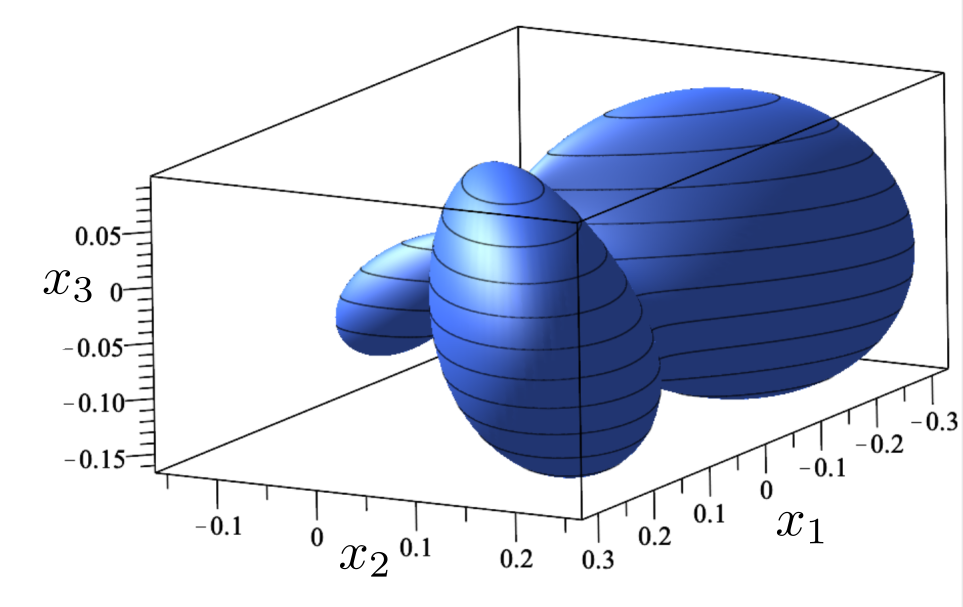}
		\caption{$\mu=-1/2$}
	\end{subfigure}
	\quad
	\begin{subfigure}[c]{0.34\linewidth}
		\centering
		\includegraphics[width=\linewidth]{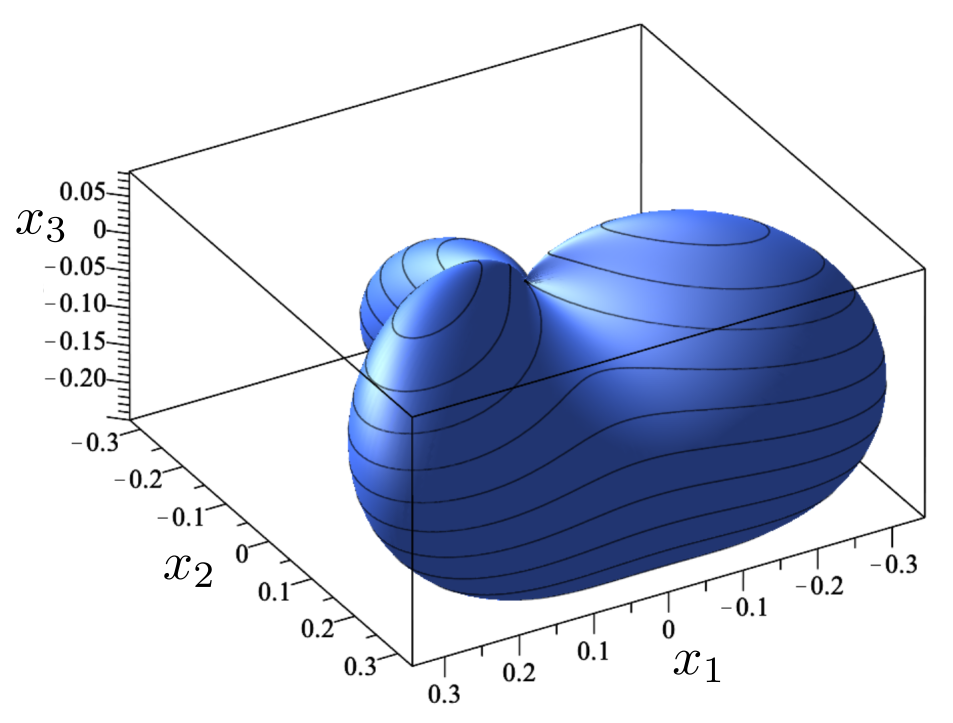}
		\caption{$\mu=-1$}
	\end{subfigure}\\
	\begin{subfigure}[c]{0.34\linewidth}
		\centering
		\includegraphics[width=\linewidth]{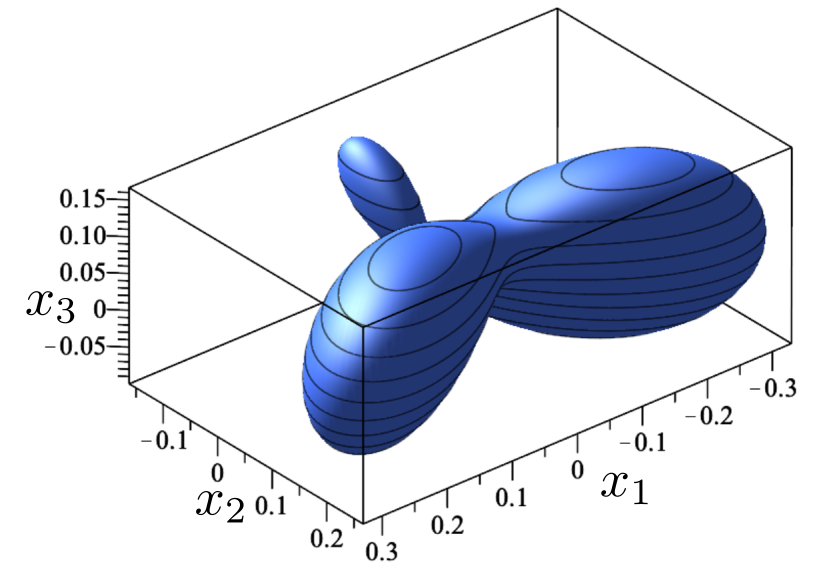}
		\caption{$\mu=1/2$}
	\end{subfigure}
	\quad
	\begin{subfigure}[c]{0.34\linewidth}
		\centering
		\includegraphics[width=\linewidth]{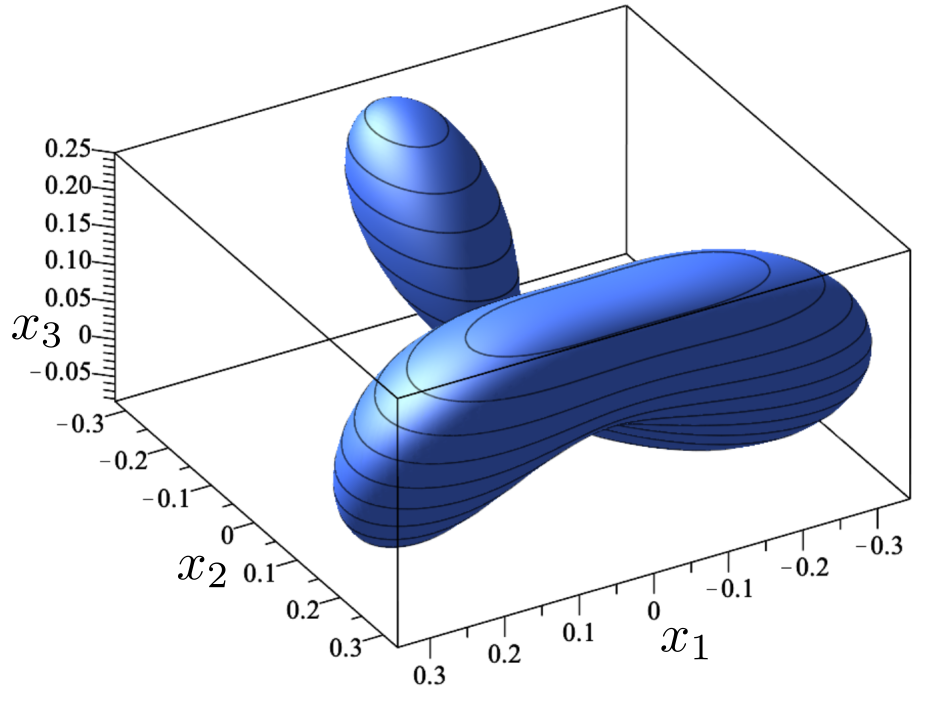}
		\caption{$\mu=1$}
	\end{subfigure}\\
	\begin{subfigure}[c]{0.34\linewidth}
		\centering
		\includegraphics[width=\linewidth]{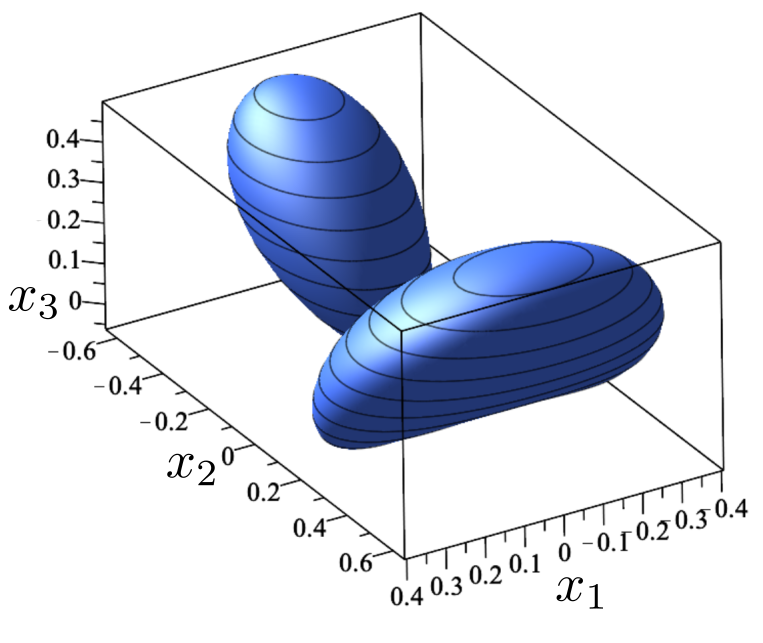}
		\caption{$\mu=2$}
	\end{subfigure}
	\quad
	\begin{subfigure}[c]{0.34\linewidth}
		\centering
		\includegraphics[width=\linewidth]{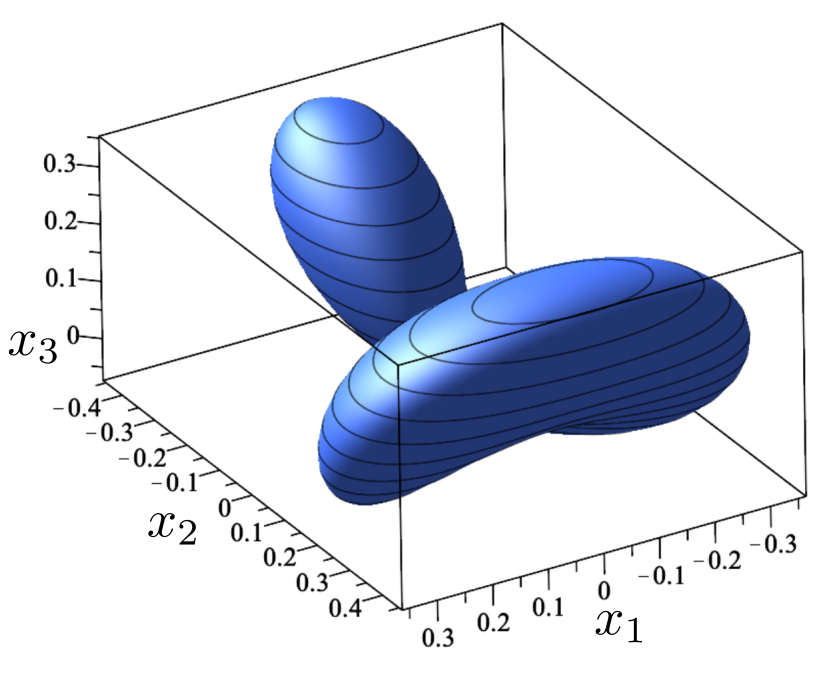}
		\caption{$\mu=\sqrt{2}$}
	\end{subfigure}
	\caption{Polar plots of the potential $\pott$ corresponding to the contour plots in \fref{fig:T_contour_plots}. Here the protruding lobes designate maxima (and invaginated minima). The origin is $p_1$, a degenerate saddle for all values of $\mu$, accompanied by a non-degenerate saddle lying on the $x_2$-axis (either $p_2$ or $p_3$ in \eref{eq:three_roots}, depending on the sign of $\mu$).}
	\label{fig:T_polar_plots}
\end{figure}

These results can be confirmed by computing numerically the index for the different critical points; such computations are summarized in \tref{tab:critical_points}, showing the index of critical points $p_2$-$p_5$ for different intervals of values of $\mu$ (recall that $p_4$ and $p_5$ only exist for $0<|\mu| \le \sqrt{2}$).
\begin{table}[h]
	\caption{\label{tab:critical_points}Critical points of the potential $\pott$ in \eref{eq:potential_mu} for $A_2=1$ and different values of $\mu$. For $|\mu|>\sqrt{2}$, neither $p_4$ nor $p_5$ exists.} 
	\begin{indented}
		\lineup
		\item[]\begin{tabular}{@{}*{5}{c}}
			\br                              
				point & $\mu < - \sqrt{2}\ $ & $- \sqrt{2} < \mu < 0\ $ &
			$0 < \mu < \sqrt{2}\ $ & $ \sqrt{2} < \mu $\cr
			\br
			$p_2$ & $1$ & $-1$ & $1$ & 1\cr 
			\mr
			$p_3$ & $1$ & $1$ & $-1$ & 1\cr 
			\mr
			$p_4$ & --- & $1$ & $1$ & --- \cr 
			\mr
			$p_5$ & --- & $1$ & $1$ & --- \cr
			\br
		\end{tabular}
	\end{indented}
\end{table}
\begin{remark}
	As for the degenerate critical point $p_1$, we have not computed directly its index, but the Poincar\'e-Hopf theorem requires $p_1$ to be a saddle, as the total index of all critical points must be $\iota=+2$ on the whole sphere $\sphere$.
\end{remark}
In \fref{fig:T_contour_plots} we present the contour plots of $\pot_\mathrm{t}^+$ for different values of $\mu$. They are accompanied in \fref{fig:T_polar_plots} by the corresponding polar plots.

\subsection{Full potential}
We want now to consider a general potential $\pot_\mathrm{s}$, i.e., the superposition of a traceless potential $\pot$, see equation \eref{eq:tracelesspotential}, and of a trace type potential $\pot_\mathrm{t}$, see equation \eref{eq:octupolar_potential_t}.

The rich phenomenology displayed by the traceless part $\Phi$ can only be enriched by considering also a trace type part; a complete analysis would most likely lead to a rather complicate discussion. Here our study will be confined to a simple, explanatory case.

The most striking feature arising from the analysis of traceless tensor potentials is maybe the presence of an exactly tetrahedral phase \cite{gaeta:octupolar}; we wonder if such a phase can also exist in the presence of a pure trace type contribution.
The general traceless potential $\pot$ is written as in \eref{eq:tracelesspotential}.
The situation in which it enjoys full tetrahedral symmetry is obtained in section~7.2 of \cite{gaeta:octupolar} for
\beq
\a_0=0,\ \a_1=0,\ \a_2= \pm\frac{1}{\sqrt{2}},\ \a_3=1,\ \b_3=- \frac12.
\eeq
In this way the oriented traceless potential $\pot$ reads as
\beq\label{eq:pot_tetra}
\pot_\mathrm{T} := x_3^3 - \frac32\( x_1^2  +  x_2^2 \)x_3 + \frac{1}{\sqrt{2}}\(x_2^2 - 3 x_1^2 \)x_2,
\eeq
where we have 
set $\a_2 = 1/\sqrt{2}$, for definiteness. 

The four maxima of the potential $\pot_\mathrm{T}$ are located at the vertices of a regular tetrahedron, and more specifically at the points given in three-dimensional Cartesian coordinates by (see \cite{gaeta:octupolar})
\begin{equation}\label{eq:tetrahedron_vertices}
	\fl
	\quad\cases{
	p_1= ( 0,0,1 ),\\
	p_2= \( 0 , \frac{2 \sqrt{2}}{3} , - \frac13 \), \
	p_3 = \( - \sqrt{\frac{2}{3}} , - \frac{\sqrt{2}}{3} , - \frac13  \), \
	p_4 = \(  \sqrt{\frac23} , - \frac{\sqrt{2}}{3} , - \frac13  \). 
}
	\end{equation}
\subsubsection{Perturbation approach.}
Here we only consider an  extreme case, i.e., that in which one of the two parts (the traceless one) can be considered as dominant, and the other one (the pure trace one) as a perturbation.

It should be stressed that we cannot arbitrarily orient both the traceless and the trace type part of the potential at the same time: we can only orient one of these (or their sum). We find it more convenient to orient the traceless part, in particular when considering the trace type part as a perturbation.

As mentioned above, we want to investigate if a potential $\pot_\mathrm{s}$ including both the traceless and the pure trace parts can have maxima at the same critical points \eref{eq:tetrahedron_vertices}, i.e., display a tetrahedral symmetry for the physical states (identified by maxima; see \cite{gaeta:octupolar}. It should be noted that we only require the locations of the maxima to be mapped unto one another by  the action of the tetrahedral group $T_d$: we are not requiring, in general, that the values of these maxima are the same (as was the case in the tetrahedral potential). Thus, letting $\pot_\mathrm{T}$ be as in \eref{eq:pot_tetra}, we shall write $\pot_\mathrm{s}=\pot_\mathrm{T}+\pot_\mathrm{t}$ as in \eref{eq:octupolar_potential_s} with coefficients given by
\begin{equation}\label{eq:parameters_pot_s}
	\fl
	\quad\cases{
	\a_0 = \eps(\de \a_0), \ 
	\a_1 = \eps (\de \a_1), \ 
	\a_2 = \frac{1}{\sqrt{2}} + \eps(\de \a_2), \ \
	\a_3  =  1  + \eps (\de \a_3),& \\
	\b_1 =  \eps (\de \b_1), \ 
	\b_2=\eps (\de \b_2), \ 
	\b_3 = - \frac12 + \eps (\de\b_3),& \\
	A_1 = \eps(\de A_1), \ 
	A_2  = \eps (\de A_2), \ 
	A_3 = \eps (\de A_3).&
}
\end{equation}
Here $\eps$ is a small parameter, all the other newly introduced parameters are expected to be of order one.\footnote{In general,   the prescription $\a_1=\b_1=\b_2=0$, which  ensures the orientation of the traceless potential, can be violated.}

We then look for critical points of $\pot_\mathrm{s}$ at first order in $\eps$, and require the points $p_1$ - $p_4$ in \eref{eq:tetrahedron_vertices} to be still critical points (and, by a perturbation argument, hence necessarily maxima) for $\pot_\mathrm{s}$.
Through some standard algebra, we find that this is the case, provided that  
\begin{eqnarray}
	\fl\quad\cases{
	\de \a_0 = \frac{\sqrt{2}}{3}\de \a_1, \\
	\de \b_1 = \frac13  \de \a_1 , \
	\de \b_2 = 0 , \\
	\de A_1 = 4 \de \a_1 , \
	\de A_2 = 2 \de \a_2 +  \frac{1}{\sqrt{2}}  \de \a_3 + 3  \sqrt{2}  \de \b_3, \
	\de A_3 = - \sqrt{2} \de \a_2 + \frac52 \de \a_3  + 3  \de \b_3 ,
}
 \end{eqnarray}
where $\de \a_1$, $\de \a_2$, $\de \a_3$, and $\de \b_3$  are free parameters.
We can afford being a bit more restrictive and consider only perturbations of the traceless part that preserve the orientation of this latter by keeping the constraints
\begin{equation}\label{eq:orienting_conditions}
\a_1= \b_1 = \b_2 = 0.
\end{equation}
We thus arrive at the conditions
\begin{equation}\label{eq:parameters_three_stars}
	\cases{
	\de A_1 =0,& \\
	\de A_2 = 2\de \a_2 + \frac{1}{\sqrt{2}}\de \a_3 \ + 3 \sqrt{2} \, \de \b_3,& \\
	\de A_3 = - \sqrt{2} \de \a_2  + \frac52\delta\alpha_3 +3\delta\beta_3,&
}
\end{equation}
where $\de \a_2$ , $\de \a_3$, and $\de \b_3$ are arbitrary constants. 
\begin{remark}
We have only required \emph{maxima} to remain at the same points. If we extend this requirement to all critical points, it turns out that $\pot_\mathrm{s}$ must be proportional to $\pot_\mathrm{T}$, thus neutralizing any contribution from a pure trace tensor.
\end{remark}

Making use of \eref{eq:parameters_three_stars}, \eref{eq:orienting_conditions}, and \eref{eq:parameters_pot_s} in \eref{eq:octupolar_potential_s}, we can easily express the symmetric potential $\pot_\mathrm{s}=\pot_\mathrm{T}+\pot_\mathrm{t}$ in the form 
\begin{equation}\label{eq:octupolar_potential_s_perturbation}
	\fl\eqalign{
	\pot_\mathrm{s}&= \frac{1}{\sqrt{2}}\ x_2 \left(x_2^2-3
	x_1^2\right)-\frac
	{3}{2} \left(x_1^2-x_2^2\right)
	x_3+x_3 \left(x_3^2-3
	x_2^2\right) \\
	&+ \eps
	\left[\left(2
	\de \a_{2}+\frac{1}{\sqrt{2}}
	\de \a_{3}+3 \sqrt{2}
	\de \b_{3}\right) x_2
	x_1^2+\de \a_{2} x_2
	\left(x_2^2-3
	x_1^2\right) \right. \\
	  & \quad\left.+\left(-
		\sqrt{2}\de \a_{2}+\frac52
		\de \a_{3}+3 
		\de \b_{3}\right)x_2^2 x_3+3 \de \b_{3}
	\left(x_1^2-x_2^2\right)
	x_3+\de \a_{3} x_3 \left(x_3^2-3
	x_2^2\right)\right],
}
\end{equation}
which is \emph{not} equivalent to $\pot_\mathrm{T}$. In cases where the (observable) physics is only described by the  \emph{location} of the maxima (or the minima) of the octupolar potential, we thus have that a perturbation of the tetrahedral potential $\pot_\mathrm{T}$ by a combination of the potentials associated to a traceless and to a trace type tensors can still describe the same physics.
\begin{remark}
Physics could also depend on the (relative or absolute) levels of the maxima of the octupolar potential in \eref{eq:octupolar_potential_s_perturbation}; thus it matters if they are at the same level or not. A tedious, but easy calculation shows that requiring all maxima of $\pot_\mathrm{s}$ in \eref{eq:octupolar_potential_s_perturbation} to be equal reduces $\pot_\mathrm{s}$ to a multiple of $\pot_\mathrm{T}$.
In other words, the only way to have degenerate maxima at tetrahedral points is with a pure tetrahedral potential.
\end{remark}
\begin{remark}
We might ask for a smaller degeneration, i.e., require that the potential $\pot_\mathrm{s}$ in \eref{eq:octupolar_potential_s_perturbation} takes the same value at the points $p_2$, $p_3$, and $p_4$ in \eref{eq:tetrahedron_vertices}, albeit this is allowed to be different from the value taken at the point $p_1$. In this case we have to require
\begin{equation}
\de \b_3  =  -  \frac{1}{\sqrt{6}}  \( 2  \sqrt{2} \de \a_2  + \de \a_3 \).
\end{equation}
With this prescription, we get
\begin{equation}
	\fl 
\pot_\mathrm{s} (p_1) = 1 + \eps \de \a_3 , \quad  \pot_\mathrm{s} (p_2) = \pot_\mathrm{s}  (p_3) = \pot_\mathrm{s} (p_4) = 1 + \eps \( \frac{8}{9}\sqrt{2} \de \a_2 + \frac{1}{9} \de \a_3 \).
\end{equation}
We can further set $\de \a_3 = 0$, so that the value of the potential at the orienting maximum (in the North pole) is unchanged; in this case, setting also $\de \a_2 = 1$, we get
\begin{equation}
\pot_\mathrm{s} (p_1) = 1,  \quad  \pot_\mathrm{s} (p_2) = \pot_\mathrm{s} (p_3) = \pot_\mathrm{s} (p_4) = 1 +  \eps\frac{8}{9}\sqrt{2}   ,
\end{equation}
and, more generally,
\begin{equation}
\eqalign{\pot_\mathrm{s} = \pot_\mathrm{T} + \eps \Bigg[& 2 \sqrt{2} x_2^2\sqrt{1-x_1^2-x_2^2} +\left(x_2^2-3 x_1^2\right) x_2\\&+\sqrt{2}\left(x_1^2-x_2^2\right)
\sqrt{1-x_1^2-x_2^2}\Bigg].}
\end{equation}
\end{remark}

\subsubsection{Non-perturbation approach.}
We can  also proceed non-perturbatively.
To this end, we consider a general superposition $\pot_\mathrm{s}$ of a traceless potential $\pot$ as in \eref{eq:tracelesspotential} and a trace type potential $\pot_\mathrm{t}$ as in \eref{eq:octupolar_potential_t}.

We then consider the gradient of $\pot_\mathrm{s}$, evaluate it at the points $p_1$-$p_4$ in \eref{eq:tetrahedron_vertices}, and require it vanishes there.
Through standard computations we obtain that this is the case provided some relations hold between the different parameters characterizing the potential. These are as follows\footnote{The similarity between the last two equations in \eref{eq:parameters_four_stars} and \eref{eq:parameters_three_stars} should be heeded.}
\begin{equation}\label{eq:parameters_four_stars}
	\cases{
	\a_0 =\sqrt{2} \b_{1}, \quad  \a_1 =  3
	\b_{1},\quad \b_2 = 0,& \\
	A_1 = 12\b_{1},&\\
	A_2 = 2 \a_{2}+\frac{1}{\sqrt{2}}
	\a_{3}+3 \sqrt{2}
	\b_{3},& \\
	A_3 = -\sqrt{2}
		\a_{2}+\frac52 
		\a_{3}+3 \sqrt{2}
		\b_{3}.&
	}
\end{equation}
Here $\a_2$, $\a_3$, $\b_1$, and $\b_3$ are free parameters.
\begin{remark}
Equations \eref{eq:parameters_four_stars} guarantee that  the tetrahedral points are critical. If we also require that $p_2$-$p_4$ in \eref{eq:tetrahedron_vertices}  are all on the same level set of $\pot_\mathrm{s}$,  we are led to the equations 
\beq\label{eq:b3tetra}
\alpha_0=\alpha_3=\beta_1=0,\quad\b_3 = - \frac16  ( 2 \sqrt{2}  \a_2  +  \a_3 ) ,
\eeq
which still leave $\a_2$ and $\a_3$ as free parameters. By \eref{eq:parameters_four_stars},  the coefficients of the pure trace part $\pot_\mathrm{t}$ then become\footnote{Note that  we are in the degenerate case $A_2 = 0$.} 
\beq\label{eq:Acoeff}
A_1  = 0, \quad  A_2  =  0, \quad A_3  = 2 \(\a_{3}-
\sqrt{2} \a_{2}\).
\eeq
By use of both \eref{eq:b3tetra} and \eref{eq:Acoeff}, we give $\pot_\mathrm{s}$ the special form 
\beq\label{eq:psi3+1}
\fl\qquad\quad
\pot_\mathrm{s}=  \frac{1}{2} \a_{3} x_3
\left(2
x_3^2-x_1^2-x_2^2\right)-\a_{2}
\left[\left(3 x_2+\sqrt{2}
x_3\right) x_1^2+x_2^2
\left(\sqrt{2}
x_3-x_2\right)\right]. \eeq
\end{remark}
\begin{remark}
If now we also require that $\pot_\mathrm{s}$ in \eref{eq:psi3+1} has in $p_1$ the same value as in $p_2$-$p_4$,  we easily see that  it must be 
\begin{equation}\label{eq:parameters_three_vertical_stars}
\a_3 = \sqrt{2}  \a_2, 
\end{equation}
which, by \eref{eq:Acoeff}, makes $A_3$ vanish as well, so that $\pot_\mathrm{s}$ is eventually proportional (through $\alpha_3$) to $\pot_\mathrm{T}$. 
\end{remark}
\begin{remark}
Consider now the  potential $\pot_\mathrm{s}$ in \eref{eq:psi3+1}, \emph{without} assuming \eref{eq:parameters_three_vertical_stars}. We have seen that it has critical points at the tetrahedral points \eref{eq:tetrahedron_vertices}. It should be noted that while working perturbatively we were guaranteed the critical points at $p_1$-$p_4$ were, for $\eps$ sufficiently small, still maxima, in the present case this is not guaranteed. To this end, we compute as in \sref{sec:nature_critical_points} the eigenvalues of the Hessian matrix $\mathsf{H}$ of $\pot_\mathrm{s}$ in \eref{eq:psi3+1}:
\begin{equation}
	\cases{
	p_1  :  \la_1 = \la_2 =  -  2 \( \sqrt{2}  \a_2 +  2 \a_3 \) ; \\
	p_2 \textit{-} p_4  :  \la_1 =  - 6 \sqrt{2} \a_2  , \quad
	 \la_2  =  -  6  \( 5  \sqrt{2}  \a_2  +  4  \a_3 \)  .
	}
 \end{equation}
These are all real, and we want all of them to be negative. It is easily seen that this is the case, provided that 
\begin{equation}
\a_2 > 0, \quad  \a_3 > - \frac{\a_2}{\sqrt{2}}.
\end{equation}
Correspondingly, we obtain that $\pot_\mathrm{s}$ in \eref{eq:psi3+1} satisfies 
\begin{equation}
\pot_\mathrm{s}(p_1)= \a_3  , \quad \pot_\mathrm{s}(p_2) = \pot_\mathrm{s} (p_3) = \pot_\mathrm{s}(p_4)  = 
\frac19 \( 8 \sqrt{2}  \a_2 + \a_3 \).
\end{equation}
So the three degenerate maxima in the Southern hemisphere are higher than the maximum at the North pole if
\begin{equation}\label{eq:inequality}
	\a_3  < \sqrt{2} \a_2  ,
\end{equation}
and lower than  the maximum at the North pole if in \eref{eq:inequality} the inequality is reversed. 
\end{remark}
	
\subsubsection{Invariance of the combined potential.}
To illustrate the subtleties that may be hidden in the fully symmetric potential $\pot_\mathrm{s}$, we consider here an invariance property that determines uniquely the traceless potential $\pot$, but fails to determine the combined potential resulting from adding to it a traceless component $\pot_\mathrm{t}$.
	In  \cite{gaeta:symmetries}, we studied  the action of the tetrahedron group $T_d$ on traceless type potentials $\pot$. Here, we discuss the action of the same on the particular potential $\pot_\mathrm{s}$ given by \eref{eq:psi3+1}.

We will use the notation of \cite{gaeta:symmetries}, in particular the representation of $T_d$ as a group of matrices acting in $\mathbb{R}^3$. It was shown there that the maximal subgroup $G$ of $T_d$ leaving the North pole fixed is made of the matrices $\{ M_1 , M_2 , M_3 , M_{13}, M_{14}, M_{15} \}$, in the notation adopted there. Here we will rename these as $M_1$-$M_6$, which  read explicitly as
	\begin{eqnarray}
	\fl\qquad	\cases{
		M_1 = \pmatrix{1&0&0\cr0&1&0\cr0&0&1},&$ M_2 =  \pmatrix{-1/2& \sqrt{3}/2& 0 \cr - \sqrt{3}/2 & -1/2& 0 \cr 0& 0& 1}$, \\
	M_3  =  \pmatrix{-1/2& -\sqrt{3}/2& 0 \cr \sqrt{3}/2& - 1/2& 0 \cr 0& 0& 1}, 
	&$M_4 = \pmatrix{-1&0&0\cr0&1&0\cr0&0&1}$, \\  M_5  = \pmatrix{1/2& -\sqrt{3}/2& 0 \cr - \sqrt{3}/2 & -1/2& 0 \cr 0& 0& 1} , 
	&$M_6 =  \pmatrix{1/2& \sqrt{3}/2& 0 \cr \sqrt{3}/2& - 1/2& 0 \cr 0& 0& 1}$.
}
	\end{eqnarray}
	The commutation relation among these can be read from \cite{gaeta:symmetries}; the only nontrivial subgroup is $G_0 = \{ M_1,M_2,M_3\}$.
	
	It is a simple matter to check that for $\pot_\mathrm{s}$ in \eref{eq:psi3+1},
	\begin{equation}
		\pot_\mathrm{s}(M \xb) = \pot_\mathrm{s}(\xb ) \qquad \forall M \in G. 
	\end{equation}
	That is, $\pot_\mathrm{s}$ is $G$-invariant.
	One might wonder if the converse is also true, i.e., if the requirement of being $G$-invariant does uniquely select $\pot_\mathrm{s}$ in \eref{eq:psi3+1}.
	
	To discuss this matter, we start from the general expression for $\pot_\mathrm{s}$ in \eref{eq:octupolar_potential_s} and require
\begin{equation}
	\pot_\mathrm{s}(M_i \xb ) =  \pot_\mathrm{s} (\xb ),
\end{equation}
	for $i = 1,\dots,6$.
	Some elementary algebra shows that this amounts to enforcing the conditions
	\begin{equation}
		\cases{
			\a_0 = 0  , \quad \a_1 = 0  , \quad \b_1 = 0 , \quad \b_2 = 0,\\
		A_1 = 0  , \quad  A_2 = 0 , \quad A_3 = 3 (\a_3 + 2 \b_3).
	}
	\end{equation}
	However, by direct computation we see that this choice of parameters does \emph{not} make $\pot_\mathrm{s}$ in \eref{eq:octupolar_potential_s} agree with \eref{eq:psi3+1}, unless we set
	\begin{equation}
	\b_3 = - \frac16  \(2 \sqrt{2} \a_2 + \a_3 \)  , 
     \end{equation}
	which, incidentally, implies the third of \eref{eq:Acoeff}. 
	We thus conclude that the condition of $G$-invariance does \emph{not}  determine uniquely $\pot_\mathrm{s}$. 
	\begin{remark}
		The $G$-invariance condition  determines uniquely the traceless potential $\pot$, whereas the trace potential $\pot_\mathrm{t}$ is determined up to a multiplicative factor $A_3$.
	\end{remark}
	\begin{remark}
	It could also be mentioned that the degeneration of  values  at the critical points $p_2$-$p_4$ does not only apply to the whole potential $\pot_\mathrm{s}$, but also to its traceless and trace components separately, although these are not separately invariant.
	\end{remark}
	
\section{Other Approaches}\label{sec:other_approaches}
So far we have privileged descriptions of the properties of an octupolar tensor $\oct$ based upon the octupolar potential $\pot$ introduced in \eref{eq:proto_potential} and the several variants encountered above. Other approaches to these properties have been proposed in the literature. We devote this section to some of these, trying to establish connections with ours.

\subsection{Maxwell multipoles}\label{sec:maxwell}
This approach to octupolar tensors in rooted in Maxwell's multipole representation of spherical harmonics \cite[pp.\,179--214]{maxwell:treatise} (see also \cite[pp.\,514--522]{courant:methods}). Our account, phrased in a modern language, follows \cite{zou:maxwell} (see also \cite{dennis:canonical} for a broader perspective).
A theorem due to Sylvester \cite{sylvester:note} (alternative proofs of which can also be found in \cite{backus:geometrical} and \cite{zou:maxwell}) put Maxwell's method on a solid mathematical ground.

There is a one-to-one correspondence between a completely symmetric tensor $\Oct\in\tspace(r,\Space)$ and a homogeneous polynomial $P_r(\x)$ of degree $r$ in $\x\in\space$ with $\dim\Space=3$ as
\begin{equation}
	\label{eq:sylvester_polynomial}
	P_r(\x)=\sum_{i_1i_2\dots i_r=1}^3A_{i_1i_2\dots i_r}x_{i_1}x_{i_2}\dots x_{i_r}.
\end{equation} 
Sylvester's theorem says that, given a real homogeneous polynomial $P_r(\x)$ of degree $r\geqq2$, there are $r$ vectors $\av_1,\av_2,\dots,\av_r\in\Space$ and a real homogeneous polynomial $P_{r-2}(\x)$ of degree $r-2$ such that
\begin{equation}
	\label{eq:sylvester_theorem}
	P_r(\x)=\prod_{s=1}^r(\av_s\cdot\x)+(\x\cdot\x)P_{r-2}(\x).
\end{equation}
Building on classical results, Zou and Zheng \cite{zou:maxwell} proved that every tensor $\Dev^{(m)}$ in the decomposition of a fully symmetric tensor $\Oct$ in \eref{eq:A_decomposition_symmetric} can be represented as
\begin{equation}
	\label{eq:sylvester_symmetric_decomposition}
	\Dev^{(m)}=A_m\irr{\av_1\otimes\av_2\cdots\av_m},
\end{equation}
where, for $m\leqq r$, $A_m>0$ is a scalar and $\av_1,\av_2,\dots\av_m$ are vectors on the unit sphere $\sphere$ in $\Space$ determined uniquely by $\Dev^{(m)}$, to within a change of sign in an even number of them. 

The poles designated on $\sphere$ by these vectors are called Maxwell's \emph{multipoles}. The connection thus established between fully symmetric traceless tensors and spherical harmonics justifies calling \emph{harmonic} these tensors, as well as the decomposition in \eref{eq:A_decomposition} for a generic tensor. This connection is further explored in \cite{applequist:traceless}. Harmonic tensors also play a role in reconstructing the \emph{crystallite orientation function} for poly-crystalline materials \cite{roe:description,adams:group,guidi:tensorial}; for this topic the reader is referred to the comprehensive review of Man~\cite{man:crystallographic}, in particular, to Chapt.~17.
\begin{remark}
	\label{rmk:slippery}
	As shown in the early work of Backus~\cite{backus:geometrical}, the multipole representation of $\Oct$, by its very geometric interpretation, can be effective in identifying the symmetries of $\Oct$. For more recent contributions to the role played by harmonic decomposition of a tensor $\Oct$ in identifying all symmetry classes it may belong to, the reader is referred to the works \cite{baerheim:harmonic,baerheim:classification,forte:symmetry_1996,forte:symmetry,auffray:decomposition,auffray:geometrical}.
This is, however, a slippery terrain, as witnessed for example by the disagreement between \cite{geymonat:classes} and \cite{zou:symmetry}, which for the piezoelectric tensor (see \sref{sec:piezoelectricity}) found with different methods $14$ and $15$  symmetry classes, respectively.
\end{remark}   

When applied to $\irr{\Oct}\in\tspace(r,\Space)$, the representation formula in \eref{eq:sylvester_symmetric_decomposition} reads as
\begin{equation}
	\label{eq:sylvester_symmetric_decomposition_irreducible}
	\irr{\Oct}=A_r\irr{\av_1\otimes\av_2\cdots\av_r},
\end{equation}
which easily identifies both the number $N(r)$ of independent parameters needed to represent $\irr{\Oct}$ and all the invariants  allowed in an isotropic scalar-valued function of $\irr{\Oct}$.

Customarily, for $r\geqq2$ and $\dim\Space>2$, $N(r)$ is given by a combinatoric argument as the difference between two binomial coefficients (see, for example, \cite[p.\,56]{snider:irreducible}),
\begin{equation}
	\label{eq:N_combinatorial_formula}
	N(r)={r+2\choose r}-{r\choose r-2},
\end{equation}	
the first representing the number of symmetric arrangements (with repetitions)  of $r$ symbols out of a pool of $3$, and the second the number of symmetric arrangements (with repetitions)  of $r-2$ symbols out of the same pool (these latter corresponding to the number of traces). A simple calculation shows that $N(r)=2r+1$. The same conclusion is reached far more easily  from \eref{eq:sylvester_symmetric_decomposition_irreducible} by remarking that $2r$ parameters are needed to represent the vectors $\av_1,\av_2,\dots\av_r$ on $\sphere$ and one more for $A_r$.
\begin{remark}
	The case $\dim\Space=2$ is special. $N(r)$ is no longer given by \eref{eq:N_combinatorial_formula}, but $N(r)=2$ for all $r$. Moreover, \eref{eq:sylvester_symmetric_decomposition_irreducible} is replaced by
	\begin{equation}
		\label{eq:sylvester_symmetric_decomposition_n=2}
		\irr{\Oct}=A_r\irr{\underbrace{\ev\otimes\cdots\otimes\ev}_{r\ \mathrm{times}}},
	\end{equation}
where all vectors $\av_i$ are the same vector $\ev$ on the unit circle $\mathbb{S}^1$ (see \cite{virga:octupolar_2D} for an explicit construction of $\ev$ when $r=3$).
\end{remark}

Similarly, since $\irr{\Oct}$ is fully determined by $A_r$ and the multipoles $\av_1,\av_2,\dots,\av_r$, the classical theorem of Cauchy~\cite{cauchy:memoire} (see also \cite[p.\,29]{truesdell:non-linear}) for the representation of isotropic scalar-valued functions depending on a finite number of vectors requires that the complete list of invariants consists of $A_r$ and the following $r(r-1)/2$ scalars
\begin{equation}\label{eq:scalar_products}
\alpha_{ij}:=\av_i\cdot\av_j\quad 1\leqq i<j\leqq r.
\end{equation}	
Then, in the special case where $r=3$, the total number of scalar invariants of $\irr{\oct}$ is immediately seen to be $4$, in agreement with \cite{smith:isotropic} (see remark~\ref{rmk:invariants}). Although the multipole representation of $\irr{\Oct}$ in \eref{eq:sylvester_symmetric_decomposition_irreducible} can more easily determine the \emph{number} of invariants, their explicit identification may be more difficult.
\begin{remark}
	\label{rmk:invariants_multipole}
	The number of isotropic invariants for symmetric traceless tensors $\Dev^{(3)}$ and $\Dev^{(4)}$, of rank $3$ and $4$ in three space dimensions, were derived in \cite{smith:isotropic} and \cite{boehler:polynomial} from the determination of the appropriate integrity bases, and found to be $4$ and $9$, respectively. As shown in \cite{zou:maxwell}, the direct derivation of this number from the harmonic decomposition in \eref{eq:sylvester_symmetric_decomposition_irreducible} agrees with that in \cite{smith:isotropic} for $\Dev^{(3)}$, but it does \emph{not} with that in \cite{smith:isotropic,boehler:polynomial} for $\Dev^{(4)}$, as it would predict $7$ invariants for the latter instead of $9$. This suggests that the invariants in the integrity bases for $\Dev^{(4)}$ in \cite{smith:isotropic,boehler:polynomial} are  not independent. A table of other inconsistencies similar to this can be found in \cite{zou:maxwell}. A large number of studies are devoted to this issue (which is not central to our review). Some, such as \cite{vannucci:polar,ahmad:invariants}, are especially relevant to the mechanics of composite materials.
\end{remark} 

The octupolar potential $\pot$ has played a special role in our review. We now wish to show how $\pot$ would be expressed for the harmonic representation in \eref{eq:sylvester_symmetric_decomposition_irreducible} for an octupolar tensor $\irr{\oct}$ in three space dimensions. A direct computation based on \eref{eq:A_partial_symmetries_2_components} shows that
\begin{equation}
	\label{eq:harmonic_octupolar_potential}
	\fl\qquad\eqalign{\pot(\x)&=A_3\Big\{(\av_1\cdot\x)(\av_2\cdot\x)(\av_3\cdot\x)\\&-\frac15(\x\cdot\x)\big[(\av_1\cdot\x)(\av_2\cdot\av_3)+(\av_2\cdot\x)(\av_3\cdot\av_1)+(\av_3\cdot\x)(\av_1\cdot\av_2)\big]\Big\}.}
\end{equation}
It is instructive to see what becomes of $\pot$ in \eref{eq:harmonic_octupolar_potential} for special choices of the unit vectors $\av_1,\av_2,\av_3$ and how these special forms of $\pot$ relate to those described above in our analysis.

First, given a Cartesian frame $\framee$, we consider the case where $\av_1=\av_2=\av_3=\ev_3$. It follows from  \eref{eq:harmonic_octupolar_potential} that $\pot$ then reduces to 
\begin{equation}
	\label{eq:eq:harmonic_octupolar_potential_case_1}
	\pot(\x)=x_3\left(x_3^2-\frac32x_1^2-\frac32x_2^2 \right),
\end{equation}
where we have set $A_3=5/2$ so that $\pot(\ev_3)=1$. With this normalization, the polar plot of $\pot$ is just the same as the one in \fref{fig:centre_polar_plot}. 

\begin{figure}[h]
	\centering
	\includegraphics[width=.4\linewidth]{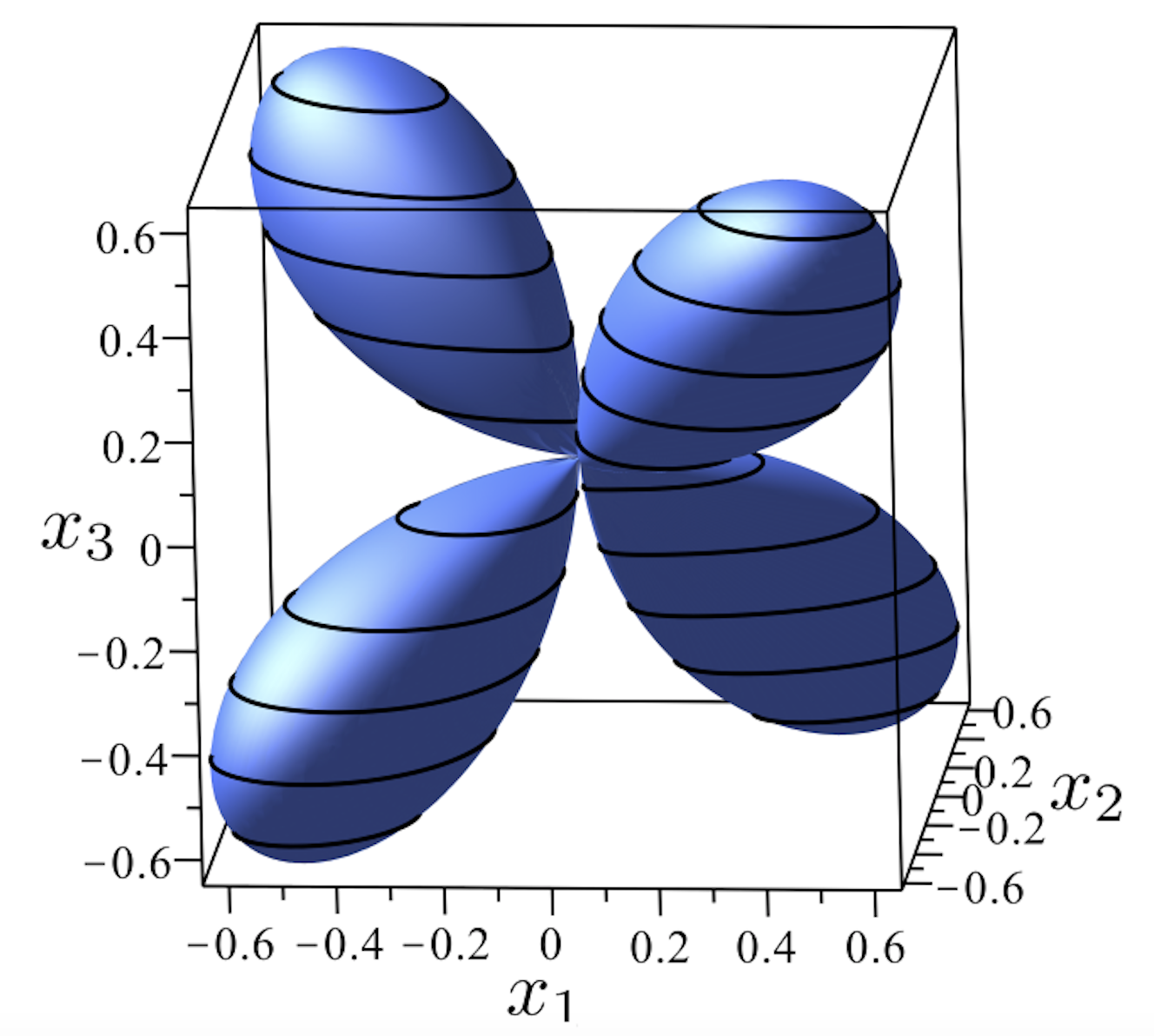}
	\caption{Polar plot of the octupolar potential in \eref{eq:harmonic_octupolar_potential_case_2}. Its maxima and minima fall on the tetrahedral vectors $\n_\alpha$ defined in \eref{eq:tetrahedral_vectors} and shown in \fref{fig:tetrahedron}.}
	\label{fig:tetrahedron_plot}
\end{figure}

If we now take $\av_i=\ev_i$, for $i=1,2,3$, \eref{eq:harmonic_octupolar_potential} delivers
\begin{equation}
	\label{eq:harmonic_octupolar_potential_case_2}
	\pot(\x)=3\sqrt{3}x_1x_2x_3,
\end{equation}
where we have chosen $A_3=3\sqrt{3}$ so that the maximum value of $\pot$ on $\sphere$ be $\pot=1$. Figure~\ref{fig:tetrahedron_plot}
illustrates the polar plot of the function in \eref{eq:harmonic_octupolar_potential_case_2} in the conventional representation adopted here: it only differs by a rigid rotation from the polar plot shown in \fref{fig:tetra_polar_plot} corresponding to the tetrahedral symmetry $T_d$ studied in \sref{sec:T_d}.

The minima and maxima of $\pot$ in \eref{eq:harmonic_octupolar_potential_case_2} are attained at the unit vectors $\n_\alpha$ defined in \eref{eq:tetrahedral_vectors} and illustrated in \fref{fig:tetrahedron}. This can also be seen by considering the \emph{tetrahedral} tensor
\begin{equation}
	\label{eq:tetrahedral_tensor}
	\mathbf{T}:=T\sum_{\alpha=1}^{4}\n_\alpha\otimes\n_\alpha\otimes\n_\alpha,
\end{equation}
with $T$ a normalizing scalar.
Since $\sum_{\alpha=1}^4\n_\alpha=\bm{0}$, $\mathbf{T}$ is a symmetric traceless octupolar tensor. The octupolar potential $\pot_\mathrm{T}$ associated with it is given by
\begin{eqnarray}
	\label{eq:tetrahedron_potential}
	\fl\qquad\pot_\mathrm{T}(\x)&:=\mathbf{T}\cdot(\xthree)=T\big[(\n_1\cdot\x)^3+(\n_2\cdot\x)^3+(\n_3\cdot\x)^3+(\n_4\cdot\x)^3\big]\nonumber\\\fl
	&=-\frac{8T}{\sqrt{3}}x_1x_2x_3,
\end{eqnarray}
which reduces to \eref{eq:harmonic_octupolar_potential_case_2} for $T=-9/8$. Comparing \eref{eq:harmonic_octupolar_potential_case_2} and \eref{eq:pot_tetra} would also be instructive.

\subsection{Curie potential}\label{sec:curie}
We have often said that the octupolar potential $\pot$, which has been our major tool in this review, identifies completely an octupolar tensor only if this is fully symmetric. Thus, for example, the octupolar potential associated with a piezoelectric tensor $\oct$, as defined in \sref{sec:piezoelectricity}, would fail to capture all its details. In particular, the definition of generalized eigenvalues and eigenvectors of $\oct$ given in \sref{sec:generalized} would be missed. A strategy has recently been developed in \cite{chen:c-eigenvalue,li:c-eigenvalues,li:c-eigenvalue,wang:new} to overcome this difficulty and to attempt at providing a similar treatment for both piezoelectric and fully symmetric octupolar tensors. Here, we briefly present this strategy, following mainly \cite{chen:c-eigenvalue}.

We start by defining the \emph{Curie} potential $\pot_\mathrm{C}$ of a piezoelectric tensor $\oct$ in three space dimensions,\footnote{The theory summarized in \cite{chen:c-eigenvalue} applies to a general piezoelectric tensor $\Oct\in\tspace(3,\Space)$ with $\dim\Space=n$. Here, we present the simplified version for $n=3$, as it is more germane to the rest of our analysis.}
\begin{equation}\label{eq:Curie_potential}
\pot_\mathrm{C}(\x,\y):=\oct\cdot\x\otimes\y\otimes\y,
\end{equation}	
which is a mapping $\pot_\mathrm{C}:\sphere\times\sphere\to\reals$. In components relative to a Cartesian frame $\framee$, equation \eref{eq:Curie_potential} reads as
\begin{equation}
	\label{eq:Curie_potential_components}
	\pot_\mathrm{C}(\x,\y)=A_{ijk}x_iy_jy_k.
\end{equation}
Reasoning as in \sref{sec:generalized} (see also equations \eref{eq:Phi_lambda} and \eref{eq:Phi_c}), the critical points of $\pot_\mathrm{C}$ on $\sphere\times\sphere$ can be viewed as critical points of the unconstrained potential 
\begin{eqnarray}
	\label{eq:Curie_potential_unconstrained}
	\pot_{\lambda,\mu}(\x,\y):=\pot_\mathrm{C}(\x,\y)-\frac12\lambda\left(x_1^2+x_2^2+x_3^2\right)-\mu\left(y_1^2+y_2^2+y_3^2\right),
\end{eqnarray}
where $\lambda$ and $\mu$ are independent Lagrange multipliers. Such unconstrained critical points are solutions of the following system of equations,
\begin{equation}
	\label{eq:C-eigenvalue_system}
	\cases{A_{ijk}y_jy_k=\lambda x_i,\\A_{ijk}x_iy_j=\mu y_k.}
\end{equation}
Multiplying by $x_i$ the first and by $y_k$ the second (summing over repeated indices), and enforcing the constraints that require both $\x$ and $\y$ on $\sphere$, we easily conclude that $\lambda=\mu$ and their common value is precisely the value of $\pot_\mathrm{C}$ at the corresponding critical point. A pair $(\x,\y)$ that solves \eref{eq:C-eigenvalue_system} is said to consist of a \emph{left} and a \emph{right C-eigenvector} of $\oct$, respectively, and $\lambda=\mu$ is the corresponding \emph{C-eigenvalue} (here the prefix \emph{C} stands for \emph{Curie}) \cite{chen:c-eigenvalue}.

A number of facts have been established for the C-eigenvalues of a piezoelectric octupolar tensor $\oct$ (see Theorems 2.3 and 2.5 of \cite{chen:c-eigenvalue}):
\begin{enumerate}[(I)]
	\item\label{item:chen_1} C-eigenvalues and associated left and right C-eigenvectors of $\oct$ do exist.
	\item\label{item:chen_2} If $\lambda$ is a C-eigenvalue of $\oct$ and $(\x,\y)$ are the corresponding left and right C-eigenvectors, then
	\begin{equation}\label{eq:item_chen_2}
		\oct\cdot\x\otimes\y\otimes\y=\lambda.
	\end{equation}	
Moreover, the triples $(\lambda,\x,-\y)$, $(-\lambda,-\x,\y)$, and $(-\lambda,-\x,-\y)$ also designate C-eigenvalues and corresponding C-eigenvectors of $\A$.
\item\label{item:chen_3} Let $\lambda_{1}$ denote the largest C-eigenvalue of $\oct$ and let $(\x_{1},\y_{1})$ denote the corresponding left and right C-eigenvectors. Then
\begin{equation}
	\label{eq:item_chen_3_1}
	\lambda_{1}=\max\{\oct\cdot\x\otimes\y\otimes\y:\x,\y\in\sphere\}.
\end{equation}
Moreover, $\lambda_{1}\x_{1}\otimes\y_{1}\otimes\y_{1}$ is the rank-one tensor that best approximates $\oct$, that is, it solves the following optimization problem,
\begin{equation}
	\label{eq:item_chen_3_2}
	\min\{\Vert\oct-\lambda\x\otimes\y\otimes\y\Vert^2:\lambda\in\mathbb{R},\ \x,\y\in\sphere\},
\end{equation}
where
\begin{equation}
	\label{eq:item_chen_3_3}
	\Vert\oct\Vert:=\sqrt{A_{ijk}A_{ijk}}
\end{equation}
designates the Frobenius norm.
\item \label{item:chen_4} If a  piezoelectric tensor $\oct$ has finitely many classes of C-eigenvalues in the complex field $\mathbb{C}$, their number counted with multiplicity is $13$.\footnote{This applies to $\oct\in\tspace(3,\Space)$ with $\dim\Space=3$. In general, for $\dim\Space=n$, this number is $(3^n-1)/2$, which is derived in \cite{che:C-eigenvalue} by an extension of \eref{eq:d_formula}.}
\end{enumerate}

Property \eref{item:chen_3} establishes a connection between C-eigenvalues and the best one-rank approximation of $\oct$. We can think of applying recursively this approximation algorithm, as suggested in \cite{zhang:rank-one} (where it was called \emph{incremental} rank-one approximation), so that the second iterate would deliver the best one-rank approximation $\lambda_2\x_2\otimes\y_2\otimes\y_2$ to $\oct_1:=\oct-\lambda_1\x_1\otimes\y_1\otimes\y_2$, and so on; the existence of C-eigenvalues established in \eref{item:chen_1} guarantees that this task can be accomplished at each step, up to the $p$-th iterate, when $\oct_p$  is itself rank-one.

According to the definition given in \cite{zhang:rank-one}, a piezoelectric tensor $\oct$ is said to be \emph{orthogonally decomposable} if it can be written as the following finite sum,
\begin{equation}
	\label{eq:orthogonally_decomposable_definition}
	\eqalign{\oct=&\sum_{i=i}^p\lambda_i\x_i\otimes\y_i\otimes\y_i,\ \lambda_i>0,\ \x_i,\y_i\in\sphere\quad\\&\mathrm{with}\ \x_i\cdot\x_j=\y_i\cdot\y_j=0\quad \forall i\neq j.}
\end{equation}
It was proved in \cite{zhang:rank-one} that an orthogonally decomposable tensor $\oct$ possesses a unique decomposition \eref{eq:orthogonally_decomposable_definition} and this is correctly identified by the incremental rank-one approximation algorithm (a different proof of this result can also be found in \cite{leurgans:decomposition}).

\begin{remark}
	\label{rmk:svd}
	The \emph{singular value decomposition} of a second-rank tensor $\bm{L}\in\tspace(2,\Space)$ with $\dim\Space=n$ amounts to represent it in the form
	\begin{equation}
		\label{eq:singular_value_decomposition}
		\bm{L}=\bm{USV}\trans,
	\end{equation}
where
\begin{equation}
	\label{eq:S_expansion}
	\bm{S}=\sum_{i=1}^n\sigma_i\ev_i\otimes\ev_i\quad\mathrm{with}\quad\sigma_i\geqq0\quad \mathrm{and}\quad \ev_i\cdot\ev_j=\delta_{ij}
\end{equation}
and $\bm{U}$, $\bm{V}$ are orthogonal tensors (such that $\bm{UU}\trans=\bm{VV}\trans=\bm{I}$). This result has a long history (neatly recounted in \cite{stewart:early}) that started with the works of Beltrami~\cite{beltrami:funzioni} and Jordan~\cite{jordan:memoire,jordan:reducion}. What makes it relevant to our topic is that \eref{eq:singular_value_decomposition} was also proved in \cite{eckart:approximation} as resulting from a rank-one approximation of $\bm{L}$, much in the same spirit as \eref{eq:orthogonally_decomposable_definition}, which could thus be seen as a possible extension of \eref{eq:singular_value_decomposition}.
\end{remark}
\begin{remark}
	\label{rmk:general_octupolar}
	The appropriate version of \eref{eq:orthogonally_decomposable_definition} valid for a generic orthogonally decomposable octupolar tensor $\oct$ is
	\begin{equation}
		\label{eq:orthogonally_decomposable_general_definition}
		\eqalign{\oct=&\sum_{i=i}^p\lambda_i\x_i\otimes\y_i\otimes\z_i,\ \lambda_i>0,\ \x_i,\y_i,\z_i\in\sphere\\ &\mathrm{with}\ \x_i\cdot\x_j=\y_i\cdot\y_j=\z_i\cdot\z_j=0\ \forall i\neq j.}
	\end{equation}
The applicability of the incremental rank-one approximation algorithm to establish \eref{eq:orthogonally_decomposable_general_definition} was also proved in \cite{zhang:rank-one}
\end{remark}
\begin{remark}
	Introducing for a general octupolar tensor $\oct$ the generalized potential
	\begin{equation}
		\label{eq:_octupolar_generalized_potential}
		\pot_\mathrm{G}(\x,\y,\z):=\oct\cdot\x\otimes\y\otimes\z,
	\end{equation}
one could easily justify the rank-one approximation algorithm delivering \eref{eq:orthogonally_decomposable_general_definition} as resulting from the search for the maximum of $\pot_\mathrm{G}$ over $\sphere\times\sphere\times\sphere$, which entrains a further generalized notion of eigenvalues and associated eigenvectors $(\lambda,\x,\y,\z)$ of $\oct$.
\end{remark}
\begin{remark}
	Even when the orthogonal decompositions in \eref{eq:orthogonally_decomposable_definition} and \eref{eq:orthogonally_decomposable_general_definition} do not apply, the one-rank approximation algorithm is still meaningful. In that case, the orthogonality conditions in both \eref{eq:orthogonally_decomposable_definition} and \eref{eq:orthogonally_decomposable_general_definition} fail to hold   and the decompositions formally delivered by these equations no longer represent $\oct$; they feature the best approximations to $\oct$ provided by its generalized eigenvalues and eigenvectors. 
\end{remark}

\section{Selected Applications}\label{sec:applications}
The applications of octupolar tensors in physics are countless. Apart from the specific fields that in \sref{sec:motivation} served as our motivation for this review, other fields have witnessed new or renewed formulations of theories that use octupolar (as well as higher-rank tensors). Here we give short accounts on just exemplary few of these fields, pausing longer on liquid crystal science, which is where our interest on the topic of this review originated.

\subsection{Gravitation}
In this context, octupolar tensors appear in the description of cubic-order spin effects in the dynamics of gravitational waves \cite{marsat:cubic}. Also, they feature in  computing invariants connected with tidal interactions that influence the late dynamics of compact binary systems, which have the potential of constituting the prime targets of a network of gravitational-wave detectors \cite{bini:gravitational}. 

\subsection{Spin states}
Majorana~\cite{majorana:atomi} introduced a geometrical picture to represent quantum states. In this representation, a pure spin-$j$ state is mapped onto $2j$ points on the unit sphere $\sphere$ (which is in this context is also called the \emph{Bloch sphere}). Recently, a generalization of this picture was proposed in \cite{giraud:tensor}, which applies to both pure and mixed spin-$j$ states; this extended representation employs a symmetric tensor of rank $2j$ in dimension $4$ (which is thus an octupolar tensor for fermions with $j=3/2$). Along the same lines, the reader will find it useful to consult the works \cite{giraud:classicality,bohnet-waldraff:tensor,qi:regularly}.

\subsection{Liquid crystals}
 In classical liquid crystal theory, the nematic director field $\vn$ describes the average orientation of the molecules that constitute the medium;  the elastic distortions of $\vn$ are locally measured by its gradient $\grad\vn$, which may become singular where the director exhibits \emph{defects} arising from a degradation of molecular order. The orientation of $\n$ should be physically indistinguishable from the orientation of $-\n$; this notion of invariance embodies the \emph{nematic} symmetry. In this short account we follow  \cite{pedrini:liquid}, to which the reader is referred for any further details.
 
 The two main descriptors, $\vn$ and $\grad\vn$, can be combined into  the third-rank octupolar tensor
\begin{equation}\label{eq:octupolar_tensor}
	\tn{A}:=\irr{\grad\vn\otimes\vn},
\end{equation}

It is worth noticing that $\oct$ defined in \eref{eq:octupolar_tensor} is invariant under the change of orientation of $\n$, and so it duly enjoys the nematic symmetry, which makes it a good candidate for measuring intrinsically the local distortions of a director field.

Selinger \cite{selinger:interpretation}, extending earlier work \cite{machon:umbilic},
suggested a new interpretation of the elastic modes for nematic liquid crystals described by the Oseen-Frank elastic free energy, which penalizes in a quadratic fashion the distortions of $\n$ away from any uniform state. The Oseen-Frank energy density $W_\mathrm{OF}$ is defined as (see, e.g., \cite[Chap.\,3]{deGennes:physics} and \cite[Chap.\,3]{virga:variational})
\begin{equation}\label{eq:frank_energy}
\eqalign{W_\mathrm{OF} &:= \frac{1}{2}K_{11}(\dv\vn)^{2} + \frac{1}{2}K_{22}(\vn\cdot\curl\vn)^{2} + \frac{1}{2}K_{33}|\vn\times\curl\vn|^{2} \\ &\ + K_{24}[\tr(\grad\vn)^{2}-(\dv\vn)^{2}],}
\end{equation}
where $K_{11}$, $K_{22}$, $K_{33}$, and $K_{24}$ are the \emph{splay}, \emph{twist}, \emph{bend}, and \emph{saddle-splay} constants,  respectively, each associated with a corresponding elastic mode.\footnote{The saddle-splay term  is a null Lagrangian \cite{ericksen:nilpotent} and an integration over the bulk  reduces it to a surface energy. Here, however, the surface-like nature of $K_{24}$ will not be exploited.} 

The decomposition of $W_\mathrm{OF}$ in independent elastic modes proposed in \cite{selinger:interpretation} is achieved through a new decomposition of $\gradn$. If we denote by $\tn{P}(\vn)$ and $\tn{W}(\vn)$ the projection onto the plane orthogonal to $\vn$ and the skew-symmetric tensor with axial vector $\vn$, respectively, then 
\begin{equation}\label{eq:grad_n}
	\grad\vn  = -\bend\otimes \vn + \frac{1}{2}T\tn{W}(\vn) + \frac{1}{2}S\tn{P}(\vn) + \bsplay,
\end{equation}
where $\bend := -(\grad\vn)\vn = \vn\times\curl\vn$ is the \emph{bend} vector, $T := \vn\cdot\curl\vn$ is the \emph{twist} (a pseudoscalar), $S := \dv\vn$ is the \emph{splay} (a scalar), and $\bsplay$ is a symmetric tensor such that $\bsplay\n=\bm{0}$ and $\tr\bsplay=0$. The properties of $\bsplay$ guarantee that when $\bsplay\neq\bm{0}$ it can be represented as
\begin{equation}\label{eq:biaxial_splay_reoresentation}
	\bsplay=q\left(\n_1\otimes\n_1-\n_2\otimes\n_2\right),
\end{equation}
where $q$ is the \emph{positive} eigenvalue of $\bsplay$. We shall call $q$ the \emph{octupolar splay} for a reason that shall soon be clear. The choice of sign for $q$ identifies (to within the orientation) the eigenvectors $\n_1$ and $\n_2$ of $\bsplay$ orthogonal to $\n$. Since $\tr\bsplay^2=2q^2$, we easily obtain from \eref{eq:grad_n} that
\begin{equation}\label{eq:q_formula}
	2q^2 = \tr(\grad\vn)^{2} + \frac{1}{2}T^{2} - \frac{1}{2}S^{2}.
\end{equation}
$W_\mathrm{OF}$ can then be given the form
\begin{equation}\label{eq:frank_energy_selinger}
	W_\mathrm{OF}= \frac{1}{2}(K_{11}-K_{24})S^{2} + \frac{1}{2}(K_{22}-K_{24})T^{2} + \frac{1}{2}K_{33}b^2 + K_{24}(2q^2),
\end{equation}
where all quadratic contributions are independent from one another.

The first advantage of such an expression is that it explicitly shows when the free energy  is positive semi-definite; this is the case when the following inequalities, due to  Ericksen \cite{ericksen:inequalities}, are satisfied,
\begin{equation}\label{eq:ericksen}
	K_{11} \geqq K_{24} \geqq 0,
	\quad
	K_{22} \geqq K_{24} \geqq 0,
	\quad
	K_{33} \geqq 0.
\end{equation} 
Whenever $q>0$, the frame $\dframe$ is identified to within a change of sign in either $\n_1$ or $\n_2$; requiring that $\n=\n_1\times\n_2$, we reduce this ambiguity to a simultaneous change in the orientation of $\n_1$ and $\n_2$. In this frame, 
\begin{equation}
	\label{eq:P_and_W}
	\tn{P}(\vn)=\tn{I}-\vn\otimes\vn\quad \mathrm{and}\quad \tn{W}(\vn)=\n_2\otimes\vo-\vo\otimes\n_2.
\end{equation}
Since $\bend\cdot\n\equiv0$, we can represent $\bend$ as $\bend=b_1\n_1+b_2\n_2$. The frame $\dframe$ is called the \emph{distortion frame} and $(S,T,b_1,b_2,q)$ the \emph{distortion characteristics} of the director field $\n$ \cite{virga:uniform}. In terms of these,  \eref{eq:grad_n} can also be written as
\begin{equation}\label{eq:grad_in_basis}
	\fl\qquad\eqalign{\grad\vn  &= \left(\frac{S}{2}+q\right)\vo\otimes\vo + \left(\frac{S}{2}-q\right)\n_2\otimes\n_2  - b_{1}\vo\otimes\vn -  b_{2}\n_2\otimes\vn\\ &+\frac12T\left(\n_2\otimes\vo -\vo\otimes\n_2\right).}
\end{equation}
Both \eref{eq:grad_n} and \eref{eq:grad_in_basis} show an intrinsic
decomposition of $\grad\vn$ into four genuine bulk contributions, namely, bend,  splay, twist, and octupolar splay.

The octupolar tensor $\oct$ defined in \eref{eq:octupolar_tensor} revealed itself as a convenient tool to illustrate director distortions \cite{pedrini:liquid}.   Having, however, symmetrized $\oct$, we have implicitly renounced to represent $T$, so no sign of twist will be revealed by $\oct$. This is the only piece of lost information.
\begin{remark}$T$ is a measure of chirality, and so it cannot be associated with a symmetric tensor. By  forming the completely skew-symmetric part of $\gradn\otimes\n$, one would  obtain the tensor $-\frac16T\bm{\epsilon}$, where $\bm{\epsilon}$ is Ricci's alternator, the most general skew-symmetric, third-rank tensor in three dimensions.
\end{remark}

Letting $\bm{x}=x_{1}\vo+x_{2}\n_2+x_{3}\vn$ be a point on the unit sphere $\sphere$ referred to the distortion frame $\dframe$, with the aid of \eref{eq:grad_in_basis}, the octupolar potential $\pot$ defined by \eref{eq:octupolar_potential_definition} can be written for $\oct$ in \eref{eq:octupolar_tensor} as follows
\begin{equation}\label{eq:extended_potential}
	\eqalign{\pot(\bm{x}) &=\left(\frac{S}{2}+q\right)x_{1}^{2}x_{3}+ \left(\frac{S}{2}-q\right)x_{2}^{2}x_{3} -b_{1} x_{1}x_{3}^{2} - b_{2} x_{2}x_{3}^{2}  \\  &+ \frac{1}{5}\big(x_{1}^{2}+x_{2}^{2}+x_{3}^{2}\big)\big(b_{1} x_{1}+b_{2} x_{2}-Sx_{3}\big).}
\end{equation}
As expected,  $\pot$ does not depend on the twist $T$, but it does depend on the  octupolar splay $q$.

A thorough analysis of $\pot$ in \eref{eq:extended_potential} is performed in \cite{pedrini:liquid}. Here, we only describe  the  very special cases where one and only one elastic mode is exhibited. 

\subsubsection*{Splay.}
\begin{figure}[h]
	\centering
	\begin{subfigure}[b]{0.3\textwidth}
		\centering
		\includegraphics[width=\textwidth]{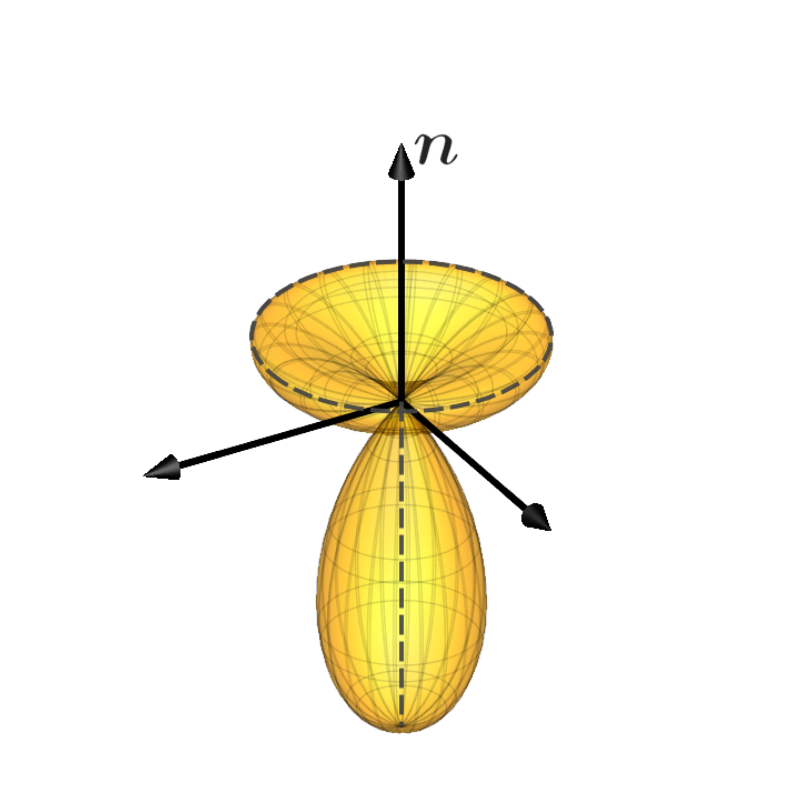}
		\caption{Pure splay}
		\label{fig:octu_s}
	\end{subfigure}
	$\quad$
	\begin{subfigure}[b]{0.3\textwidth}
		\centering
		\includegraphics[width=\textwidth]{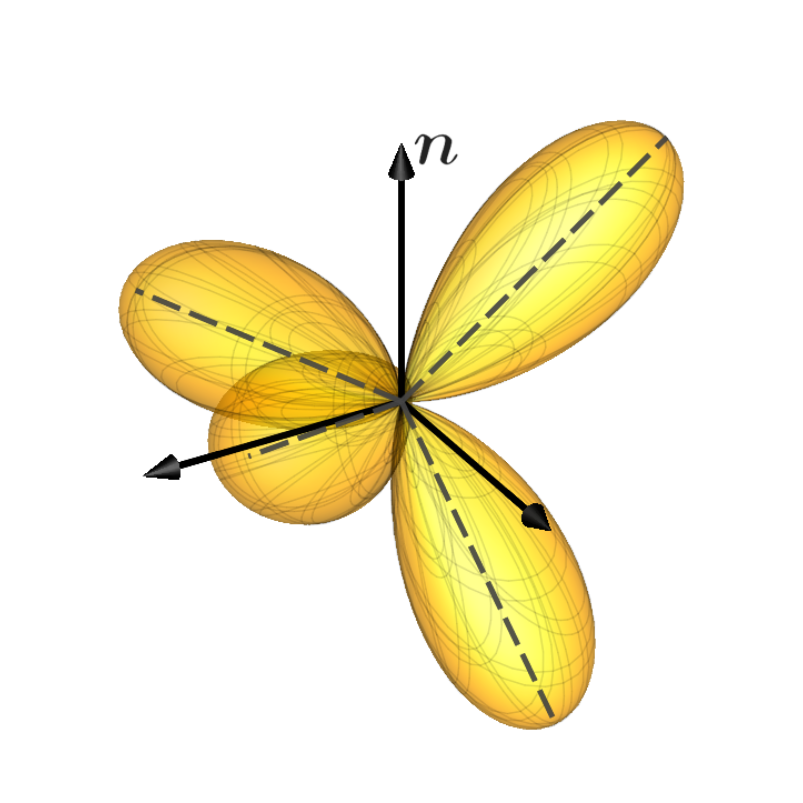}
		\caption{Pure octupolar splay}
		\label{fig:octu_d}
	\end{subfigure}
	$\quad$
	\begin{subfigure}[b]{0.3\textwidth}
		\centering
		\includegraphics[width=\textwidth]{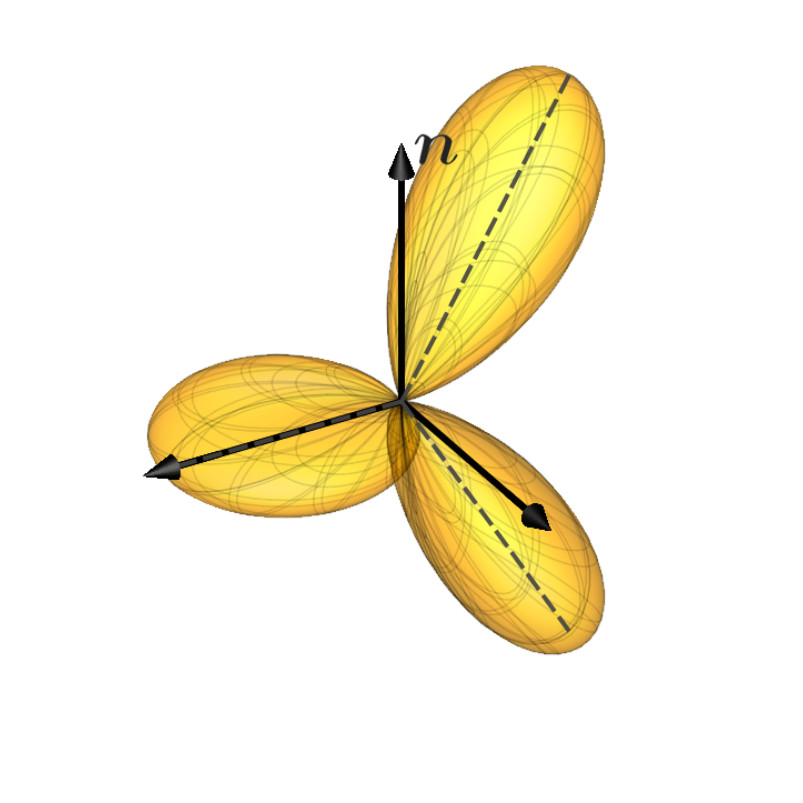}
		\caption{Pure bend}
		\label{fig:octu_b}
	\end{subfigure}
	\caption{Polar plots of the octupolar potential $\pot$ in \eref{eq:extended_potential} for pure elastic modes. Dashed lines are associated with maxima (and conjugated minima). Reprinted with permission from \cite{pedrini:liquid}.}
	\label{fig:pure_modes}
\end{figure}

When splay is the only active mode, the choice of $\vo$ and $\n_2$ in the plane orthogonal to $\vn$ is arbitrary. This fact reverberates in the symmetries of the octupolar potential and also in its critical points. In this case,
\begin{equation}
	\Phi(\bm{x}) = \frac{1}{10}S\big(3x_{1}^{2}x_{3} + 3x_{2}^{2}x_{3} - 2x_{3}^{3}\big).
\end{equation}
Graphically, $\pot(\bm{x})$ is depicted in \fref{fig:octu_s}, which is nothing but \fref{fig:centre_polar_plot}  turned upside down.

\subsubsection*{Octupolar splay.}
When both $S=0$ and $b=0$, but $q>0$, the potential $\pot$ reduces to
\begin{equation}
	\Phi(\bm{x}) = q(x_{1}^{2}-x_{2}^{2})x_{3}.
\end{equation}
Figure~\ref{fig:octu_d} shows that $\Phi(\bm{x})$ has four identical lobes, spatially distributed at the vertices of a regular tetrahedron; this is just the same plot as in figures~\ref{fig:tetra_polar_plot} and \ref{fig:tetrahedron_plot}, but differently oriented in the reference frame. Accordingly, its maxima are the four points
\begin{equation}
	\label{eq:four_points}
	\x_{1,2}=\frac{1}{\sqrt{3}}(\pm\sqrt{2}\vo + \vn)\quad\mathrm{and}\quad \x_{3,4}=\frac{1}{\sqrt{3}}(\pm\sqrt{2}\n_2 - \vn),
\end{equation} 
each with  value $2q/(3\sqrt{3})$.
\subsubsection*{Bend.}
For pure bend, we can choose $\vo$ and $\n_2$ such that $\bend=b\vo$ with $b>0$. Then the potential,
\begin{equation}
	\Phi(\bm{x}) = \frac{1}{5}bx_{1}\left(x_{1}^{2}+x_{2}^{2}-4x_{3}^{2}\right) = \frac{b}{5}x_{1}\left(1-5x_{3}^{2}\right),
\end{equation}
has three lobes: two larger, with equal  height $16b/(15\sqrt{15})$ at
\begin{equation}
\label{eq:two_points}
\x_{1,2}=\frac{1}{\sqrt{15}}\left(-2\vo\pm\sqrt{11}\vn\right),
\end{equation}
and one smaller at $\bm{x}_3=\vo$ with height $b/5$. As shown in \fref{fig:octu_b}, the polar plot of $\Phi$ is invariant under both a rotation by angle $\pi$ around $\vo$ and the mirror symmetry with respect to the plane containing $(\n_1,\n)$.
\begin{remark}\label{rmk:selinger_tensor}
	In a phenomenological theory for modulated nematic liquid crystal phase recently proposed in \cite{rosseto:modulated}, octupolar order plays a central role, as molecules are envisioned as stretched tetrahedra. Motivated in part by the properties of the distortion tensor $\tn{D}$ in \eref{eq:biaxial_splay_reoresentation}, the authors of this study describe octupolar order through a third-rank tensor, which in our formalism can be written as
	\begin{equation}
		\label{eq:octupolar_selinger}
		\oct=\bm{\Omega}\otimes\n,
	\end{equation}
where $\n$ is the nematic director and $\bm{\Omega}$ is a second-rank symmetric traceless tensor that annihilates $\n$. The tensor $\A$ in \eref{eq:octupolar_selinger} falls in yet another category of third-rank tensors, which we have not explicitly considered, but is amenable to the method outlined here. In three space dimensions, this tensor is represented by $4$ scalar parameters; it can be associated with the following octupolar potential on $\sphere\times\sphere$,
\begin{equation}
	\label{eq:potential_selinger}
	\pot_\mathrm{O}(\x,\y):=\oct\cdot\x\otimes\x\otimes\y=(\x\cdot\bm{\Omega}\x)(\n\cdot\y).
\end{equation}
\end{remark}

\section{Conclusion}\label{sec:conclusions}
Strictly speaking, an \emph{octupolar tensor} $\oct$ is a third-rank symmetric traceless tensor, which is also called a \emph{harmonic} tensor in some literature. There is an impressive body of works devoted to this special class of tensors and their application to diverse fields of physics. Here, we endeavoured to review some of these works in an attempt to broaden the scope where this specific mathematical tool can be placed. 

Not only have we considered fully symmetric tensors, but also partly symmetric ones and fully general tensors. Of course, the more general was the setting, the less simple were the results.

In the diverse territories we have traversed we found guidance in the unifying concept of \emph{octupolar potential} $\pot$, which, being a scalar-valued function representable on the unit sphere, added geometrical charm to a somewhat algid algebra.

Seeing diverse approaches displayed before us, a number of questions come naturally to mind, none necessarily with an easy answer. Many---we are sure---have already been heeded by the reader. Here, we mention just two of these, which have especially attracted our attention. 

First, one wonders whether there is a systematic way to relate the generalized eigenvectors of $\oct$ to its multipoles. Second, one would like to explore further the geometric properties enjoyed by the octupolar potential $\pot$ defined for a non-symmetric $\oct$.

We hope that these and other  issues  may be  addressed in the future as a result of our attempt to put octupolar tensors  within a unifying setting. We trust that practitioners from the diverse fields touched upon in this review may take even a modest advantage from the perspectives we have offered. Should  this be  the case, our effort would not have been completely in vain. 

\ack
We are grateful to Rebecca Gillan from IOP for having invited this review and for her patience in tolerating the long delays that this project has suffered from various interferences; her kind perseverance has been one of the major drives  for the completion of this work. Both authors are members of the Italian \emph{Gruppo Nazionale per la Fisica Matematica} (GNFM), an articulation of the Italian \emph{Istituto Nazionale di Alta Matematica} (INdAM). G.G. thanks the \emph{Santa Marinella Research Institute} (SMRI),
where his part of the present work was carried out.

\section*{References}

\end{document}